\def\gsim{\mathrel{\scriptstyle{\buildrel > \over \sim}}}

\documentclass[aps,prb,preprint,superscriptaddress]{revtex4}

\usepackage{amsmath}
\usepackage{epsfig,bm} 

\begin{document}

\title{Superconductivity by Hidden Spin Fluctuations in Electron-Doped Iron Selenide}

\author{J.P. Rodriguez}

\affiliation{Department of Physics and Astronomy,
California State University, Los Angeles, California 90032}

%\date{\today}

\begin{abstract}
Berg, Metlitski and Sachdev, Science {\bf 338}, 1606 (2012),
 have shown that the exchange of hidden spin fluctuations 
by conduction electrons with two orbitals
 can result in high-temperature superconductivity
 in copper-oxide materials.
We introduce a similar model for high-temperature iron-selenide superconductors
that are electron doped.
Conduction electrons carry the minimal $3 d_{xz}$ and $3 d_{yz}$ iron-atom orbitals.
Low-energy hidden spin fluctuations at the checkerboard wavevector ${\bm Q}_{\rm AF}$
result from nested Fermi surfaces
 at the center and at the corner of the unfolded (one-iron) Brillouin zone.
Magnetic frustration from super-exchange interactions via the selenium atoms
stabilize hidden spin fluctuations at ${\bm Q}_{\rm AF}$ versus true spin fluctuations.
At half filling,
Eliashberg theory based purely on the exchange of hidden spin fluctuations reveals
a Lifshitz transition to electron/hole Fermi surface pockets
 at the corner of the folded (two-iron) Brillouin zone,
but with vanishing spectral weights.
The underlying hidden spin-density wave groundstate is therefore a Mott insulator.
Upon electron doping,
Eliashberg theory finds that the spectral weights of the hole Fermi surface pockets remain vanishingly small,
while the spectral weights of the larger electron Fermi surface pockets become appreciable.
This prediction is therefore consistent with the observation of electron Fermi surface
pockets alone in electron-doped iron selenide 
by angle-resolved photoemission spectroscopy (ARPES).
Eliashberg theory also finds an instability to $S^{+-}$ superconductivity at electron doping,
with isotropic Cooper pairs that alternate in sign between the visible electron Fermi surface pockets
and the faint hole Fermi surface pockets.
Comparison with the isotropic energy gaps observed in electron-doped iron selenide
by ARPES and by scanning tunneling microscopy (STM) is
% made.
consistent with short-range hidden magnetic order.
\end{abstract}

\maketitle

\section{Introduction}
Electron-doped iron selenides
are perhaps the most interesting class of materials inside the
family of iron-based superconductors\cite{qian_11,xu_12,zheng_11,yu_11}.
% represent a class of materials
%that are among the most interesting 
%in condensed matter physics.
By contrast with bulk FeSe, which is a low-temperature superconductor,
a monolayer of FeSe on a doped strontium-titanate substrate becomes
a high-temperature superconductor\cite{xue_12}, with a critical temperature
in the range $40$-$50$ K\cite{zhang_14,deng_14},
and possibly higher\cite{ge_15}.
Angle-resolved photoemission spectroscopy (ARPES)
reveals that the substrate injects electrons into the FeSe monolayer
that bury the hole bands at the center of the Brillouin zone 
below the Fermi level\cite{liu_12}.
ARPES also reveals an energy gap at the remaining
electron-type Fermi surface pockets at the corner of the folded (two-iron) Brillouin zone\cite{peng_14,lee_14}.
It agrees with the energy gap found by
scanning tunneling microscopy (STM)\cite{xue_12,fan_15}.
On the other hand,
transport studies find perfect conductivity
below the critical temperature  where the gap opens in ARPES and in STM\cite{zhang_14,ge_15}.
These probes provide compelling evidence for a superconducting state at high temperature.
Electron doping of FeSe layers can also be achieved by other means,
such as by alkali-atom intercalation\cite{qian_11,xu_12,zheng_11,yu_11}, 
and by organic-molecule intercalation\cite{zhao_16,niu_15,yan_15}.
%by dosing with alkali atoms\cite{miyata_15,wen_16},
% and by the application of a gate voltage\cite{lei_16,hosono_16}.
These systems show the same Lifshitz transition of the Fermi surface topology,
where the hole bands are buried below the Fermi level at the $\Gamma$-point,
but where electron Fermi surface pockets at the corner of the folded Brillouin zone  remain.
These systems also show high-temperature superconductivity,
with an isotropic energy gap that opens at the electron-type Fermi surface pockets.

The coincidence of high-temperature superconductivity with the absence of Fermi surface nesting 
in electron-doped FeSe is puzzling.  
%By contrast,
%For example,
 High-temperature superconductivity
 in   iron-pnictide materials occurs only when
 the Fermi surfaces 
exhibit partial nesting,
for example.
In particular,
the end-member compound KFe$_2$As$_2$ of the series of iron-pnictide compounds\cite{budko_prb_13}
(Ba$_{1-x}$K$_x$)Fe$_2$As$_2$ shows hole-type Fermi surface pockets at the
center of the Brillouin zone,
but no electron Fermi surface pockets at the corner of the folded
Brillouin zone to nest with\cite{sato_prl_09}.  
It is a low-temperature superconductor with $T_c\cong 3$ K, however.
Early theoretical responses to the puzzle posed by electron-doped iron selenide
proposed a nodeless $D$-wave superconducting state\cite{maier_11,wang_11}, 
with a full gap over each electron pocket that alternates in sign between them.
In the case of bulk electron-doped iron-selenide materials,
it was argued that strong dispersion of the energy bands along the $c$ axis
results in inner and outer electron Fermi surface pockets
at the corner of the folded Brillouin zone
because of hybridization due to the two inequivalent iron sites,
and that true $D$-wave nodes appear as a result\cite{mazin_11}.
%after zone-folding the one-iron Brillouin zone
In the case of an isolated layer of FeSe,
the two electron Fermi surface pockets at the corner of the two-iron Brillouin zone cross,
but do not hybridize, because of glide-reflection symmetry\cite{Lee_Wen_08}.
Inner and outer electron Fermi surface pockets due to hybridization are predicted when
strong enough spin-orbit coupling is included, however,
%\cite{cvetkovic_vafek_13}
and true $D$-wave nodes are again predicted\cite{agterberg_17,eugenio_vafek_18}.
These considerations led to
% another proposed gap symmetry for electron-doped iron selenide called 
the proposal of the anti-bonding $S^{+-}$ state.
%  It is an $S$-wave state on
It is an isotropic Cooper pair state
that alternates in sign between  the inner and the outer
electron Fermi surfaces at the corner of the folded (two-iron) Brillouin zone
\cite{mazin_11,khodas_chubukov_12}.
ARPES finds no sign of nodes in the gap and no sign of hybridization on 
the electron Fermi surface pockets\cite{peng_14,lee_14}, however.
Further, STM and the dependence of the specific heat and of nuclear magnetic resonance (NMR) on temperature
are consistent with a gap over the entire
Fermi surface\cite{zheng_11,yu_11,xue_12}.
These measurements could then rule out the nodeless $D$-wave state and the anti-bonding $S^{+-}$ state
in high-$T_c$  iron-selenide superconductors.

Below, we will show
that a spin-fermion model over the square lattice
that is similar to that introduced by Berg, Metlitski and Sachdev
in the context of copper-oxide high-temperature superconductors\cite{BMS_12}
harbors an alternative solution to the puzzling isotropic gap shown by electron-doped iron selenide.
The non-interacting electrons are in the principal
$3d_{xz}$ and $3d_{yz}$ iron orbitals,
and they form a semi-metallic Fermi surface that is  nested  by the checkerboard wavevector
${\bm Q}_{\rm AF} = (\pi/a, \pi/a)$.
The latter can result in hidden magnetic order
% nearby in the phase diagram
 when magnetic frustration is present,
at weak enough Hund's Rule coupling\cite{jpr_rm_18}.
Such hidden antiferromagnetism is characterized by the most symmetric  of three possible order parameters
for a hidden spin density wave (hSDW):
$${1\over{2{\cal N}}}\sum_i e^{i {\bm Q}_{\rm AF}\cdot{\bm r}_i}
\sum_{s=\uparrow,\downarrow} ({\rm sgn}\, s)
i\langle c_{i,d_{xz},s}^{\dagger} c_{i,d_{yz},s} - c_{i,d_{yz},s}^{\dagger} c_{i,d_{xz},s}\rangle,$$
where ${\cal N}$ is the number of site-orbitals.
It is isotropic with respect to rotations of the orbitals about the $z$ axis,
and it therefore does not couple to nematicity. (Cf. refs. \cite{xu_muller_sachdev_08,kang_fernandez_16}.)
Based on the interaction of such fermions with the corresponding hidden spin fluctuations,
an Eliashberg theory for $S$-wave pairing over the two bands of electrons is developed.
Hopping matrix elements are chosen so that 
perfect nesting exists 
at half filling\cite{jpr_rm_18}.
As the interaction grows strong,
we find ({\it i}) a Lifshitz transition to electron-type and hole-type Fermi surface pockets at the corner
of the folded (two-iron) Brillouin zone.
The new Fermi surfaces remain perfectly nested by ${\bm Q}_{\rm AF}$,
but their spectral weights are vanishingly small due to strong wavefunction renormalization.
Upon electron doping,
the hole Fermi surfaces remain faint,
while the electron Fermi surfaces become visible because of only moderate wavefunction renormalization.
This prediction agrees with previous calculations based on the corresponding local-moment model\cite{jpr_17}.
The Eliashberg theory also reveals ({\it ii}) an instability at the renormalized Fermi surface
to $S^{+-}$ pairing that alternates in sign between the visible electron-type Fermi surfaces
and the faint hole-type Fermi surfaces.
We  shall now provide details of how such hidden $S^{+-}$ superconductivity emerges 
from the two-band Eliashberg theory.

\section{Bare Nested Fermi Surfaces and Hidden Spin Fluctuations}
The conditions for perfect nesting  of the Fermi surfaces
in two-orbital hopping Hamiltonians for iron selenide 
will be determined in what follows\cite{jpr_rm_18}.
Also,
the space of hSWD states generated by rotations of the two isospin degrees of freedom,
the $d+$ and $d-$ orbitals, will be discussed.

\subsection{Electron Hopping}
The electronic kinetic energy is governed by the hopping Hamiltonian
\begin{equation}
H_{\rm hop} =
-\sum_{\langle i,j \rangle} (t_1^{\alpha,\beta} c_{i, \alpha,s}^{\dagger} c_{j,\beta,s} + {\rm h.c.})
-\sum_{\langle\langle  i,j \rangle\rangle} (t_2^{\alpha,\beta} c_{i, \alpha,s}^{\dagger} c_{j,\beta,s} + {\rm h.c.}),
\label{hop}
\end{equation}
where  the repeated indices $\alpha$ and $\beta$ are summed over the iron ${d+}$ and ${d-}$ orbitals,
where the repeated index $s$ is summed over electron spin,
and where $\langle i,j\rangle$ and $\langle\langle i,j\rangle\rangle$
represent nearest neighbor (1) and next-nearest neighbor (2) links on the
square lattice of iron atoms.
Above, $c_{i, \alpha,s}$ and $c_{i, \alpha,s}^{\dagger}$
denote annihilation and creation operators for an
electron of spin $s$ in orbital $\alpha$ at site $i$.
We keep only the $3d_{xz}$ and $3d_{yz}$ orbitals of the iron atoms,
which are the principal ones in iron selenide.
In particular,
let us work in the isotropic basis of orbitals
$d-=(d_{xz}-id_{yz})/{\sqrt 2}$ and $d+=(d_{xz}+id_{yz})/{\sqrt 2}$.
The reflection symmetries shown by a single layer of iron selenide imply
that the above intra-orbital and inter-orbital hopping matrix elements
show $s$-wave and $d$-wave symmetry, respectively\cite{Lee_Wen_08,raghu_08,jpr_mana_pds_14}.
In particular,
nearest neighbor hopping matrix elements satisfy
\begin{eqnarray}
t_1^{\pm \pm} ({\hat {\bm x}}) &=& t_1^{\parallel} = t_1^{\pm \pm} ({\hat {\bm y}}) \nonumber\\
t_1^{\pm\mp} ({\hat {\bm x}}) &=& t_1^{\perp} = -t_1^{\pm \mp} ({\hat {\bm y}}),
\label{t1}
\end{eqnarray}
where $t_1^{\parallel}$ and $t_1^{\perp}$ are real,
while next-nearest neighbor hopping matrix elements satisfy
\begin{eqnarray}
t_2^{\pm \pm} ({\hat {\bm x}}+{\hat {\bm y}}) = \; t_2^{\parallel} &=& t_2^{\pm \pm} ({\hat {\bm y}}-{\hat {\bm x}}) \nonumber\\
t_2^{\pm \mp} ({\hat {\bm x}}+{\hat {\bm y}}) = \pm t_2^{\perp} &=& -t_2^{\pm \mp} ({\hat {\bm y}}-{\hat {\bm x}}),
\label{t2}
\end{eqnarray}
where $t_2^{\parallel}$ is real, and where $t_2^{\perp}$ is pure imaginary.

\begin{figure}
\hspace*{1cm}
\includegraphics[scale=1.00, angle=0]{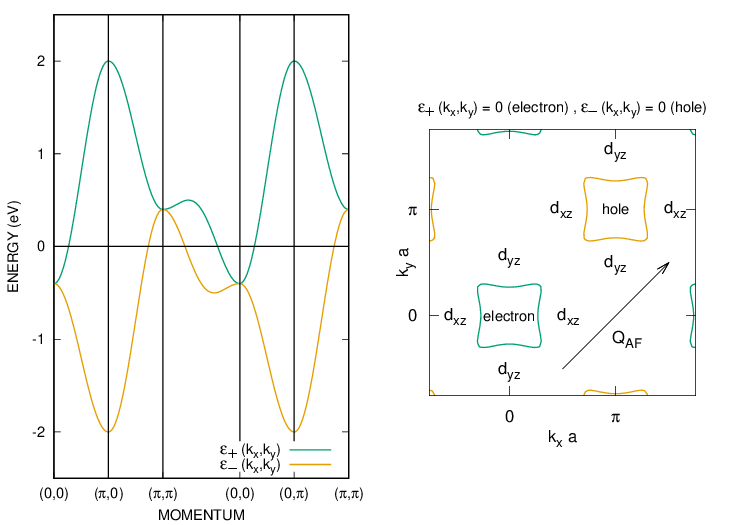}
\caption{Band structure with perfectly nested Fermi surfaces at half filling:
$\varepsilon_+({\bm k}) = 0$ and  $\varepsilon_-({\bm k}) = 0$,
with hopping matrix elements
$t_1^{\parallel} = 100$ meV, $t_1^{\perp} = 500$ meV, $t_2^{\parallel} = 0$,
and $t_2^{\perp} / i = 100$ meV.} 
\label{FS0}
\end{figure}

The above hopping Hamiltonian $H_{\rm hop}$ then has
intra-orbital and inter-orbital matrix elements 
%
%%\begin{eqnarray}
\begin{subequations}
\begin{align}
\label{mtrx_lmnt_a}
\varepsilon_{\parallel}({\bm k}) =& -2 t_1^{\parallel} (\cos k_x a + \cos k_y a)
-2 t_2^{\parallel} (\cos k_+ a + \cos k_- a) \\
\label{mtrx_lmnt_b}
\varepsilon_{\perp}({\bm k}) =& -2 t_1^{\perp} (\cos k_x a - \cos k_y a)
-2 t_2^{\perp} (\cos k_+ a - \cos k_- a)
\end{align}
\end{subequations}
%%\end{eqnarray}
%
with  $k_{\pm} = k_x \pm k_y$.
It is easily diagonalized by
plane waves of $d_{x(\delta)z}$ and $i d_{y(\delta)z}$ orbitals that are rotated
with respect to the principal axes by an angle $\delta({\bm k})$:
\begin{eqnarray}
|{\bm k}, d_{x(\delta)z}\rangle\rangle &=&
{\cal N}^{-1/2} \sum_i e^{i{\bm k}\cdot{\bm r}_i}
[e^{i\delta({\bm k})} |i, d+\rangle + e^{-i\delta({\bm k})} |i, d-\rangle], \nonumber\\
i|{\bm k}, d_{y(\delta)z}\rangle\rangle &=&
{\cal N}^{-1/2} \sum_i e^{i{\bm k}\cdot{\bm r}_i}
[e^{i\delta({\bm k})} |i, d+\rangle - e^{-i\delta({\bm k})} |i, d-\rangle],
\label{plane_waves}
\end{eqnarray}
where ${\cal N} = 2 N_{\rm Fe}$ is the number of iron site-orbitals.
The phase shift $\delta({\bm k})$ is set by
%%
%\begin{equation}
$\varepsilon_{\perp}({\bm k}) = |\varepsilon_{\perp}({\bm k})| e^{i 2 \delta({\bm k})}$.
%\label{phase_shift}
%\end{equation}
%%
Specifically,
%
%%\begin{eqnarray}
\begin{subequations}
\begin{align}
\label{c_2dlt}
\cos\,2\delta({\bm k}) =& {-t_1^{\perp}(\cos\, k_x a - \cos\, k_y a)\over
{\sqrt{t_1^{\perp 2}(\cos\, k_x a - \cos\, k_y a)^2 +
|2 t_2^{\perp}|^2 (\sin\, k_x a)^2 (\sin\, k_y a)^2}}}, \\
\label{s_2dlt}
\sin\,2\delta({\bm k}) =& {2 (t_2^{\perp} / i)(\sin\, k_x a) (\sin\, k_y a)\over
{\sqrt{t_1^{\perp 2}(\cos\, k_x a - \cos\, k_y a)^2 +
|2 t_2^{\perp}|^2 (\sin\, k_x a)^2 (\sin\, k_y a)^2}}}.
\end{align}
\end{subequations}
%%\end{eqnarray}
%
The phase shift is notably singular at ${\bm k} = 0$ and ${\bm Q}_{\rm AF} = (\pi/a,\pi/a)$,
where the matrix element $\varepsilon_{\perp}({\bm k})$ vanishes.
The energy eigenvalues  of the bonding ($+$) and anti-bonding ($-$) plane waves (\ref{plane_waves})
are respectively given by
$\varepsilon_+({\bm k}) = \varepsilon_{\parallel}({\bm k}) + |\varepsilon_{\perp}({\bm k})|$ and
$\varepsilon_-({\bm k}) = \varepsilon_{\parallel}({\bm k}) - |\varepsilon_{\perp}({\bm k})|$.

\begin{figure}
%\hspace*{1cm}
\includegraphics[scale=1.00, angle=0]{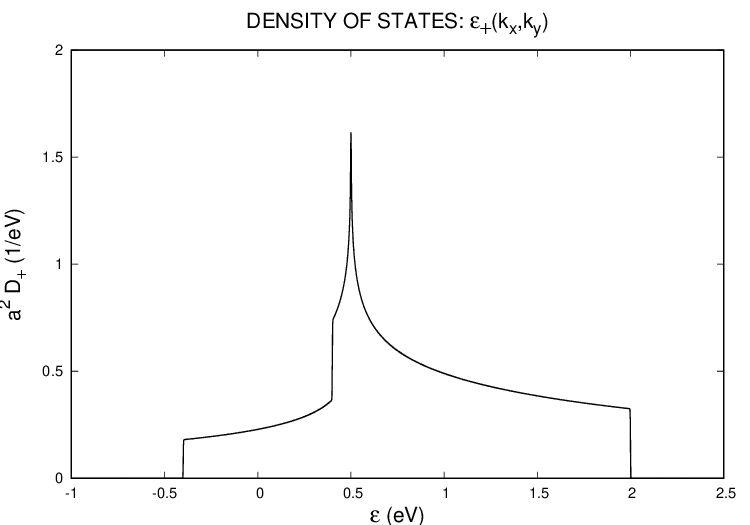}
\caption{Density of states of the bonding band evaluated numerically
at hopping parameters that are listed in the caption to Fig. \ref{FS0}:
$a^2 D_+(\varepsilon) = N_{\rm Fe}^{-1}\sum_{\bm k} \delta[\varepsilon-\varepsilon_+({\bm k})]$.
The unfolded (one-iron) Brillouin zone is divided into a $10,000\times 10,000$ grid,
while the $\delta$-function is approximated by
$(4 k_B T_0)^{-1} {\rm sech}^2 (\varepsilon/2 k_B T_0)$.
Here, $k_B T_0$ is $3$ parts in $10,000$ of the bandwidth.}
\label{DoS}
\end{figure}

Henceforth, we shall turn off next-nearest neighbor intra-orbital hopping: $t_2^{\parallel} = 0$.
Notice that the above energy bands now satisfy the perfect nesting condition
\begin{equation}
\varepsilon_{\pm}({\bm k}+{\bm Q}_{\rm AF}) = - \varepsilon_{\mp}({\bm k}),
\label{prfct_nstng}
\end{equation}
where ${\bm Q}_{\rm AF} = (\pi/a,\pi/a)$ is the wavevector
for the checkerboard on the square lattice of iron atoms.
The Fermi level at half filling therefore lies at $\epsilon_{\rm F} = 0$.
Figure \ref{FS0} displays
perfectly nested electron-type and hole-type Fermi surfaces for hopping
parameters $t_1^{\parallel} = 100$ meV, $t_1^{\perp} = 500$ meV, $t_2^{\parallel} = 0$
and $t_2^{\perp} / i=  100$ meV.
Figure \ref{DoS} shows the density of states of the bonding ($+$) band.
%The sharp peak marks a transition point in the topology of the constant-energy contours. (CHECK!!!)

%

\subsection{Extended Hubbard Model}\label{hubbard_model}\label{xtndd_hbbrd_mdl}
The Hamiltonian of the underlying extended Hubbard model\cite{jpr_rm_18}  has three parts:
$H = H_{\rm hop} + H_U + H_{\rm sprx}$.
On-site Coulomb repulsion is counted by the second term\cite{2orb_Hbbrd},
\begin{eqnarray}
H_U =  \sum_i &&[U_0 n_{i,\alpha,\uparrow} n_{i,\alpha,\downarrow}
                +J_0 {\bm S}_{i, d-}\cdot {\bm S}_{i, d+} \nonumber \\
                &&+U_0^{\prime} n_{i,d+} n_{i,d-}
                +J_0^{\prime} (c_{i,d+,\uparrow}^{\dagger}c_{i,d+,\downarrow}^{\dagger}
                            c_{i,d-,\downarrow}c_{i,d-,\uparrow}+ {\rm h.c.})].
\label{U}
\end{eqnarray}
where $n_{i,\alpha,s}$ is the occupation operator,
and where $n_{i,\alpha} = n_{i,\alpha,\uparrow} + n_{i,\alpha,\downarrow}$.
Also, ${\bm S}_{i,\alpha}$ is the spin operator.
Above, $U_0>0$ is the intra-orbital on-site Coulomb repulsion energy,
while $U_0^{\prime} > 0$ is the inter-orbital one.
It is worth pointing out the following expression for the sum of 
these two on-site repulsion terms in (\ref{U}):
\begin{equation}
U_0 n_{i,\alpha,\uparrow} n_{i,\alpha,\downarrow} +
U_0^{\prime} n_{i,d+} n_{i,d-} =
(U_0 - U_0^{\prime}) n_{i,\alpha,\uparrow} n_{i,\alpha,\downarrow} +
{1\over 2} U_0^{\prime}\, n_i (n_i -1) ,
\label{on-site_rplsn}
\end{equation}
where $n_i = n_{i,d+} + n_{i,d-}$ is the net occupation per iron site $i$.
Above also,
$J_0$ is the Hund's Rule exchange coupling constant,
which has a ferromagnetic (negative) sign,
while $J_0^{\prime}$ is the matrix element for on-site Josephson tunneling between orbitals.
The third and last term in the Hamiltonian represents super-exchange interactions
among the iron spins via the selenium atoms:
\begin{eqnarray}
H_{\rm sprx} =
\sum_{\langle i,j \rangle} && J_1^{({\rm sprx})} ({\bm S}_{i, d-} + {\bm S}_{i, d+})
\cdot ({\bm S}_{j, d-} + {\bm S}_{j, d+}) \nonumber \\
&&+\sum_{\langle\langle  i,j \rangle\rangle} J_2^{({\rm sprx})} ({\bm S}_{i, d-} + {\bm S}_{i, d+})
\cdot ({\bm S}_{j, d-} + {\bm S}_{j, d+}).
\label{sprx}
\end{eqnarray}
Above, $J_1^{({\rm sprx})}$ and $J_2^{({\rm sprx})}$ are positive
super-exchange coupling constants
over nearest neighbor and next-nearest neighbor iron sites.

\begin{table}
%\hspace*{3cm}
\begin{tabular}{|c|c|c|}
\hline
$I^{(3)}$ $|$ $S^{(z)}/\hbar$ & isospin triplet ($I=1$), spin singlet ($S=0$) &  isospin singlet ($I=0$), spin triplet ($S=1$) \\
\hline
$+1$ & ${d+}_1 {d+}_2 {1\over{\sqrt{2}}} (\uparrow_1\downarrow_2-\downarrow_1\uparrow_2)$
     & ${1\over{\sqrt{2}}} ({d+}_1 {d-}_2 - {d-}_1 {d+}_2) \uparrow_1\uparrow_2$ \\
 $0$ & ${1\over{\sqrt{2}}} ({d+}_1 {d-}_2 + {d-}_1 {d+}_2) 
        {1\over{\sqrt{2}}} (\uparrow_1\downarrow_2-\downarrow_1\uparrow_2)$ 
     & ${1\over{\sqrt{2}}} ({d+}_1 {d-}_2 - {d-}_1 d+_2)
        {1\over{\sqrt{2}}} (\uparrow_1\downarrow_2+\downarrow_1\uparrow_2)$ \\
$-1$ & ${d-}_1 {d-}_2 {1\over{\sqrt{2}}} (\uparrow_1\downarrow_2-\downarrow_1\uparrow_2)$
     & ${1\over{\sqrt{2}}} ({d+}_1 {d-}_2 - {d-}_1 {d+}_2) \downarrow_1\downarrow_2$ \\
\hline
\end{tabular}
\caption{Atomic iron states at half filling. The isospin triplet states violate Hund's Rule,
while the isospin singlet states obey Hund's Rule.}
\label{atomic_states}
\end{table}

It is instructive to uncover the energy spectrum of
the Hamiltonian $H_U$ at a single iron site $i$, at half filling with two electrons.
Table \ref{atomic_states} gives the corresponding six-dimensional Hilbert space
in the singlet-triplet/spin-isospin basis.
Here, the $d+$ and the $d-$ orbitals comprise the isospin-1/2 states.
Specifically,
the isospin operators
along the axes $n=1,2,3$
 for a single electron have the form
$I^{(n)} = +{1\over 2}|u_n\rangle\langle u_n|- {1\over 2}|v_n\rangle\langle v_n|$,
with orbitals 
$(u_1, v_1) = (d_{xz}, d_{yz})$, 
$(u_2, v_2) = (d_{x^{\prime}z}, d_{y^{\prime}z})$, and
$(u_3, v_3) = ({d+}, {d-})$.
Here, $x^{\prime} = (x+y)/\sqrt{2}$ and $y^{\prime} = (-x+y)/\sqrt{2}$.
The eigenstates of (\ref{U}) at a single iron site are the product of spin ($S=1$) triplet states
with the isospin ($I=0$) singlet state
\begin{equation}
\phi_0(1,2) = {1\over{\sqrt{2}}} [u_n(1)v_n(2)-v_n(1)u_n(2)],
\label{iso_singlet}
\end{equation}
%
%with $u=d+$ and $v=d-$,
and the product of the spin ($S=0$) singlet state
with the isospin ($I=1$) triplet states
\begin{equation}
\phi_n(1,2) = {1\over{\sqrt{2}}} [u_n(1)v_n(2)+v_n(1)u_n(2)], \quad n = 1, 2, 3.
\label{iso_triplet}
\end{equation}
%
%%
%%%\begin{eqnarray}
%\begin{subequations}
%\begin{align}
%\label{iso_triplet_1}
%\phi_1(1,2) &= {1\over{\sqrt{2}}} [d_{xz}(1) d_{yz}(2) + d_{yz}(1) d_{xz}(2)] , \\
%\label{iso_triplet_2}
%\phi_2(1,2) &= {1\over{\sqrt{2}}} [d_{x^{\prime}z}(1) d_{y^{\prime}z}(2) +
%                                    d_{y^{\prime}z}(1) d_{x^{\prime}z}(2)] , \\
%\label{iso_triplet_3}
%\phi_3(1,2) &= {1\over{\sqrt{2}}} [{d+}(1) {d-}(2) + {d-}(1) {d+}(2)] .
%\end{align}
%\end{subequations}
%%%\end{eqnarray}
%%
Recall that the isospin singlet pair state $\phi_0(1,2)$ is {\it unique} up to a phase factor.
The orbital pair states (\ref{iso_singlet}) and (\ref{iso_triplet}) above
% carry isospin $I=1$, on the other hand, and they
satisfy $I^{(n)} \phi_0 = 0$ and $I^{(n)} \phi_n = 0$, where $I^{(n)} = I^{(n)}(1) + I^{(n)}(2)$.
% are respectively null after the application of the isospin operators $I^{(n)}$.
% $I^{(1)}$, $I^{(2)}$, and $I^{(3)}$.
%(See Appendix \ref{ppndx_isospin_operator}.)
And why do the pair states (\ref{iso_singlet}) and (\ref{iso_triplet})
%(\ref{iso_triplet_1})-(\ref{iso_triplet_3})
make up the energy spectrum of $H_U$?
First, observe that the spin singlet and spin triplet states listed in Table \ref{atomic_states}
are all eigenstates of the Hund's Rule
% (second)
 term in (\ref{U}),
with energy splitting $E^{(0)}_{\rm singlet} - E^{(0)}_{\rm triplet} = -J_0$.
Second, notice that all six pair  states listed in Table \ref{atomic_states}
are eigenstates of the sum (\ref{on-site_rplsn})
 of the intra-orbital and inter-orbital on-site repulsion 
%(first and third)
 terms in (\ref{U}),
with energy splitting
between the doubly occupied and singly occupied $d+$ and $d-$ orbitals,
$E^{(0)}_{d(1)d(2)} - E^{(0)}_{d(1) {\bar d}(2)} = U_0 - U_0^{\prime}$.
Third, notice that the pair states $\phi_1(1,2)$ and $\phi_2(1,2)$ are
odd and even superpositions of ${d+}(1) {d+}(2)$ and ${d-}(1) {d-}(2)$.
The former pair states, hence, are eigenstates of the on-site Josephson tunneling terms in (\ref{U}),
with energy splitting
$E^{(0)}_1 - E^{(0)}_2 = -2 J_0^{\prime}$.
The remaining pair states $\phi_0 (1,2)$ and $\phi_3 (1,2)$
%become null after the application of
do not participate in on-site Josephson tunneling.
Table \ref{atomic_energies} lists the atomic energies of these pair states compared to
the one along the isospin $I^{(3)}$ axis.

\begin{table}
%\hspace*{3cm}
\begin{tabular}{|c|c|c|c|}
\hline
Isospin Axis of Pair State $(n)$ &\ $S$ \ &\ $I$ \ &   $\Delta H_U$ \\
\hline
any $(0)$   &    $1$   &   $0$   &    $J_0$ \\
$I^{(1)}$    &    $0$   &   $1$   & \  $U_0-U_0^{\prime}-J_0^{\prime}$ \ \\
$I^{(2)}$    &    $0$   &   $1$   & \  $U_0-U_0^{\prime}+J_0^{\prime}$ \ \\
$I^{(3)}$    &    $0$   &   $1$   &    $0$ \\
\hline
\end{tabular}
\caption{Relative energy $H_U$ of atomic pair states, $\phi_n(1,2)$,
compared to that of $\phi_3(1,2)$.
Recall that $J_0 < 0$.}
\label{atomic_energies}
\end{table}

Last, we point out that both
the on-site Josephson tunneling terms in (\ref{U})
and the first term in (\ref{on-site_rplsn}) for the on-site repulsion
break isospin rotation invariance.  
Such symmetry-breaking contributions in the on-site Hamiltonian $H_U$ are
consolidated by the Hamiltonian
\begin{equation}
H_{U}^{\prime}  = \sum_i 2 [+J_0^{\prime}\, I_{i,\uparrow}^{(1)} I_{i,\downarrow}^{(1)}
 - J_0^{\prime}\, I_{i,\uparrow}^{(2)} I_{i,\downarrow}^{(2)} +
              (U_0-U_0^{\prime}) I_{i,\uparrow}^{(3)} I_{i,\downarrow}^{(3)}] ,
\label{break_iso_symm}
\end{equation}
where ${\bm I}_{i,\uparrow}$ and ${\bm I}_{i,\downarrow}$
are the respective isospin operators
for spin-$\uparrow$ and spin-$\downarrow$ electrons at iron site $i$.
(See Appendix \ref{ppndx_isospin_operator}.)
They each represent $2 \times 2$ isospin operators
acting on the $d+$ and $d-$ orbitals
for an electron of such spin.

% encoded by the effective Hamiltonian for a half filled iron site $i$,
%%
%\begin{equation}
%H_{i}^{\prime} (1,2) = +J_0^{\prime}\, I_i^{(1)} I_i^{(1)} - J_0^{\prime}\, I_i^{(2)} I_i^{(2)} + 
%              (U_0-U_0^{\prime}) I_i^{(3)} I_i^{(3)} ,
%\label{break_iso_symm}
%\end{equation}
%%
%where ${\bm I}_i = {\bm I}_i (1) + {\bm I}_i (2)$.
%Here, ${\bm I}_i (1)$ and ${\bm I}_i (2)$
%represent the $2 \times 2$ isospin operators for electron $1$ and for electron $2$
%acting on their respective ``up'' and ``down'' orbitals, $d+$ and $d-$.
%(See also Appendix \ref{ppndx_isospin_operator}.)

\subsection{Hidden Magnetic Order}
\begin{figure}
%\hspace*{2cm}
\includegraphics[scale=1.00, angle=0]{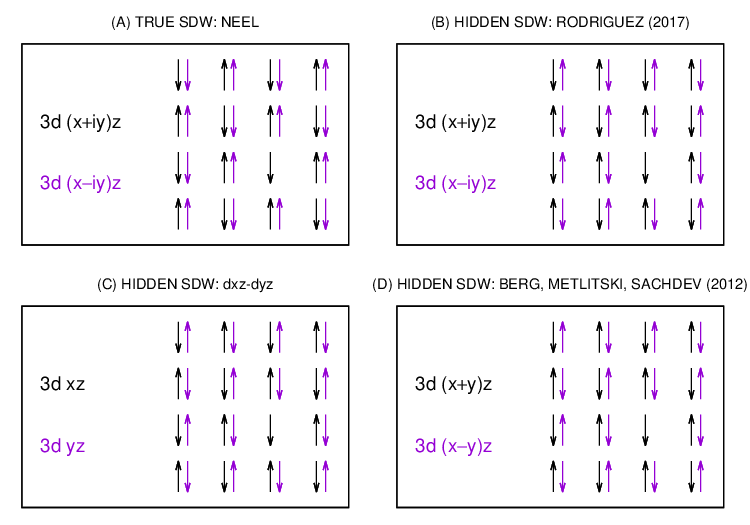}
\caption{Spin/orbital configurations
for (a) conventional N\'eel order, and for hidden N\'eel order among orbitals
along (b) the $I^{(3)}$ isospin axis, along (c) the $I^{(1)}$ isospin axis,
and along (d) the $I^{(2)}$ isospin axis.}
\label{sdw_hsdw_states}
\end{figure}

The true electronic spin at an iron site $i$ is measured by the operator 
${\bm S}_{i} = {\bm S}_{i,d+} + {\bm S}_{i,d-}$,
with  ${\bm S}_{i,\alpha} = ({\hbar}/2) \sum_{s,s^{\prime}}
c_{i,\alpha,s}^{\dagger}{\boldsymbol{\sigma}}_{s,s^{\prime}} c_{i,\alpha,s^{\prime}}$,
where ${\boldsymbol{\sigma}}$ denote the Pauli matrices.
In the present case, we keep only the principal $d-$ and $d+$ orbitals, $\alpha$. 
Hidden spin excitations must be orthogonal to true spin excitations.  
Hidden spin excitations  then correspond to ``pion'' excitations
of the latter isospin degrees of freedom.  
Table \ref{isospin} lists these spin excitations explicitly, which carry isospin $I=1$.
% equal to unity.
They are isospin components of the tensor product
$({\bm S}\otimes{\bm I})_{i} = (\hbar/4) \sum_{\alpha,\alpha^{\prime}}\sum_{s,s^{\prime}}
c_{i,\alpha,s}^{\dagger}{\boldsymbol{\sigma}}_{s,s^{\prime}}
{\boldsymbol{\tau}}_{\alpha,\alpha^{\prime}} c_{i,\alpha^{\prime},s^{\prime}}$,
where ${\boldsymbol{\tau}}$ also denote the Pauli matrices.
(See Appendix \ref{ppndx_isospin_operator}.)
Notice that hidden spin excitations generated by the ($\pi^0$) operator
$2({\bm S}\otimes I^{(3)})_{i} = {\bm S}_{i,d+} - {\bm S}_{i,d-}$
% ({\hbar}/2) \sum_{s,s^{\prime}}
%(c_{i,d+,s}^{\dagger}{\boldsymbol{\sigma}}_{s,s^{\prime}} c_{i,d+,s^{\prime}} -
%c_{i,d-,s}^{\dagger}{\boldsymbol{ \sigma}}_{s,s^{\prime}} c_{i,d-,s^{\prime}})$
are the most symmetric ones, showing isotropy about the orbital $z$ axis.
This is displayed explicitly by Table \ref{hsw},
in the row corresponding to the isospin quantization axis $I^{(3)}$,
where  $2({\bm S}\otimes I^{(3)})_{i}$  is written
in terms of $d_{xz}$ and $d_{yz}$ orbitals.
Figure \ref{sdw_hsdw_states} shows  three different
hidden spin-density orderings made up, respectively, of the three magnetic moments
$2({\bm S}\otimes I^{(n)})_{i}$ over the square lattice $i$,
with $n=1,2,3$.
Such hSDW groundstates have been introduced recently
%in the literature
in the context of copper-oxide high-$T_c$ superconductors\cite{BMS_12},
of heavy fermion compounds\cite{riseborough_12},
and of iron-selenide high-$T_c$ superconductors\cite{jpr_17,jpr_rm_18}.

\begin{table}
%\hspace*{3cm}
\begin{tabular}{|c|c|c|c|c|}
\hline
spin operator & meson analog & $\ I\ $ &\ $I^{(3)}$\ & type of spin \\
\hline
$c_{i,d+}^{\dagger} {\boldsymbol{\sigma}}\, c_{i,d+} + c_{i,d-}^{\dagger} {\boldsymbol{\sigma}}\, c_{i,d-}$ & $\omega$ & $0$ & $0$ & true \\
$c_{i,d+}^{\dagger} {\boldsymbol{\sigma}}\, c_{i,d+} - c_{i,d-}^{\dagger} {\boldsymbol{\sigma}}\, c_{i,d-}$ & $\pi^0$ & $1$ & $0$ & hidden \\
$c_{i,d+}^{\dagger} {\boldsymbol{\sigma}}\, c_{i,d-}$ & $\pi^+$ & $1$ & $+1$ & hidden \\
$c_{i,d-}^{\dagger} {\boldsymbol{\sigma}}\, c_{i,d+}$ & $\pi^-$ & $1$ & $-1$ & hidden \\
\hline
\end{tabular}
\caption{List of spin-excitation operators according to isospin.  Above, ${\boldsymbol{\sigma}}$
denotes the Pauli matrices for spin, and $I$ and $I^{(3)}$ denote the isospin quantum numbers.
Summation over spin indices is implicit.  Meson analogs are obtained by identifying the
$d+$ orbital with the $u$ quark and the $d-$ orbital with the $d$ quark.
See Appendix \ref{ppndx_isospin_operator} for a definition of the isospin operator.}
\label{isospin}
\end{table}

In the last case,
perfect nesting of electron-type and hole-type Fermi surfaces
exists at half filling and  $t_2^{\parallel} = 0$,
which is displayed by Fig. \ref{FS0}.
This implies an instability to a spin-density wave at the wavevector corresponding to
N\'eel antiferromagnetic order, ${\bm Q}_{\rm AF} = (\pi/a,\pi/a)$.
The atomic limit discussed at the end of the previous subsection 
becomes a useful  guide 
to determine the relative stability of the
four checkerboard spin density waves displayed by Fig. \ref{sdw_hsdw_states}
in the limit of strong on-site repulsion.
First, it is important to point out that the on-site pair states that compose the hSDW groundstates
displayed by Figs. \ref{sdw_hsdw_states}(b)-(d) can be expressed as even and odd superpositions of
spin singlet and  spin triplet states,
\begin{equation}
{1\over{\sqrt{2}}}\phi_n(1,2) {1\over{\sqrt{2}}}(\uparrow_1\downarrow_2-\downarrow_1\uparrow_2)
%\quad {\rm and} \quad
\pm
{1\over{\sqrt{2}}}\phi_0(1,2) {1\over{\sqrt{2}}}(\uparrow_1\downarrow_2+\downarrow_1\uparrow_2) ,
\label{slater_determinant_id}
\end{equation}
where $\phi_0(1,2)$ and $\phi_n(1,2)$ are orbital pair states 
given by (\ref{iso_singlet}) and (\ref{iso_triplet}).
Such atomic pair states violate Hund's Rule.  The corresponding hSDW groundstates
therefore compete with the conventional SDW groundstate displayed by Fig. \ref{sdw_hsdw_states}(a)
in the regime of weak Hund's Rule coupling.
Local-moment Heisenberg models find, in particular,
%and an analysis of the extended Hubbard model within the random phase approximation (RPA)
that such hSDW states are more stable than both the conventional checkerboard and stripe SDW states
in the presence of magnetic frustration (\ref{sprx}),
 at weak enough Hund's Rule coupling\cite{jpr_10}.
%,jpr_17,jpr_20a}.
(See Fig. \ref{phase_diagram}.)
And which of the three hSDW states displayed by Figs. \ref{sdw_hsdw_states}(b)-(d)
is the most energetically favorable one?
Contrasting 
the corresponding atomic pair states (\ref{slater_determinant_id})
with the atomic spectrum listed by Table \ref{atomic_energies}
% discussed at the end of the previous subsection
indicates that the hSDW displayed by Fig. \ref{sdw_hsdw_states}(b),
which corresponds to the $I^{(3)}$ isospin axis ($n=3$),
is the lowest in energy at sufficiently large intra-orbital on-site repulsion:
$U_0 - U_0^{\prime} > |J_0^{\prime}|$.
Notice that both the sum of the on-site repulsion terms (\ref{on-site_rplsn})
and the on-site Josephson tunneling terms in the Hamiltonian (\ref{U})
break $SU(2)$ isospin rotation invariance. % in such case.
These terms are collected (\ref{break_iso_symm}) by $H_U^{\prime}$.
Last,
it is worth re-emphasizing here
that the hSDW corresponding to atomic pair states (\ref{slater_determinant_id}) with $n=3$
is notably isotropic with respect to rotations of the orbitals about the $z$ axis.

\begin{table}
\begin{tabular}{|c|c|c|}
\hline
hidden spin operator & isospin quantization axis & reference \\
\hline
$c_{i,d_{xz}}^{\dagger} {\boldsymbol{\sigma}}\, c_{i,d_{xz}} - c_{i,d_{yz}}^{\dagger} {\boldsymbol{\sigma}}\, c_{i,d_{yz}}$ & $I^{(1)}$ & none \\
%\footnote{Equivalent to Berg, Metlitski and Sachdev (2012) when the orbital axes are rotated by $45^{\circ}$: $x=(x^{\prime}+y^{\prime})/\sqrt{2}$ and $y=(y^{\prime}-x^{\prime})/\sqrt{2}$.} \\
$c_{i,d_{xz}}^{\dagger} {\boldsymbol{\sigma}}\, c_{i,d_{yz}} + c_{i,d_{yz}}^{\dagger} {\boldsymbol{\sigma}}\, c_{i,d_{xz}}$ & $I^{(2)}$ & Berg, Metlitski and Sachdev (2012) \\
$i(c_{i,d_{xz}}^{\dagger} {\boldsymbol{\sigma}}\, c_{i,d_{yz}} - c_{i,d_{yz}}^{\dagger} {\boldsymbol{\sigma}}\, c_{i,d_{xz}})$ & $I^{(3)}$ & Rodriguez (2017) \\
\hline
\end{tabular}
\caption{List of hidden spin-excitation operators by isospin quantization axis. 
Summation over spin indices is implicit.
Examples of where such hidden spin excitations appear in the literature are also listed. % under ``references''.
Note: the spin operator in the second row ($I^{(2)}$) is diagonal in the orbital basis
rotated by 45 degrees about the $z$ axis;
$c_{i,d_{x^{\prime}z}}^{\dagger} {\boldsymbol{\sigma}}\, c_{i,d_{x^{\prime}z}} - c_{i,d_{y^{\prime}z}}^{\dagger} {\boldsymbol{\sigma}}\, c_{i,d_{y^{\prime}z}}$,
where $x^{\prime} = (x+y)/\sqrt{2}$ and $y^{\prime} = (-x+y)/\sqrt{2}$.
See Appendix \ref{ppndx_isospin_operator} for a definition of the isospin operator.}
\label{hsw}
\end{table}

The long-range hidden N\'eel order shown  by the hSDW state
(Fig. \ref{sdw_hsdw_states}b) implies low-energy spinwave excitations
that collapse to zero energy at the ordering wavevector ${\bm Q}_{\rm AF}$.
These hidden spinwaves emerge from the dynamics between the bulk spin,
${\bm S}_i = {\bm S}_{i,d-} + {\bm S}_{i,d+}$,
and the hidden ordered magnetic moment\cite{jpr_rm_18},
${\bm m}_i(\pi) = {\bm S}_{i,d-} - {\bm S}_{i,d+}$.
It is yet another example of antiferromagnetic dynamics
first discovered by Anderson \cite{anderson_52,halperin_hohenberg_69,forster_75}.
%The previous are conjugate dynamical variables that satisfy the commutation relations
%$[S_i, m_j(\pi)] = i \hbar \epsilon_{i,j,k} m_k (\pi)$.
%%$[S_i, m_j(\pi)] = i\epsilon_{i,j} (2 s_1) \hbar^2$
%%for $i$ and $j$ equal to $x$ or $y$.
%At the long-wavelength limit, they lead to dynamics governed by the non-linear $\sigma$-model,
%which is a functional of the unit vector field ${\bm n} = {\bm m} (\pi) / |{\bm m} (\pi)|$.
%Its Lagrangian is given by
%$L = \int d^2 r {1\over 2}(\chi_{\perp} |\dot{\bm n}|^2 - \rho_s |{\bm\nabla} {\bm n}|^2)$,
%and where $\rho_s$ is the spin rigidity of the hSDW.
The dynamical propagator for hidden spinwaves can then be defined as
$iD({\bm q},\omega) =
\langle {1\over{\sqrt{2}}} m^{+}(\pi)
{1\over{\sqrt{2}}} m^{-}(\pi)\rangle |_{{\bm q},\omega}$,
where
$m^{\pm} (\pi) = m_x (\pi) \pm i\, m_y (\pi)$.
Here, we  have assumed that the hSDW spontaneously breaks symmetry along the $z$ axis.
%The previous antiferromagnetic dynamics then yields the following form for this propagator:
%It can be read off directly from the Lagrangian of the non-linear $\sigma$-model,
Within the random phase approximation (RPA)
of the two-orbital extended Hubbard model,
recent calculations of the dynamical spin susceptibility in the hSDW state,
Fig. \ref{sdw_hsdw_states}b,
yield the universal form\cite{halperin_hohenberg_69,forster_75}
\begin{equation}
D({\bm q},\omega) = {(2 s_1)^2\over{\chi_{\perp}}}
[\omega^2 - \omega_b^2({\bm q})]^{-1}
\label{D}
\end{equation}
at long wavelength and low frequency\cite{jpr_20a}. 
Above, $2 s_1 \hbar$ is the magnitude of the hidden ordered magnetic moment ${\bm m}(\pi)$
at an iron site,
while $\chi_{\perp}$ is the spin susceptibility of the hSDW 
for external magnetic field applied  perpendicular to
% the hidden ordered magnetic moment, 
${\bm m}(\pi)$.
The poles in frequency in (\ref{D}) disperse as
\begin{equation}
\omega_b ({\bm q}) = (c_b^2 |{\bar{\bm q}}|^2 + \Delta_b^2)^{1/2},
\label{w_b}
\end{equation}
where ${\bm q} = {\bar{\bm q}} + {\bm Q}_{\rm AF}$.
Above, the velocity of the hidden spinwaves is given by $c_b = (\rho_s/\chi_{\perp})^{1/2}$,
where $\rho_s$ is the spin stiffness of the hSDW,
while the spin gap $\Delta_b$ is null when the hSDW state shows long-range order.
It can be demonstrated that the spin $s_1$ is equal to the spin per orbital in the local-moment limit
described by the two-orbital Heisenberg model\cite{jpr_10} (\ref{hund_heisenberg}).
%Also, an analysis of
%within the random phase approximation (RPA)
%finds that $s_1$ is equal to the sub-lattice magnetization per orbital in the hSDW state\cite{jpr_20a}.

%

\section{Eliashberg Theory}
After adding on-iron-site Coulomb repulsion (\ref{U}) and 
magnetic frustration from super-exchange via the selenium atoms (\ref{sprx})
to the electron hopping Hamiltonian (\ref{hop}),
the author and Melendrez recently showed that the hSDW state,
with opposing N\'eel antiferromagnet order over the square lattice of iron atoms per $d\pm$ orbital,
is stable 
within the mean-field approximation
at perfect nesting\cite{jpr_rm_18}. (See Fig. \ref{FS0}.)
And after developing an Eliashberg theory in the particle-hole channel,
these authors then showed that coupling to hidden spin fluctuations,
(\ref{D}) and (\ref{w_b}),
 shifts the two electronic bands by an equal and opposite energy,
 leading to electron/hole Fermi surface pockets at the corner of the folded (two-iron) Brillouin zone.
They also notably found that the
% wavefunction renormalization
spectral weight, $1/Z$ ,
tends to zero at the new Fermi surface pockets.

Berg, Metlitski and Sachdev have performed determinant quantum Monte Carlo (DQMC) simulations
on a similar model\cite{BMS_12} that includes weak nesting of Fermi surfaces by
the N\'eel wavevector ${\bm Q}_{\rm AF}$
and coupling to hidden spin fluctuations
 with isospin quantum number $I^{(2)} = 0$.
(See Table \ref{hsw} and Fig. \ref{sdw_hsdw_states}d.)
%that are a superposition of those
% numbers $I^{(3)} = +1$ and $-1$.
%(See Tables \ref{isospin} and \ref{hsw}.)
They find a quantum-critical phase transition at low temperature 
between a hSDW and a $D$-wave
superconductor, with Cooper pairs
on nominal $x$ versus $y$ orbitals that alternate in sign between them.
Below, we will show that a similar quantum-critical phase transition exists upon electron doping
of the hSDW state considered here,
with isospin quantum number $I^{(3)} = 0$ instead.
In particular, an Eliashberg theory in 
the conventional particle-particle channel\cite{eliashberg_60,eliashberg_61,schrieffer_64,scalapino_69}
 will be revealed for electron-doped states that exhibit only short-range hSDW order.
It predicts Cooper pairs that show $S$-wave symmetry, however.

\subsection{Hidden Spin Fluctuations and Interaction with Electrons}
In the hidden N\'eel state considered here,
% at long range order,
with spontaneous symmetry breaking along the $z$ axis,
the propagator for spinwaves is given by
\begin{equation}
\biggl\langle {1\over{\sqrt{2}}} m^{+}(\pi)
{1\over{\sqrt{2}}} m^{-}(\pi)\biggr\rangle \bigg|_{{\bm q},\omega} = iD({\bm q},\omega),
\label{xy}
\end{equation}
with its form set by (\ref{D}) and (\ref{w_b}).
We shall henceforth assume that the spin gap $\Delta_b$ grows in a continuous fashion from zero
upon crossing the quantum critical point.
Electron doping from half filling shall be one of the principal
tuning parameters for the quantum-critical phase transition.
(Cf. Fig. \ref{phase_diagram}.)
Spin isotropy is recovered upon crossing the quantum critical point, however.
It dictates the form
\begin{equation}
\biggl\langle {1\over{\sqrt{2}}} m^{(z)}(\pi)
{1\over{\sqrt{2}}} m^{(z)}(\pi)\biggr\rangle \bigg|_{{\bm q},\omega} = {1\over 2} iD({\bm q},\omega),
\label{z}
\end{equation}
for the nature of hidden spin fluctuations along the $z$ axis at $\Delta_b > 0$.

As was mentioned earlier,
the extended Hubbard model over the square lattice of iron atoms in FeSe
that  was introduced in subsection \ref{hubbard_model}
at perfect nesting of the Fermi surfaces (Fig. \ref{FS0})
harbors a hSDW state when magnetic frustration is present\cite{jpr_rm_18}.
A mean field theory approximation of the extended Hubbard model implies an
isotropic interaction between spin fluctuations and electrons of the form
$H_{\rm e-hsw} = -\sum_i \sum_{\alpha} U(\pi) {\bm m}_{i,\alpha} \cdot 2{\bm S}_{i,\alpha}$,
where
\begin{equation}
U(\pi) = U_0 + {1\over 2}J_0.
\label{U_pi}
\end{equation}
Here, $U_0$ is the on-site repulsive energy cost for the formation
of a spin singlet on the $d+$ orbital or on the $d-$ orbital,
while $J_0$ is the (ferromagnetic) Hund's Rule spin-exchange coupling constant
between these two orbitals.
The transverse contributions yield the interaction
$H_{\rm e-hsw}^{(xy)} = -\sum_i \sum_{\alpha} U(\pi) 
(m_{i,\alpha}^+ S_{i,\alpha}^- + m_{i,\alpha}^- S_{i,\alpha}^+)$,
while the longitudinal contributions yield the interaction
$H_{\rm e-hsw}^{(z)} = -\sum_i \sum_{\alpha} U(\pi) m_{i,\alpha}^{(z)} 2 S_{i,\alpha}^{(z)}$.
In the basis of electron energy bands,
they yield the following contribution to the Hamiltonian due to the
interaction of electrons with hidden spin fluctuations:
\begin{eqnarray}
H_{\rm e-hsw}^{(xy)} = -{1\over{\sqrt{2}}}{U(\pi)\over{a {\cal N}^{1/2}}}
 \sum_{\bm k} \sum_{{\bm k}^{\prime}}\sum_n
[m^+(\pi,{\bm q})
c_{\downarrow}^{\dagger}({\bar n},{\bar{\bm k}}^{\prime})
c_{\uparrow}(n,{\bm k}) & {\cal M}_{n,{\bm k};{\bar n},{\bar{\bm k}}^{\prime}} \nonumber \\
&+{\rm h.c.}]
\label{e-hsw_xy}
\end{eqnarray}
and
\begin{eqnarray}
H_{\rm e-hsw}^{(z)} = -{1\over{\sqrt{2}}}{U(\pi)\over{a {\cal N}^{1/2}}}
 \sum_{\bm k} \sum_{{\bm k}^{\prime}}\sum_n\sum_s
 m^{(z)}(\pi,{\bm q}) &
c_{s}^{\dagger}({\bar n},{\bar{\bm k}}^{\prime})
c_{s}(n,{\bm k}) \cdot \nonumber \\
& \cdot {\cal M}_{n,{\bm k};{\bar n},{\bar{\bm k}}^{\prime}} \,({\rm sgn}\,s),
\label{e-hsw_z}
\end{eqnarray}
where ${\bm q} = {\bm k}-{\bar{\bm k}}^{\prime}$ is the momentum transfer,
with ${\bar{\bm k}}^{\prime} = {\bm k}^{\prime} + {\bm Q}_{\rm AF}$.
Above, $c_s^{\dagger} (n,{\bm k})$ and $c_s(n,{\bm k})$
are electron creation and destruction  operators
for plane-wave states (\ref{plane_waves}).
The band indices $n=1$ and $n=2$ correspond, respectively,
to anti-bonding ($-$) planewaves
in the $d_{y(\delta)z}$ orbital
and to bonding ($+$) planewaves in the $d_{x(\delta)z}$ orbital.
Also, ${\bar n}$ denotes the opposite band.
The orbital matrix element that appears in (\ref{e-hsw_xy})
 and in (\ref{e-hsw_z}) is given by\cite{jpr_rm_18}
\begin{equation}
{\cal M}_{n,{\bm k};{\bar n},{\bar{\bm k}}^{\prime}} =
\pm \sin[\delta({\bm k})+\delta({\bm k}^{\prime})].
\label{M_E}
\end{equation}
(See Appendix \ref{ppndx_m_e}.)
Above, intra-band transitions are neglected because they do not show nesting.

We shall now apply the Nambu-Gorkov formalism 
for paired states\cite{schrieffer_64,scalapino_69,nambu_60,gorkov_58}.
It then becomes useful to  write the above electron-hidden-spinwave interactions in terms of spinors:
\begin{eqnarray}
H_{\rm e-hsw}^{(xy)} = \mp{1\over{\sqrt{2}}}{U(\pi)\over{{a \cal N}^{1/2}}}
 \sum_{\bm k} \sum_{{\bm k}^{\prime}}
[m^+(\pi,{\bm q})
C_{n}^{\dagger}({\bar{\bm k}}^{\prime}) \tau_3 {\bar C}_{{\bar n}}({\bm k})
 \sin[\delta({\bm k}) & + \delta({\bm k}^{\prime})]  \nonumber\\
& +{\rm h.c.}], 
\label{E-hSW_xy}
\end{eqnarray}
and
\begin{eqnarray}
H_{\rm e-hsw}^{(z)} = \mp{1\over{\sqrt{2}}}{U(\pi)\over{{a \cal N}^{1/2}}}
 \sum_{\bm k} \sum_{{\bm k}^{\prime}}\sum_n
m^{(z)}(\pi,{\bm q})
& C_{\bar n}^{\dagger}({\bar{\bm k}}^{\prime}) \tau_0 C_{n}({\bm k}) \cdot \nonumber \\
& \cdot \sin[\delta({\bm k}) + \delta({\bm k}^{\prime})],
\label{E-hSW_z}
\end{eqnarray}
with 
\begin{equation}
C_n({\bm k}) =
\left[ {\begin{array}{c}
c_{\uparrow}(n,{\bm k}) \\ c_{\downarrow}^{\dagger}(n,-{\bm k})
\end{array} } \right]
\label{spinor_u_d}
\end{equation}
and
\begin{equation}
{\bar C}_n({\bm k}) =
\left[ {\begin{array}{c}
c_{\downarrow}(n,{\bm k}) \\ c_{\uparrow}^{\dagger}(n,-{\bm k})
\end{array} } \right].
\label{spinor_d_u}
\end{equation}
Above, $\tau_3$ is the Pauli matrix along the $z$ axis,
and $\tau_0$ is the $2\times 2$ identity matrix.
Also, the explicit matrix
element (\ref{M_E})
 ${\cal M}_{n,{\bm k};{\bar n},{\bar{\bm k}}^{\prime}}$
has been substituted in. (See Appendix \ref{ppndx_m_e}.)
It is important to point out that the band index $n$ is {\it fixed} 
in expression (\ref{E-hSW_xy}) for  $H_{\rm e-hsw}^{(xy)}$ above.
The $n=1$ and the $n=2$ expressions are equivalent.

\subsection{Electron Propagators and Eliashberg Equations}
Let $C_n({\bm k},t)$ and ${\bar C}_n({\bm k},t)$  denote the time evolution of
the Nambu-Gorkov spinors,
%(\ref{spinor_u_d}) 
$C_n({\bm k})$ and 
%(\ref{spinor_d_u}) 
${\bar C}_n({\bm k})$,
and let $C_n^{\dagger}({\bm k},t)$ and ${\bar C}_n^{\dagger}({\bm k},t)$ denote the time evolution
of their conjugates,
% creation/destruction operators 
$C_n^{\dagger}({\bm k})$ and ${\bar C}_n^{\dagger}({\bm k})$.
The Nambu-Gorkov electron propagators are then
the Fourier transforms
$i G_n({\bm k},\omega) = \int d t_{1,2} e^{i \omega t_{1,2}}
\langle T[C_n({\bm k},t_1) C_n^{\dagger}({\bm k},t_2)]\rangle$
and
$i {\bar G}_n({\bm k},\omega) = \int d t_{1,2} e^{i \omega t_{1,2}}
\langle T[{\bar C}_n({\bm k},t_1) {\bar C}_n^{\dagger}({\bm k},t_2)]\rangle$,
where $t_{1,2} = t_1 - t_2$, and where $T$ is the time-ordering operator.
They are $2 \times 2$  matrices.
In the absence of interactions, their matrix inverses are then given by
\begin{equation}
G_{0 n}^{-1}({\bm k},\omega) =
\omega\, \tau_0 - [\varepsilon_n({\bm k}) - \mu_0] \, \tau_3.
\label{1/G0}
\end{equation}
%
%where $\tau_0$ is the $2 \times 2$ identity matrix,
%and where $\tau_3$ is the Pauli matrix along the $z$ axis.
Following the standard prescription\cite{schrieffer_64,scalapino_69},
 let us next assume that the matrix inverse of the Nambu-Gorkov Greens function takes the form
\begin{equation}
G_n^{-1}({\bm k},\omega) =
Z_n({\bm k},\omega)  \omega\, \tau_0
- [\varepsilon_n({\bm k})-\mu_n]\, \tau_3
-Z_n({\bm k},\omega) \Delta_n({\bm k})\,\tau_1.
\label{1/G}
\end{equation}
Here, $Z_n({\bm k},\omega)$ is the wavefunction renormalization,
$\Delta_n({\bm k})$ is the quasi-particle gap,
and $\mu_n - \mu_0$ is the shift in the energy band.
%renormalized chemical potential.
Matrix inversion of (\ref{1/G}) yields the 
Nambu-Gorkov Greens function\cite{schrieffer_64,scalapino_69,nambu_60,gorkov_58}
$G = \sum_{\mu = 0}^{3} G^{(\mu)} \tau_{\mu}$, with components
\begin{eqnarray}
G_n^{(0)} &=& {1\over{2 Z_n}}
\Biggl({1\over{\omega-E_n}} + {1\over{\omega+E_n}}\Biggr),\nonumber \\
G_n^{(1)} &=& {1\over{2 Z_n}}
\Biggl({1\over{\omega-E_n}} - {1\over{\omega+E_n}}\Biggr)
{\Delta_n\over E_n} , \nonumber \\
G_n^{(3)} &=& {1\over{2 Z_n}}
\Biggl({1\over{\omega-E_n}} - {1\over{\omega+E_n}}\Biggr)
{\varepsilon_n-\mu_n\over Z_n E_n},
\label{Green}
\end{eqnarray}
and $G_n^{(2)} = 0$. Above, the excitation energy is
\begin{equation}
E_n({\bm k},\omega) =
\sqrt{\Biggl[{\varepsilon_n({\bm k})-\mu_n\over{Z_n({\bm k},\omega)}}\Biggr]^2
+ \Delta_n^2({\bm k})}.
\label{Energy}
\end{equation}
Last, because the spinors (\ref{spinor_u_d}) and (\ref{spinor_d_u}) are related by spin flip,
and because we assume spin singlet Cooper pairs,
then ${\bar G}$ is obtained from $G$ by the replacement $\Delta_n\rightarrow -\Delta_n$.
This yields
${\bar G}_n^{(0)} = G_n^{(0)}$, ${\bar G}_n^{(1)} = - G_n^{(1)}$, ${\bar G}_n^{(2)} = - G_n^{(2)}$,
and ${\bar G}_n^{(3)} = G_n^{(3)}$.

\begin{figure}
%\hspace*{2cm}
\includegraphics[scale=1.00, angle=0]{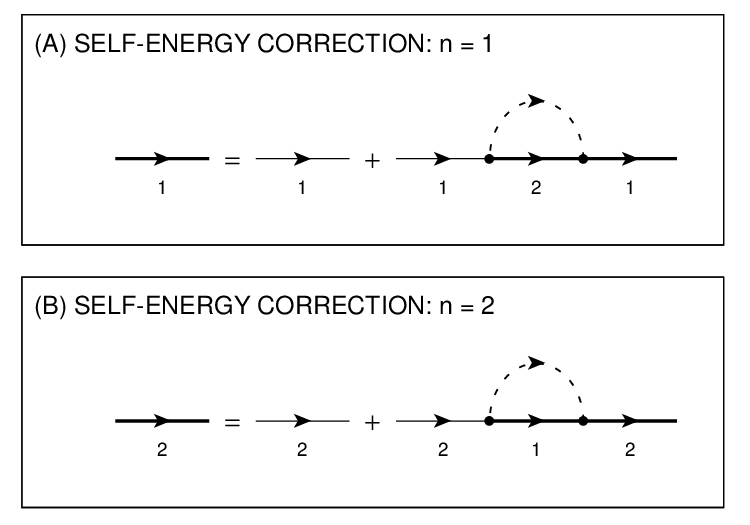}
\caption{Feynman diagrams for electron propagator
with purely inter-band scattering, in the absence of vertex corrections.}
\label{EFD}
\end{figure}

To obtain the Eliashberg equations,
recall first the definition of the self-energy correction per band:
$G_n^{-1} = G_0^{-1} - \Sigma_n$.
Comparison of the inverse Greens functions
(\ref{1/G0}) and (\ref{1/G})
then yields the following expression for it\cite{schrieffer_64,scalapino_69}:
\begin{equation}
\Sigma_n({\bm k},\omega) =
[1-Z_n({\bm k},\omega)] \omega\, \tau_0
-(\mu_n-\mu_0)\, \tau_3
+Z_n({\bm k},\omega) \Delta_n({\bm k})\,\tau_1 .
\label{Sigma}
\end{equation}
Next, we neglect vertex corrections from 
the electron-hidden-spinwave interactions, (\ref{E-hSW_xy}) and (\ref{E-hSW_z}).
Figure \ref{EFD} displays the resulting self-consistent approximation.
This approximation will be justified {\it a posteriori} in the next section.
The self-energy correction is then given by % the self-consistent approximation:
\begin{eqnarray}
\Sigma_n({\bm k},\omega) = i
\int_{\rm BZ} {d^2 k^{\prime}\over{(2\pi)^2}}  \int_{-\infty}^{+\infty}{d\omega^{\prime}\over{2\pi}}
&& {U^2(\pi)\over 2} \sin^2[\delta({\bm k})+\delta({\bm k}^{\prime})] D({\bm q},q_0) \cdot \nonumber\\
&& \cdot [\tau_3 {\bar G}_{\bar n}({\bar{\bm k}}^{\prime},\omega^{\prime})\tau_3
+{1\over 2} G_{\bar n}({\bar{\bm k}}^{\prime},\omega^{\prime})],
\label{self-energy}
\end{eqnarray}
with $q_0 = \omega - \omega^{\prime}$,
and with ${\bm q} = {\bm k} - {\bar{\bm k}}^{\prime}$.
Observe, finally, that
$\tau_3 \tau_{\mu} \tau_3 = {\rm sgn}_{\mu}\tau_{\mu}$,
where ${\rm sgn}_0 = +1 = {\rm sgn}_3$, and
where ${\rm sgn}_1 = -1 = {\rm sgn}_2$.
Identifying expressions (\ref{Sigma}) and (\ref{self-energy}) for
the self-energy corrections then yields the following self-consistent
Eliashberg equations at zero temperature:
{\begin{eqnarray}
-[Z_n({\bm k},\omega)-1] \omega &= + &
\int_{\rm BZ} {d^2 k^{\prime}\over{(2\pi)^2}} \, i \int_{-\infty}^{+\infty}{d\omega^{\prime}\over{2\pi}}
 {U^2(\pi)\over 2}  \sin^2[\delta({\bm k})+\delta({\bm k}^{\prime})] \cdot \nonumber\\
&& \cdot D({\bm q},q_0) [{\bar G}_{\bar n}^{(0)}({\bar{\bm k}}^{\prime},\omega^{\prime}) 
+{1\over 2} G_{\bar n}^{(0)}({\bar{\bm k}}^{\prime},\omega^{\prime})], \nonumber\\
\quad\qquad\qquad \mu_0-\mu_n &= + &
\int_{\rm BZ} {d^2 k^{\prime}\over{(2\pi)^2}} \, i \int_{-\infty}^{+\infty}{d\omega^{\prime}\over{2\pi}}
 {U^2(\pi)\over 2}  \sin^2[\delta({\bm k})+\delta({\bm k}^{\prime})] \cdot \nonumber\\
&& \cdot D({\bm q},q_0) [{\bar G}_{\bar n}^{(3)}({\bar{\bm k}}^{\prime},\omega^{\prime})
+{1\over 2} G_{\bar n}^{(3)}({\bar{\bm k}}^{\prime},\omega^{\prime})], \nonumber\\
 Z_n({\bm k},\omega) \Delta_n({\bm k},\omega) &= - &
\int_{\rm BZ} {d^2 k^{\prime}\over{(2\pi)^2}} \, i \int_{-\infty}^{+\infty}{d\omega^{\prime}\over{2\pi}}
 {U^2(\pi)\over 2}  \sin^2[\delta({\bm k})+\delta({\bm k}^{\prime})] \cdot \nonumber\\
&& \cdot D({\bm q},q_0) [{\bar G}_{\bar n}^{(1)}({\bar{\bm k}}^{\prime},\omega^{\prime})
-{1\over 2} G_{\bar n}^{(1)}({\bar{\bm k}}^{\prime},\omega^{\prime})].\nonumber \\
\label{E_eqs_T}
\end{eqnarray}}
%\hspace*{2cm} %
The Greens functions above are listed in (\ref{Green}) and below (\ref{Energy}).

Last,
it becomes useful to write the propagator for hidden spinwaves (\ref{D}) as
\begin{equation}
D({\bm q},\omega) = {(2 s_1)^2\over{\chi_{\perp}}}
{1\over{2\omega_b({\bm q})}}\Biggl[{1\over{\omega - \omega_b({\bm q})}}
-{1\over{\omega + \omega_b({\bm q})}}\Biggr].
\label{d}
\end{equation}
The integrals over frequency in the Eliasgberg equations above (\ref{E_eqs_T}) can be evaluated
by going into the complex plane.
Specifically, make the replacement $E_n \rightarrow E_n - i\eta$ in the poles of
the electron Greens functions (\ref{Green}),
make the replacement $\omega_b({\bm q}) \rightarrow \omega_b({\bm q}) - i\eta$ in the poles
of the spin-wave propagator (\ref{d}),
and regularize the contour integrals 
by including the factor $e^{i\omega^{\prime}\tau}$ in the integrands.
Here, $\eta\rightarrow 0+$ and $\tau\rightarrow 0+$.
Application of Cauchy's residue theorem
yields the following  result, which is
equivalent to Brillouin-Wigner
second-order perturbation theory\cite{schrieffer_64}:
\begin{eqnarray}
[Z_n({\bm k},\omega)-1] \omega &=& {3\over 2} 
\int_{\rm BZ} {d^2 k^{\prime}\over{(2\pi)^2}} U^2(\pi)
{s_1^2\over{\chi_{\perp}}} {\sin^2[\delta({\bm k})+\delta({\bm k}^{\prime})]\over{Z_{\bar n}({\bar{\bm k}}^{\prime},\omega^{\prime})}} \cdot \nonumber\\
&& \cdot {1\over{2\omega_b({\bm q})}}
\Biggl[{1\over{\omega_b({\bm q})+E_{\bar n}({\bar{\bm k}}^{\prime})-\omega}}
-{1\over{\omega_b({\bm q})+E_{\bar n}({\bar{\bm k}}^{\prime})+\omega}}\Biggr], \nonumber\\
\quad\qquad \mu_0-\mu_n &=& - {3\over 2}
\int_{\rm BZ} {d^2 k^{\prime}\over{(2\pi)^2}} U^2(\pi) 
{s_1^2\over{\chi_{\perp}}} {\sin^2[\delta({\bm k})+\delta({\bm k}^{\prime})]\over{Z_{\bar n}({\bar{\bm k}}^{\prime},\omega^{\prime})}}
{\varepsilon_{\bar n}({\bar{\bm k}}^{\prime})-\mu_{\bar n}\over{Z_{\bar n}({\bar{\bm k}}^{\prime},\omega^{\prime}) E_{\bar n}({\bar{\bm k}}^{\prime})}} \cdot \nonumber\\
&& \cdot {1\over{2\omega_b({\bm q})}}
\Biggl[{1\over{\omega_b({\bm q})+E_{\bar n}({\bar{\bm k}}^{\prime})-\omega}}
+{1\over{\omega_b({\bm q})+E_{\bar n}({\bar{\bm k}}^{\prime})+\omega}}\Biggr], \nonumber\\
 Z_n({\bm k},\omega) \Delta_n({\bm k},\omega) &=& - {3\over 2} 
\int_{\rm BZ} {d^2 k^{\prime}\over{(2\pi)^2}} U^2(\pi) 
{s_1^2\over{\chi_{\perp}}} {\sin^2[\delta({\bm k})+\delta({\bm k}^{\prime})]\over{Z_{\bar n}({\bar{\bm k}}^{\prime},\omega^{\prime})}}
{\Delta_{\bar n}({\bar{\bm k}}^{\prime},\omega^{\prime})\over{E_{\bar n}({\bar{\bm k}}^{\prime})}} \cdot \nonumber\\
&& \cdot {1\over{2\omega_b({\bm q})}}
\Biggl[{1\over{\omega_b({\bm q})+E_{\bar n}({\bar{\bm k}}^{\prime})-\omega}}
+{1\over{\omega_b({\bm q})+E_{\bar n}({\bar{\bm k}}^{\prime})+\omega}}\Biggr]. \nonumber\\
\label{E_eqs}
\end{eqnarray}
Above\cite{scalapino_69}, $\omega^{\prime} = E_{\bar n}({\bar{\bm k}}^{\prime})$.
In the previous, the momentum integrals have been shifted by ${\bm Q}_{\rm AF}$ for convenience
in order to exploit perfect nesting (\ref{prfct_nstng}).
Also, the prefactors of $3/2$ above are a result of the identities 
between $G_n^{(\mu)}$ and ${\bar G}_n^{(\mu)}$ that are listed below (\ref{Energy}).
We shall now  find solutions to the Eliashberg equations.

\section{Lifshitz Transition and Pairing Instability at the Fermi Surface}
Henceforth, assume isotropic ($S$-wave) Cooper pairs.
Following the standard procedure\cite{schrieffer_64,scalapino_69},
let us  multiply both sides of the Eliashberg equations (\ref{E_eqs})
by $\delta[\varepsilon_n({\bm k}) - \mu_n] / D_n(\mu_n)$
and integrate in momentum over the first Brillouin zone.
The Eliashberg equations (\ref{E_eqs}) thereby reduce to
%
%%\begin{eqnarray}
\begin{subequations}
\begin{align}
\label{2_E_eqs_a}
(Z_n-1) \omega =&
\int_{-W_{\rm bottom}({\bar n})}^{+W_{\rm top}({\bar n})} d\varepsilon^{\prime} Z_{\bar n}^{\prime -1}
\int_0^{\infty} d\Omega\,  U^2 F_0^{(n,{\bar n})}(\Omega;\mu_n,\mu_{\bar n})\cdot \nonumber\\
& \cdot {1\over{2}}
\Biggl[{1\over{\Omega+E_{\bar n}^{\prime}-\omega}}
-{1\over{\Omega+E_{\bar n}^{\prime}+\omega}}\Biggr], \\
\label{2_E_eqs_b}
\quad \mu_0-\mu_n =&
-\int_{-W_{\rm bottom}({\bar n})}^{+W_{\rm top}({\bar n})} d\varepsilon^{\prime} Z_{\bar n}^{\prime -1}
\int_0^{\infty} d\Omega\,  U^2 F_0^{(n,{\bar n})}(\Omega;\mu_n,\mu_{\bar n})
{\varepsilon^{\prime}-\mu_{\bar n}
\over{Z_{\bar n}^{\prime} E_{\bar n}^{\prime}}}\cdot \nonumber\\
& \cdot {1\over{2}}
\Biggl[{1\over{\Omega+E_{\bar n}^{\prime}-\omega}}
+{1\over{\Omega+E_{\bar n}^{\prime}+\omega}}\Biggr], \\
\label{2_E_eqs_c}
\qquad Z_n \Delta_n =&
-\int_{-W_{\rm bottom}({\bar n})}^{+W_{\rm top}({\bar n})} d\varepsilon^{\prime} Z_{\bar n}^{\prime -1}
\int_0^{\infty} d\Omega\,  U^2 F_0^{(n,{\bar n})}(\Omega;\mu_n,\mu_{\bar n})
{\Delta_{\bar n}^{\prime}
\over{E_{\bar n}^{\prime}}}\cdot \nonumber\\
& \cdot {1\over{2}}
\Biggl[{1\over{\Omega+E_{\bar n}^{\prime}-\omega}}
+{1\over{\Omega+E_{\bar n}^{\prime}+\omega}}\Biggr],
\end{align}
\end{subequations}
%%\end{eqnarray}
%
where
\begin{eqnarray}
U^2 F_0^{(n,{\bar n})}(\Omega;\varepsilon,\varepsilon^{\prime}) = {1\over{D_{n}(\varepsilon)}}
{3\over 2} \int {d^2 k\over{(2\pi)^2}} && \int {d^2 k^{\prime}\over{(2\pi)^2}}
 U^2(\pi) {s_1^2\over{\chi_{\perp}}}
{\sin^2[\delta({\bm k})+\delta({\bm k}^{\prime})]
\over{\omega_b({\bm q})}} \cdot \nonumber \\
&& \cdot \delta[\varepsilon_n({\bm k})-\varepsilon] \delta[\varepsilon_{\bar n}({\bar{\bm k}}^{\prime})-\varepsilon^{\prime}]
\delta[\omega_b({\bm q}) - \Omega],\nonumber \\
\label{U2F}
\end{eqnarray}
and where
$$E_{\bar n}^{\prime} = ([(\varepsilon^{\prime}-\mu_{\bar n})/Z_{\bar n}^{\prime}]^2 + \Delta_{\bar n}^{\prime 2})^{1/2}.$$
Here, the wavefunction renormalization and the gap are averaged over the new Fermi surface:
$Z_n({\bm k},\omega)\rightarrow
[D_n(\mu_n)]^{-1} (2\pi)^{-2} \int_{\rm BZ} d^2 k\, Z_n({\bm k},\omega) \delta[\varepsilon_n({\bm k})-\mu_n]$,
and
$Z_n \Delta_n({\bm k},\omega)\rightarrow
[D_n(\mu_n)]^{-1} (2\pi)^{-2} \int_{\rm BZ} d^2 k\, Z_n \Delta_n({\bm k},\omega) \delta[\varepsilon_n({\bm k})-\mu_n]$.
The neglect of angular dependence is exact
for  circular Fermi surface pockets at $(\pi/a,0)$ and at $(0,\pi/a)$.
This occurs for $\mu_2$ near the upper band edge of $\varepsilon_+({\bm k})$
and for $\mu_1$ near the lower band edge of $\varepsilon_-({\bm k})$,
in the absence of nearest-neighbor intra-orbital hopping, $t_1^{\parallel} \rightarrow 0$.
Above, we have also approximated
the function $U^2 F_0^{(n,{\bar n})}(\Omega;\mu_n,\varepsilon^{\prime})$ of $\varepsilon^{\prime}$
by its value at the renormalized Fermi level, $U^2 F_0^{(n,{\bar n})}(\Omega;\mu_n,\mu_{\bar n})$.
%It is also understood in (\ref{U2F})
%that the limit implicit in the last $\delta$-function factor is taken last.

\subsection{Half Filling}
One of the central aims of this paper is to reveal a Lifshitz transition
from the Fermi surfaces depicted by Fig. \ref{FS0}
to electron/hole pockets at the corner of the folded (two-iron) Brillouin zone.
%depicted by Fig. \ref{FS1}.
Let us start at half filling: $\mu_0 = 0$.
The Fermi surfaces are then set by
$\varepsilon_-({\bm k}) = -\nu$ and $\varepsilon_+({\bm k}) = +\nu$,
where $\mu_1 = -\nu$ and $\mu_2 = +\nu$ are the 
shifts in energy 
%renormalized chemical potentials
 of the anti-bonding ($-$) band
and of the bonding ($+$) band, respectively.
Because of perfect nesting (\ref{prfct_nstng}), we have
$\varepsilon_{\pm}({\bar{\bm k}}) - \mu_{\pm} = \mu_{\mp} - \varepsilon_{\mp} ({\bm k})$.
The Eliashberg equations (\ref{E_eqs}) are then symmetric with respect to the permutation of the band indices.
We thereby have $Z_1 = Z_2$ and $\Delta_1 = -\Delta_2$.
These unknowns, in addition to $\nu$, are to be determined
by the Eliashberg equations (\ref{2_E_eqs_a})-(\ref{2_E_eqs_c}).

\begin{figure}
%\hspace*{2cm}
\includegraphics[scale=1.00, angle=0]{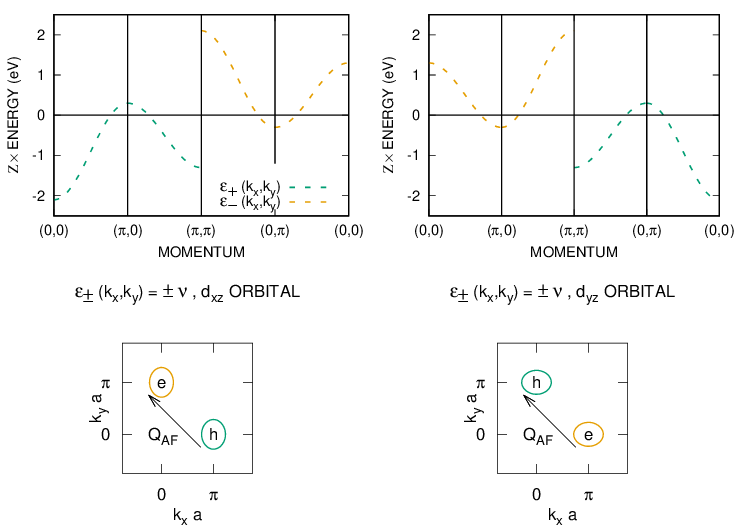}
\caption{Renormalized electron bands and Fermi surfaces at half filling after the Lifshitz transition
from Fig. \ref{FS0}.  The staggered band shift is $\nu = 1.7$ eV.
The orbital character is only approximate,
although it becomes exact as the area of the Fermi surface pockets vanishes as $U(\pi)$ diverges.}
\label{FS1}
\end{figure}

The effective spectral weight of the hidden spinwaves, $U^2 F_0^{(2,1)}(\Omega;\mu_2,\mu_1)$,
can be evaluated by choosing coordinates for the momentum of the electron,
 $(k_{\parallel}, k_{\perp})$,
that are respectively  parallel and perpendicular
to the Fermi surface of the bonding band (FS$_+$):
 $\nu = \varepsilon_+({\bm k})$.
And because of perfect nesting (\ref{prfct_nstng}),
it coincides with the Fermi surface of the anti-bonding ($-$) band
after the momentum is shifted  by ${\bm Q}_{\rm AF}$:
${\bm k}^{\prime}\rightarrow {\bar{\bm k}}^{\prime}$.
(See Figs. \ref{FS0} and \ref{FS1}.)
This yields the intermediate result
\begin{eqnarray}
U^2 F_0^{(2,1)}(\Omega;\mu_2,\mu_1) = {1\over{D_+(\nu)}}
{3\over 2} && \oint_{{\rm FS}_+} {d k_{\parallel}\over{(2\pi)^2}}
 \oint_{{\rm FS}_+} {d k_{\parallel}^{\prime}\over{(2\pi)^2}}
 U^2(\pi) {s_1^2\over{\chi_{\perp}}}
{1 \over{\Omega}} \cdot \nonumber \\
&& \cdot {\sin^2[\delta({\bm k})+\delta({\bm k}^{\prime})]\over{|{\bm v}_+({\bm k})| |{\bm v}_+({\bm k}^{\prime})|}}
\delta[\omega_b({\bm q}) - \Omega],
\end{eqnarray}
where ${\bm v}_+ = \partial\varepsilon_+ /\partial {\bm k}$ is the group velocity.
Yet
the dispersion of the spectrum of hidden spinwaves follows
$\omega_b({\bm q}) = \sqrt{c_b^2 |{\bar{\bm q}}|^2 + \Delta_b^2}$
at the long-wavelength limit.
Making the approximation 
$|{\bar{\bm q}}| \cong |k_{\parallel} - k_{\parallel}^{\prime}|$
 at small momentum transfers
then yields the following dependence on frequency
for the effective spectral weight:
$U^2 F_0^{(2,1)}(\Omega;\mu_2,\mu_1) = \epsilon_{\rm E}(\nu)/\sqrt{\Omega^2 - \Delta_b^2}$
for  $\Omega > \Delta_b$,
with a constant pre-factor
\begin{equation}
\epsilon_{\rm E}(\nu) = {1\over{D_+(\nu)}}
{3\over 2} \oint_{{\rm FS}_+} {d k_{\parallel}\over{(2\pi)^4}}
 U^2(\pi) {s_1^2\over{\chi_{\perp}}}
{[\sin\, 2\delta({\bm k})]^2\over{c_b |{\bm v}_+({\bm k})|^2}},
\label{epsilon_E}
\end{equation}
while $U^2 F_0^{(2,1)}(\Omega;\mu_2,\mu_1) = 0$ for  $0 \leq \Omega \leq \Delta_b$.

Next, let us assume 
the trivial solution for the gap equations (\ref{2_E_eqs_c}): $\Delta_n = 0$.
It will be shown {\it a posteriori} that this is indeed the case.
We can now find solutions to the remaining Eliashberg equations
 (\ref{2_E_eqs_a}) and (\ref{2_E_eqs_b}).
In particular,
assume that the equal and opposite shift in energy $\nu$ of the bands
%chemical potential per band
lies near the upper edge $W_{\rm top}$
of the bonding band $\varepsilon_+({\bm k})$
at $(\pi/a,0)$ and at $(0,\pi/a)$.
(Cf. Fig. \ref{DoS}.)
Figure \ref{FS1} displays the Fermi surfaces in such case.
Substituting in the previous result for the dependence on frequency
 of $U^2 F_0^{(2,1)}(\Omega;\mu_2,\mu_1)$ yields
the first Eliashberg equation:
\begin{eqnarray}
\omega(Z-1) = {\epsilon_{\rm E}\over 2}
\int_{\Delta_b}^{\omega_{\rm uv}} {d\Omega\over{\sqrt{\Omega^2-\Delta_b^2}}}
{\rm ln}\Biggl|{\Omega + \omega\over{\Omega - \omega}} \cdot
{W/Z + \Omega-\omega\over{W/Z + \Omega+\omega}}\Biggr|.
\label{1st_E_eq}
\end{eqnarray}
Here,
we have reversed the order of integration:
$[-W_{\rm bottom},+W_{\rm top}]$
% $[\nu - W,\nu]$
 is the range of integration over
$\varepsilon^{\prime}$  in (\ref{2_E_eqs_a}),
where $-W_{\rm bottom}$ and $+W_{\rm top}$ denote the minimum and the maximum
  of the band $\varepsilon_+ ({\bm k})$, respectively.
Its  bandwidth is then $W=W_{\rm bottom} + W_{\rm top}$.
Also, $\omega_{\rm uv}$ is an ultra-violet cutoff in frequency for the hidden spinwaves.
Expanding the integrand above to linear order in frequency $\omega$ then yields ultimately
the Eliashberg equation
for the wavefunction renormalization at the Fermi level, $\omega = 0$:
\begin{eqnarray}
Z-1 = \epsilon_{\rm E}
\int_{\Delta_b}^{\omega_{\rm uv}} {d\Omega\over{\sqrt{\Omega^2-\Delta_b^2}}}
\Biggl({1\over{\Omega}}- {1\over{W/Z + \Omega}}\Biggr).
\label{1st_e_eq}
\end{eqnarray}
Likewise, inverting the order of integration of the second Eliashberg equation (\ref{2_E_eqs_b})
for the inter-band energy shift yields
\begin{eqnarray}
\nu = \epsilon_{\rm E}
\int_{\Delta_b}^{\omega_{\rm uv}} {d\Omega\over{\sqrt{\Omega^2-\Delta_b^2}}}
\, {\rm ln}\, \Biggl|{W/Z + \Omega\over{\Omega}}\Biggr|
\label{2nd_e_eq}
\end{eqnarray}
at $\omega = 0$.

Long-range hSDW order exists at half filling because of perfect nesting (Fig. \ref{FS1}).
We must therefore approach criticality: $\Delta_b \rightarrow 0$.
The Eliashberg equations (\ref{1st_e_eq}) and (\ref{2nd_e_eq})
predict a Lifshitz transition of the topology of the Fermi surface
that is confirmed by making the following change of variables: 
$Z = \varepsilon_{\rm E} / \Delta_b$ and
$\cosh\, x = \Omega / \Delta_b$.
At criticality, $\Delta_b\rightarrow 0$, they yield Eliashberg equations
\begin{equation}
{\varepsilon_{\rm E}\over{W}} = {\epsilon_{\rm E}\over{W}} [I(0)-I(y)]
\quad {\rm and}\quad
{\nu\over{W}} = {\epsilon_{\rm E}\over{W}} J(y),
\label{e_eqs}
\end{equation}
where
%
%%\begin{eqnarray}
\begin{subequations}
\begin{align}
\label{I}
I(y) &=
\int_0^{\infty} dx {1\over{y  + \cosh\, x}}, \\
\label{J}
J(y) &= 
\int_0^{\infty} dx\, {\rm ln} \Biggl(1 + {y\over{\cosh\, x}}\Biggr),
\end{align}
\end{subequations}
%%\end{eqnarray}
%
with  $y = W / \varepsilon_E$. 
The quadratic dependence of $\epsilon_{\rm E}$ on Hubbard repulsion (\ref{epsilon_E})
implies that $\nu$ saturates to $W_{\rm top}$ as $U(\pi)$ diverges.
(See Fig. \ref{FS1}.)
Dividing the two Eliashberg equations (\ref{e_eqs}),
we then get the transcendental equation
\begin{equation}
y^{-1} {W\over{W_{\rm top}}} =
{I(0) - I(y)\over
{J(y)}}
\label{transcend}
\end{equation}
as $U(\pi)\rightarrow\infty$.
Notice that $y$ depends only on $W/W_{\rm top} = (t_1^{\parallel} + t_1^{\perp}) / t_1^{\perp}$
in such case.
The definite integrals (\ref{I}) and (\ref{J}) can be evaluated in closed form. 
(See Appendix \ref{ppndx_def_int}.)
Numerical solutions to the transcendental equation (\ref{transcend}) are listed in Table \ref{eps_chi}.

Last, what is the energy gap of the superconducting state
at half filling, approaching criticality?
Again, the antisymmetry displayed by the gap equations (\ref{2_E_eqs_c}) 
at half filling
with respect to the permutation of band indices
 implies perfect  $S^{+-}$ Cooper pairing:
$\Delta_1 = + \Delta$ and $\Delta_2 = -\Delta$.
(Cf. refs. \cite{mazin_08}, \cite{kuroki_08}, \cite{graser_09} and \cite{linscheid_16}.)
The last Eliashberg equation (\ref{2_E_eqs_c}) then reads
\begin{equation}
Z \Delta = \int_{-W_{\rm bottom}}^{+W_{\rm top}} d\varepsilon^{\prime} Z^{-1}
{\Delta^{\prime}\over{E^{\prime}}}\int_{\Delta_b}^{\omega_{\rm uv}} d\Omega
{\epsilon_{\rm E}\over{\sqrt{\Omega^2 - \Delta_b^2}}} 
{1\over{\Omega+E^{\prime}}}
\label{3rd_e_eq}
\end{equation}
at the Fermi level, $\omega = 0$,
where $E^{\prime} = \sqrt{[(\varepsilon^{\prime}-\nu)/Z]^2+\Delta^{\prime 2}}$.
After again making the change of variable
$\Omega = \Delta_b \cosh(x)$,
the first integral over $\Omega$ in (\ref{3rd_e_eq}) becomes
$${\rm lim}_{\Delta_b\rightarrow 0}
\int_0^{\infty} dx {\epsilon_{\rm E}\over{\Delta_b}}
\Bigl[\sqrt{\Bigl({\varepsilon^{\prime}-\nu\over{\varepsilon_{\rm E}}}\Bigr)^2+
\Bigl({\Delta^{\prime}\over{\Delta_b}}\Bigl)^2}+\cosh\, x\Bigr]^{-1} =
 {\epsilon_{\rm E}\over{\Delta^{\prime}}} {\rm ln}\Bigl(2 {\Delta^{\prime}\over{\Delta_b}}\Bigr).$$
Here we have used 
${\rm lim}_{y\rightarrow\infty} I(y) = y^{-1} {\rm ln}(2y)$.
(See Appendix \ref{ppndx_def_int}.)
Assume now the simple Bardeen-Cooper-Schrieffer (BCS) form
 for the frequency dependence of the gap\cite{schrieffer_64}:
\begin{equation}
\Delta(\omega) =
\begin{cases}
\Delta_0 & {\rm for} \quad |\omega| < \omega_c,\\
0 & {\rm otherwise},
\end{cases}
\label{BCS}
\end{equation}
but in the limit $\omega_c\rightarrow 0$.
It is therefore consistent with the previous solutions for $Z$ and for $\nu$ in the normal state.
The second integral over $\varepsilon^{\prime}$ in the gap equation (\ref{3rd_e_eq}) then becomes
$$\Delta_0 \int_{-\omega_c}^{+\omega_c} d\omega^{\prime}
(\omega^{\prime 2} + \Delta_0^2)^{-1/2} =
2 \Delta_0 \sinh^{-1}\Bigl({\omega_c\over{\Delta_0}}\Bigl).$$
Here, we have made the change of variable
$\omega^{\prime} = (\varepsilon^{\prime} - \nu)/Z$.
Substituting in the form of the wavefunction renormalization
$Z = \varepsilon_{\rm E}/\Delta_b$
into the left-hand side of the gap equation (\ref{3rd_e_eq})
plus some manipulation then yields
$${\Delta_0\over{\sinh^{-1}\Bigl({\omega_c\over{\Delta_0}}\Bigl)}} =
{\rm lim}_{\Delta_b\rightarrow 0}
2 {\epsilon_{\rm E}\over{\varepsilon_{\rm E}}} \Delta_b \, 
{\rm ln}\Bigl(2 {\Delta_0\over{\Delta_b}}\Bigr)
= 0.$$
As expected,
we therefore have a null gap due to superconductivity, $\Delta_0 = 0$, at half filling, at criticality.

Finally,
the Eliashberg energy scale
$\epsilon_{\rm E}$ can be easily estimated in the case of small circular
renormalized Fermi surface
pockets\cite{jpr_rm_18},
which occurs as $t_1^{\parallel} \rightarrow 0$.
In such case, it becomes convenient to re-express (\ref{epsilon_E}) as
\begin{equation}
\epsilon_{\rm E}(\nu) =  {3\over{(2\pi)^3}} { U^2\over{ D_+(\nu)}} 
{s_1^2 k_{\rm F}\over{\chi_{\perp} c_b v_{\rm F}^2}},
\label{e_over_w}
\end{equation}
where $U^2$ is the product of $U^2(\pi)$ with the average of
$\sin^2 (2\delta)$ around the hole-type Fermi surface pockets
% at the corner of the two-iron Brillouin zone
 shown in Fig. \ref{FS1}.
Here, $k_{\rm F}$ and $v_{\rm F}$ are the Fermi wavenumber and the Fermi velocity, respectively.
They are given by
$k_F = a^{-1} (2\pi x_0)^{1/2}$,
where $x_0$ denotes the concentration of electrons/holes in each Fermi surface pocket,
and by
$v_F = 2 t_1^{\perp} a^2 k_F$.
The solution to the Eliashberg equations (\ref{e_eqs})
yields $\epsilon_{\rm E}  \cong W/3$.
(See Table \ref{eps_chi}.)
Expression (\ref{e_over_w}) then implies that
 the effective interaction strength scales as
$U\propto x_0^{1/4}$.
Further, expression (\ref{s_2dlt})
yields the result
$\sin \, 2\delta({\bm k}) \cong [(t_2^{\perp}/i) / 2 t_1^{\perp}] (k_F a)^2 (\sin\, 2 \phi)$,
where $\phi$ is the angle that  ${\bm k}$ makes about the center of the Fermi surface pocket.
The Eliashberg energy scale is thereby  given explicitly by
the following expression at criticality\cite{jpr_rm_18}, as $t_1^{\parallel} \rightarrow 0$:
\begin{equation}
\epsilon_{\rm E} = {3\over 32} \Biggl({x_0\over{2\pi}}\Biggr)^{3/2}
{U^2(\pi)\over{a^2 D_+(\nu)}} {s_1^2\over{a^2 \chi_{\perp}}}
{|t_2^{\perp}|^2\over{(c_b/a) |t_1^{\perp}|^4}}.
\label{est_eps_e}
\end{equation}
The solution $\epsilon_{\rm E} \cong W/3$ listed in Table \ref{eps_chi}
%Comparing again this estimate with solutions to the Eliashberg equation
then yields that the area of the electron/hole Fermi surface pockets
shown in Fig. \ref{FS1}
is related to the Hubbard repulsion by $U(\pi)\propto x_0^{-3/4}$.
We therefore conclude that the effective interaction strength $U$ vanishes
with the strength of the Hubbard repulsion\cite{jpr_rm_18} as $U(\pi)^{-1/3}$.
In the case where the spectrum
$\omega_b ({\bm q})$
 of hidden spin fluctuations is {\it fixed},
this justifies the neglect of vertex corrections
to the self-energy corrections shown by Fig. \ref{EFD}
% to their interaction with electrons,
%%hidden spin fluctuations,
% (\ref{E-hSW_xy}) and (\ref{E-hSW_z}),
at large Hubbard repulsion, $U(\pi) \rightarrow\infty$.

As on-site repulsion $U_0$ grows strong,
the Eliashberg equations (\ref{2_E_eqs_a})-(\ref{2_E_eqs_c})
therefore predict a Lifshitz transition from unrenormalized Fermi surfaces shown in Fig. \ref{FS0}
 to renormalized Fermi surface pockets show in Fig. \ref{FS1}.
The groundstate remains  an hSDW at half filling due to nested Fermi surface pockets
at the corner of the folded Brillouin zone.
% (see Fig. \ref{FS1}),
% at strong Hubbard repulsion $U_0$.
It must be emphasized, however, that the spectral weight of the renormalized
Fermi surface pockets is vanishingly small:
$Z^{-1} = \Delta_b /\varepsilon_{\rm E} \rightarrow 0$ at criticality,
$\Delta_b\rightarrow 0$.
This implies that the hSDW state at half filling is in fact a Mott insulator.
It is also important to mention that
these results for the Lifshitz transition confirm
previous ones that start from the other side of the QCP at $\Delta_b = 0$.
They were based on an Eliashberg theory
 in the particle-hole channel
for the long-range ordered hSDW state\cite{jpr_rm_18}.

\subsection{Weak Electron Doping}
We will now obtain solutions to the Eliashberg equations (\ref{E_eqs})
at small deviations in the electron density from half filling.  
In the normal state, $\Delta_1 = 0 = \Delta_2$,
the corresponding  equations
for the wavefunction renormalizations and for the band shifts read
%
%%\begin{eqnarray}
\begin{subequations}
\begin{align}
\label{n_E_eqs_a}
\quad\qquad Z_n - 1 =& {3\over 2}
\int {d^2 k^{\prime}\over{(2\pi)^2}}  U^2(\pi) 
{s_1^2\over{\chi_{\perp}}} {\sin^2[\delta({\bm k})+\delta({\bm k}^{\prime})]\over{Z_{\bar n}\, \omega_b({\bm q})}} \cdot \nonumber \\
& \cdot {1\over{[\omega_b({\bm q})+|\varepsilon_{+}({\bm k}^{\prime})-\nu_{\bar n}| / Z_{\bar n}]^2}} ,\\
\label{n_E_eqs_b}
\nu_n-({\rm sgn}\, n)\mu_0 =&  {3\over 2}
\int {d^2 k^{\prime}\over{(2\pi)^2}}  U^2(\pi) 
{s_1^2\over{\chi_{\perp}}} {\sin^2[\delta({\bm k})+\delta({\bm k}^{\prime})]\over{Z_{\bar n}\, \omega_b({\bm q})}} \cdot \nonumber \\
& \cdot {{\rm sgn}[\nu_{\bar n}-\varepsilon_{+}({\bm k}^{\prime})]\over{\omega_b({\bm q})+|\varepsilon_+({\bm k}^{\prime})-\nu_{\bar n}| / Z_{\bar n}}} . 
\end{align}
\end{subequations}
%%\end{eqnarray}
%
Above, $\nu_1 = - \mu_1$ and $\nu_2 = + \mu_2$ are the staggered band shifts.
Also, the identity
\begin{equation}
\sin[\delta({\bm k}) + \delta({\bm k}^{\prime})] = \sin[\delta({\bar{\bm k}}) + \delta({\bar{\bm k}}^{\prime})]
\label{id}
\end{equation}
has been applied above in the case $n=1$ for the anti-bonding ($-$) band.
(See Appendix \ref{ppndx_m_e}.)
Assume, in particular, that the chemical potential is positive, but small:
$\mu_0\rightarrow 0+$.  
Assume, next, a linear response $\delta Z_1$ and $\delta Z_2$
with respect to the wavefunction renormalization at half filling, $Z_1=Z$ and $Z_2=Z$,
along with a linear response $\delta\nu_1$ and $\delta\nu_2$
with respect to the staggered band shifts at half filling, $\nu_1 = \nu$ and $\nu_2 = \nu$.
Taking a variation of (\ref{n_E_eqs_a}) yields one linear equation per band, $n = 1, 2$.
Adding and subtracting these yields the following linear relations in terms of even and odd variations
with respect to half filling:
\begin{eqnarray}
Z\, \delta Z(+) & = A\, \delta Z(+) - B\, \delta\nu(+) , \nonumber\\
\quad\qquad 0 & = A\, \delta Z(-) - B\, \delta\nu(-) ,
\label{Z}
\end{eqnarray}
where
$\delta Z (\pm) = {1\over 2}(\delta Z_2 \pm \delta Z_1)$ and
$\delta\nu (\pm) = {1\over 2}(\delta\nu_2 \pm \delta\nu_1)$ .
Here, we have constants
\begin{eqnarray}
\label{A}
A &=& {3\over 2}
\int {d^2 k^{\prime}\over{(2\pi)^2}} U^2(\pi)
{s_1^2\over{\chi_{\perp}}} {\sin^2[\delta({\bm k})+\delta({\bm k}^{\prime})]\over{\omega_b({\bm q})}}
{1\over{Z}} {|\varepsilon_+({\bm k}^{\prime})-\nu|/Z\over{[\omega_b({\bm q})+|\varepsilon_+({{\bm k}^{\prime}})-\nu| / Z]^3}} , \nonumber\\
\label{B}
B &=& {3\over 2}
\int {d^2 k^{\prime}\over{(2\pi)^2}} U^2(\pi)
{s_1^2\over{\chi_{\perp}}} {\sin^2[\delta({\bm k})+\delta({\bm k}^{\prime})]\over{\omega_b({\bm q})}}
{1\over{Z}} {{\rm sgn}[\nu-\varepsilon_+({\bm k}^{\prime})]\over{[\omega_b({\bm q})+|\varepsilon_+({\bm k}^{\prime})-\nu| / Z]^3}} . \nonumber \\
\end{eqnarray}
Likewise, taking a variation of (\ref{n_E_eqs_b}) yields a second linear equation per band, $n = 1, 2$.
Adding and subtracting these as well yields two more linear relations in terms of even and odd variations
with respect to half filling:
\begin{eqnarray}
\ \ \qquad \nu\, \delta Z(+) + Z\, \delta\nu(+) & = (E-C)\delta\nu(+) + D\, \delta Z(+) , \nonumber\\
 Z\mu_0 + \nu\, \delta Z(-) - Z\delta\nu(-) & = (E-C)\delta\nu(-) + D\, \delta Z(-) .
\label{nu}
\end{eqnarray}
Here, we have constants
\begin{eqnarray}
C &=& {3\over 2}
\int {d^2 k^{\prime}\over{(2\pi)^2}} U^2(\pi)
{s_1^2\over{\chi_{\perp}}} {\sin^2[\delta({\bm k})+\delta({\bm k}^{\prime})]\over{\omega_b({\bm q})}}
{1\over{Z}} {1\over{[\omega_b({\bm q})+|\varepsilon_+({\bm k}^{\prime})-\nu| / Z]^2}} , \nonumber\\
D &=& {3\over 2}
\int {d^2 k^{\prime}\over{(2\pi)^2}} U^2(\pi)
{s_1^2\over{\chi_{\perp}}} {\sin^2[\delta({\bm k})+\delta({\bm k}^{\prime})]\over{\omega_b({\bm q})}}
{1\over{Z}} {{\rm sgn}[\nu-\varepsilon_+({\bm k}^{\prime})] |\varepsilon_+({\bm k}^{\prime})-\nu|/Z
\over{[\omega_b({\bm q})+|\varepsilon_+({\bm k}^{\prime})-\nu| / Z]^2}} , \nonumber \\
\end{eqnarray}
and
\begin{equation}
E = {3\over 2}
\int {d^2 k^{\prime}\over{(2\pi)^2}} U^2(\pi)
{s_1^2\over{\chi_{\perp}}} {\sin^2[\delta({\bm k})+\delta({\bm k}^{\prime})]\over{\omega_b({\bm q})}}
{2\, \delta [\nu-\varepsilon_+({\bm k}^{\prime})]
\over{\omega_b({\bm q})+|\varepsilon_+({\bm k}^{\prime})-\nu| / Z}} .
\label{E}
\end{equation}
Collecting terms in (\ref{Z}) and in (\ref{nu}), we get
\begin{equation}
\delta Z(+) = -{B\over{F}}\, \delta\nu(+)\quad {\rm and}\quad \delta Z(+) = - {2 Z - E\over{G}}\,\delta\nu(+)
\label{even}
\end{equation}
in the even channel,
and we get
\begin{equation}
\delta Z(-) = {B\over{A}}\, \delta\nu(-)\quad {\rm and}\quad Z\mu_0 = E\,\delta\nu(-) - G\, \delta Z(-)
\label{odd}
\end{equation}
in the odd channel,
where $F = Z - A$ and $G = \nu - D$.  These constants are then
\begin{eqnarray}
\label{F}
F &=& {3\over 2}
\int {d^2 k^{\prime}\over{(2\pi)^2}} U^2(\pi)
{s_1^2\over{\chi_{\perp}}} {\sin^2[\delta({\bm k})+\delta({\bm k}^{\prime})]\over{Z}}
{1\over{[\omega_b({\bm q})+|\varepsilon_+({\bm k}^{\prime})-\nu| / Z]^3}} , \\
\label{G}
G &=& {3\over 2}
\int {d^2 k^{\prime}\over{(2\pi)^2}} U^2(\pi)
{s_1^2\over{\chi_{\perp}}} {\sin^2[\delta({\bm k})+\delta({\bm k}^{\prime})]\over{Z}}
{{\rm sgn}[\nu-\varepsilon_+({\bm k}^{\prime})]
\over{[\omega_b({\bm q})+|\varepsilon_+({\bm k}^{\prime})-\nu| / Z]^2}} .
\end{eqnarray}
In deriving expression (\ref{F}),
the first Eliashberg equation (\ref{n_E_eqs_a}) at half filling
$Z-1 = C$
has been approximated by $Z = C$.
This is exact at criticality, $\Delta_b\rightarrow 0$.
And in deriving expression (\ref{G}),
the second Eliashberg equation (\ref{n_E_eqs_b}) for $\nu$ at half filling
has been applied.

We shall now evaluate the constants above that determine the linear response of
the Eliashberg equations in the normal state
driven by weak electron doping with respect to half filling:
  (\ref{n_E_eqs_a}) and (\ref{n_E_eqs_b}),
as $\mu_0\rightarrow 0+$.
Criticality is again assumed at half filling: $\Delta_b\rightarrow 0$.
Let us begin by evaluating the constant $G$ (\ref{G}).
First, average it over the Fermi surface:
$G \rightarrow [D_+(\nu)]^{-1} (2\pi)^{-2} \int_{\rm BZ} d^2 k\, G \, \delta[\varepsilon_+({\bm k})-\nu]$.
Second, replace   the integrals over momentum with the product of $\Omega$ and
the spectral density (\ref{U2F}) at half filling:
$U^2 F_0 (\Omega) = \epsilon_{\rm E}/\sqrt{\Omega^2 - \Delta_b^2}$ for $\Omega > \Delta_b$,
and $U^2 F_0(\Omega) = 0$ otherwise.  This yields
\begin{equation}
G = \int_{\Delta_b}^{\omega_{\rm uv}} d\Omega {\epsilon_{\rm E}\over{\sqrt{\Omega^2 - \Delta_b^2}}}
\, \Omega \int_{-W_{\rm bottom}}^{+W_{\rm top}} d\varepsilon^{\prime} Z^{-1}
{{\rm sgn}(\nu-\varepsilon^{\prime})\over
{[\Omega + |\varepsilon^{\prime}-\nu|/Z]^2}}.
\end{equation}
Third, perform the first integral over the energy band $\varepsilon_+({\bm k})$ in the limit
of strong on-site repulsion,
$U(\pi)\rightarrow\infty$,
in which case  $\nu$ approaches the top of the band, $W_{\rm top}$.
It is equal to $\Omega^{-1} - (W/Z + \Omega)^{-1}$.
Fourth,
make the change of variable $\Omega = \Delta_b \cosh(x)$
and take  the limit $\Delta_b\rightarrow 0$.
This yields $G = \epsilon_{\rm E}\, y\, I(y)$,
where $I(y)$ is the definite integral (\ref{I}),
with $y=W/\varepsilon_{\rm E}$.
A closed-form expression for $I(y)$ is obtained in Appendix \ref{ppndx_def_int}.

\begin{table}
%\hspace*{3cm}
\begin{tabular}{|c|c|c|}
\hline
response/variation &\ $\delta Z$\ &\ $\delta \nu$\ \\
\hline
$\delta Z$\ & $A = {\varepsilon_{\rm E}\over{\Delta_b}}-F$\ {\rm with}\ $F = {1\over 2}{\epsilon_{\rm E}\over{\Delta_b}}[{\pi\over 2}-I(y)-y I^{\prime}(y)]$\ & $B = {1\over 2}{\epsilon_{\rm E}\over{\Delta_b^2}}[I^{\prime}(y)-I^{\prime}(0)]$\  \\
$\delta \nu$\ & $D = \nu-G$\ {\rm with}\ $G = \epsilon_{\rm E} y I(y)$\ & $E = \pi {\epsilon_{\rm E}\over{\Delta_b}}$\ \\
\hline
\end{tabular}
\caption{Coefficients of the linear response to weak electron doping of 
the Eliashberg equations (\ref{n_E_eqs_a}) and (\ref{n_E_eqs_b}),
at criticality $\Delta_b\rightarrow 0$: Eqs. (\ref{Z}) and (\ref{nu}).
A closed form expression for the definite integral $I(y)$
is  given in Appendix \ref{ppndx_def_int}, where  $y = W/\varepsilon_{\rm E}$.}
\label{A_G}
\end{table}

The remaining constants can be evaluated in a similar way.  In particular,
applying the same set of steps above to the expression for the constant $F$ (\ref{F})
yields the definite integral
\begin{equation}
F = {1\over 2} {\epsilon_{\rm E}\over{\Delta_b}}
\int_0^{\infty} dx {1\over{\cosh x}}\Biggl[1 - {1\over{(1 + y\, {\rm sech}\, x)^2}}\Biggr]
\label{f}
\end{equation}
at criticality, $\Delta_b\rightarrow 0$,
 where $y = W/\varepsilon_{\rm E}$.
It is shown in Appendix \ref{ppndx_def_int} that (\ref{f}) reduces to the closed-form expression
$F = {1\over 2}(\epsilon_{\rm E} / \Delta_b)[{\pi\over 2} - I(y) - y\, I^{\prime}(y)]$,
where $I^{\prime}(y)$ denotes the derivative of $I(y)$.
Likewise, performing the same set of steps on the expression for the constant $B$ (\ref{B})
yields the definite integral
\begin{equation}
B = {1\over 2} {\epsilon_{\rm E}\over{\Delta_b^2}}
\int_0^{\infty} dx \Biggl[{1\over{(\cosh x)^2}}-{1\over{(y + \cosh x)^2}}\Biggr]
\label{b}
\end{equation}
at criticality.
Comparison with the definite integral (\ref{I}) therefore yields the expression
$B = {1\over 2}(\epsilon_{\rm E} / \Delta_b^2)[I^{\prime}(y) - I^{\prime}(0)]$.
And recall that a closed-form expression for the constant $A$ is obtained
from that for $F$ above through the identity $A = Z - F$.
Last, performing the same set of steps on the expression for the constant $E$ (\ref{E})
yields the definite integral
\begin{equation}
E = 2 {\epsilon_{\rm E}\over{\Delta_b}}
\int_0^{\infty} dx {1\over{\cosh x}}
= \pi {\epsilon_{\rm E}\over{\Delta_b}}
\end{equation}
at criticality.
This completes the evaluation of the constants that determine the linear response
of the renormalized electronic structure shown in Fig. \ref{FS1}
 to weak electron doping at criticality,
$\Delta_b\rightarrow 0$.

\begin{table}
%\hspace*{3cm}
\begin{tabular}{|c|c|c|c|c|}
\hline
$W/W_{\rm top}$ &\ $\varepsilon_{\rm E}/W$\ &\ $\epsilon_{\rm E}/W$\ &\ $\chi_{\rm E}$\ &\ $X_{\rm E} W$\ \\
\hline
$1.0$ & $0.343$ & $0.366$ & $2.045$ & $8.671$ \\
$1.1$ & $0.298$ & $0.303$ & $2.120$ & $9.673$ \\
$1.2$ & $0.261$ & $0.255$ & $2.200$ & $10.780$ \\
$1.3$ & $0.231$ & $0.218$ & $2.284$ & $12.001$ \\
$1.4$ & $0.206$ & $0.189$ & $2.373$ & $13.344$ \\
$1.5$ & $0.185$ & $0.165$ & $2.467$ & $14.821$ \\
\hline
\end{tabular}
\caption{Numerical solutions of Eliashberg equations at half filling, at criticality,
in the limit $U(\pi)\rightarrow \infty$: Eq. (\ref{transcend}).
Also listed are the susceptibilities about half filling:
$\delta\mu_1 = \chi_{\rm E} \, \mu_0 = \delta\mu_2$ and
$\delta Z_2 / Z = X_{\rm E} \mu_0 = - \delta Z_1 / Z$.
Note that $W/W_{\rm top} = 1 + (t_1^{\parallel}/t_1^{\perp})$.}
\label{eps_chi}
\end{table}

In conclusion,
at weak electron doping,
the normal-state Eliashberg equations  (\ref{n_E_eqs_a}) and (\ref{n_E_eqs_b}) 
yield independent linear-response equations in the even and in the odd channels,
(\ref{even}) and (\ref{odd}). 
The coefficients of the linear response are summarized by Table \ref{A_G}.
In the even channel, we thereby get
$\delta Z(+) = 0$ and $\delta\nu(+) = 0$ if $B G \neq (2Z -E) F$.
Notice that $B$, $G$ and $F$ are positive, while
$2Z-E = (2\varepsilon_{\rm E} - \pi \epsilon_{\rm E})/\Delta_b$ is negative
by Table \ref{eps_chi}.
The former inequality is therefore valid,
and we get $\delta Z_1 = -\delta Z_2$ and $\delta\mu_1 = \delta\mu_2$.
And in the odd  channel, (\ref{odd}) yields
\begin{equation}
\delta Z(-) = \chi_{\rm E} {B\over A} \mu_0 \quad {\rm and} \quad
\delta\nu(-) = \chi_{\rm E}\, \mu_0 ,
\label{lin_rep_odd}
\end{equation}
with susceptibility
$\chi_{\rm E} = Z / (E- {B G \over{A}})$.
%We have calculated $\chi_{\rm E}$
The latter can be calculated from the previous closed-form expressions
for the constants $A$ thru $G$ that are listed in Table \ref{A_G},
and the results are listed in Table \ref{eps_chi}.
Importantly, $\chi_{\rm E}$ is positive at $W/W_{\rm top}$ between $1.0$ and $1.5$,
which corresponds to at most weak eccentricity in the electron/hole Fermi surface pockets
at the corner of the two-iron Brillouin zone.
Recall that $\delta\nu(-) = {1\over 2}(\delta\mu_1 + \delta\mu_2)$
is the average chemical-potential shift,
which is equal to $\delta\mu_1 = \delta\mu_2$.
The latter and (\ref{lin_rep_odd}) therefore imply
 a {\it rigid shift} of the renormalized electronic structure
at half filling by a chemical-potential shift
% $\delta\mu_1 = \delta\mu_2$
proportional to the electron doping.
Figure \ref{FS2} is such a rigid shift of Fig. \ref{FS1}.
Also recall that $\delta Z(-) = {1\over 2}(\delta Z_2 - \delta Z_1)$,
which is equal to $\delta Z_2 = -\delta Z_1$.
Upon electron doping,
the latter and (\ref{lin_rep_odd}) imply, on the other hand,
 that the wavefunction renormalization increases
with respect to $Z = \varepsilon_{\rm E}/\Delta_b$ on the hole Fermi surface pockets ($n=2$),
while that it decreases with respect to $Z$ on the electron Fermi surface pockets ($n=1$).
The magnitude of the equal and opposite variation in the wavefunction renormalization is
best stated as $\delta Z(-) /Z = X_{\rm E} \mu_0$, where
$X_{\rm E} = {B\over A}/(E-{B G\over A})$.
The values of $X_{\rm E} W$ listed in Table \ref{eps_chi}
 suggest that $Z_1\gsim 1$ and that $Z_2\sim 2 Z$ at electron doping greater than $x_0$.
This  will be discussed at length below and in the next section.

Yet what is the superconducting gap at weak electron doping with respect to half filling?
Inspection of the gap equations in the Eliashberg equations (\ref{E_eqs}) yields that 
they are equivalent
to the  ones {\it at} half filling to linear order in the variations
 $\delta Z_1$, $\delta Z_2$, $\delta\mu_1$ and $\delta\mu_2$,
and in the gaps
$\Delta_1$ and $\Delta_2$.
Because $\Delta_1$ and $\Delta_2$ are null at half filling,
the {\it linear} susceptibility of these quantities with electron doping $\mu_0 > 0$ is also null.
%(CLARIFY that the kernel of the gap equation COINCIDES with the kernel AT half filling to linear order!!!!!)
Any superconducting gap that opens at weak electron doping must therefore depend non-linearly on
the doping concentration. (See Fig. \ref{phase_diagram}.)

\subsection{Moderate Electron Doping}
\begin{figure}
%\hspace*{2cm}
\includegraphics[scale=1.00, angle=0]{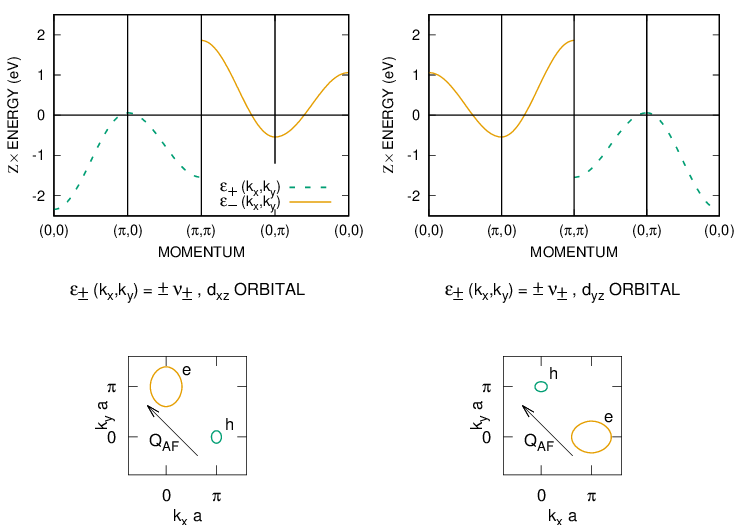}
\caption{Renormalized electron bands and Fermi surfaces at electron doping after the Lifshitz transition.
Again, the orbital character is only approximate.}
\label{FS2}
\end{figure}

Let us next seek solutions to the Eliashberg equations,
 (\ref{2_E_eqs_a}-\ref{2_E_eqs_c}),
 at moderate electron doping $x\sim x_0$.
The previous linear response due to weak electron doping predicts
a rigid shift in energy of the renormalized electronic structure at half filling displayed by Fig. \ref{FS1}.
It is depicted by Fig. \ref{FS2},
 where the top of the bonding ($+$) band lies just above the Fermi level.
The previous linear response about half filling also predicts wavefunction renormalizations
$Z_2$ and $Z_1$
for the bonding band ($n=2$) and for the anti-bonding band ($n=1$), respectively,
above and below the unique value at half filling.
What then does the third Eliashberg equation for 
the superconducting gap (\ref{2_E_eqs_c}) predict at moderate doping?

We shall follow the historical approach for the solution of the Eliashberg equations
in the case of the electron-phonon interaction\cite{schrieffer_64,scalapino_69,morel_anderson_62,mcmillan_68,ginzburg_kirzhnits_82,carbotte_90}.
In particular,
before confronting the gap equation,
it is useful first to obtain the wavefunction renormalizations of 
the two bands at the Fermi level in the normal state.
Neglecting frequency dependence, the first Eliashberg equation (\ref{2_E_eqs_a}) then yields 
the following wavefunction renormalizations at the Fermi level, $\omega = 0$:
\begin{equation}
Z_n-1 = 
\int_{\Delta_b}^{\omega_{\rm uv}} d\Omega\, U^2 F_0^{(n,{\bar n})}(\Omega,\mu_n,\mu_{\bar n})
\Biggl({1\over{\Omega}}- {1\over{W/Z_{\bar n} + \Omega}}\Biggr).
\label{1st_E_Eq}
\end{equation}
Again, the order of integration in (\ref{2_E_eqs_a}) has been reversed.
Next, assume
weak to moderate wavefunction renormalization in the anti-bonding ($-$) band
and  strong wavefunction renormalization in the bonding ($+$) band:
$\lambda_1$ such that $W/Z_1 \gg \Delta_b$, and  
$\lambda_2 \gg 1$ such that $W/Z_2 \ll \Delta_b$.
Here $\lambda_n = Z_n -1$.
Notice that the last inequality is consistent with the previous results at weak electron doping:
$\varepsilon_{\rm E}/\Delta_b < Z_2$.
The above Eliashberg equations (\ref{1st_E_Eq})
then yield the results
\begin{equation}
\lambda_2 \cong
\int_{\Delta_b}^{\omega_{\rm uv}} d\Omega\, \Omega^{-1} U^2 F_0^{(2,1)}(\Omega,\mu_2,\mu_1) ,
\label{lmbd_2}
\end{equation}
and
\begin{equation}
\lambda_1 \cong {W\over{Z_2}}
\int_{\Delta_b}^{\omega_{\rm uv}} d\Omega\, \Omega^{-2} U^2 F_0^{(1,2)}(\Omega,\mu_1,\mu_2) ,
\label{lmbd_1}
\end{equation}
or $\lambda_1 \cong (W /Z_2) \lambda_2\, {\overline{\Omega^{-1}}}$.  
The distribution in the average ${\overline{\Omega^{-1}}}$ is
normalized by the integral (\ref{lmbd_2}) because of the approximate identity
$U^2 F_0^{(1,2)}(\Omega,\mu_1,\mu_2) \cong U^2 F_0^{(2,1)}(\Omega,\mu_2,\mu_1)$.
By (\ref{U2F}),
the latter is due to the approximate identity obeyed by the density of states,
$D_-(\mu_1) \cong D_+(\mu_2)$, at  $\mu_1$ and $\mu_2$ near the bottom and near the top
of the respective bands $\varepsilon_-({\bm k})$ and $\varepsilon_+({\bm k})$.
Here, also, we have applied the identity (\ref{id}).
Because $\lambda_2\gg 1$,
we then have that $\lambda_1 \cong W\, {\overline{\Omega^{-1}}}$.
Finally,  the initial assumption of moderate $\lambda_1$ is confirmed by noting that
$W / Z_1 \cong  W /(1 + W\, {\overline{\Omega^{-1}}})\cong ({\overline{\Omega^{-1}}})^{-1}$,
which is much greater than $\Delta_b$.

We shall now show that an instability to $S$-wave Cooper pairing exists that alternates in sign
between the strong  electron-type Fermi surface of the anti-bonding ($-$) band, $n=1$,
 and the weak hole-type Fermi surface of the
bonding ($+$) band, $n=2$. (See Fig. \ref{FS2}.)
In particular, assume the simple BCS form (\ref{BCS}) for the frequency dependence 
of the respective gaps, $\Delta_1(\omega)$ and $\Delta_2(\omega)$,
 with frequency cutoffs $\omega_{c}(1)$ and $\omega_{c}(2)$.
After neglecting the frequency dependence of the wavefunction renormalizations,
the gap equations (\ref{2_E_eqs_c}) then read
\begin{eqnarray}
Z_n \Delta_n =
-2\int_{|\Delta_{\bar n}|}^{\omega_{c}({\bar n})} d E^{\prime} {\Delta_{\bar n}\over{\sqrt{E^{\prime 2}-\Delta_{\bar n}^2}}}
\int_{\Delta_b}^{\omega_{\rm uv}} d\Omega\,  U^2 F_0^{(n,{\bar n})}(\Omega;\mu_n,\mu_{\bar n})
{1\over{\Omega+E^{\prime}}} . \nonumber \\
\label{3_Eq_c}
\end{eqnarray}
Assume, further, the BCS limit: $\omega_{c}(1), \omega_{c}(2)\rightarrow 0$.
Taking the normal-state values for the wavefunction renormalizations discussed above is then valid.
Also,
the denominator above, $\Omega + E^{\prime}$, can then be replaced by $\Omega$.
After comparison with (\ref{lmbd_2}),
we thereby arrive at the gap equations
\begin{equation}
Z_n \Delta_n =
-2\int_{|\Delta_{\bar n}|}^{\omega_{c}({\bar n})} d E^{\prime} {\lambda_2
\over{\sqrt{E^{\prime 2}-\Delta_{\bar n}^2}}} \Delta_{\bar n} ,
\label{4_Eq_c}
\end{equation}
or
$\Delta_1 = - K_{1,2} \Delta_{2}$ and $\Delta_2 = - K_{2,1} \Delta_{1}$, 
with kernels
\begin{eqnarray}
K_{1,2} &=& 2{\lambda_2\over{Z_1}} \sinh^{-1}\Biggl[{\sqrt{\omega_{c}^2(2)-\Delta_2^2}\over{|\Delta_{2}|}}\Biggr]
\quad {\rm and} \quad \nonumber \\
K_{2,1} &=& 2{\lambda_2\over{Z_2}} \sinh^{-1}\Biggl[{\sqrt{\omega_{c}^2(1)-\Delta_1^2}\over{|\Delta_{1}|}}\Biggr] .\nonumber \\
\label{K}
\end{eqnarray}
Importantly, these equations imply that $\Delta_1$ and $\Delta_2$ are of opposite sign!
An $S^{+-}$ pairing instability therefore exists
between the strong and the weak Fermi
surfaces shown in Fig. \ref{FS2}.

To obtain explicit solutions of the gap equations,
it is useful to
multiply and divide these, which  yields
\begin{equation}
1 = K_{1,2} K_{2,1}
\quad {\rm and} \quad
\Bigl({\Delta_2\over{\Delta_1}}\Bigr)^2 = {K_{2,1}\over{K_{1,2}}} .
\label{gap_eqs}
\end{equation}
Taking the product of the above  then gives
$|\Delta_2 / \Delta_1| = K_{2,1} \cong 2 \sinh^{-1}[\sqrt{\omega_c^2(1)-\Delta_1^2} / |\Delta_1|]$.
Assuming $|\Delta_1|$ near $\omega_c(1)$ in turn yields
$|\Delta_2| \cong 2 \sqrt{2 \omega_c(1)}\sqrt{\omega_c(1) - |\Delta_1|}$.
Substituting  the previous into the first gap equation displayed by (\ref{gap_eqs}) then yields
\begin{equation}
1 \cong 2 {\lambda_2\over{Z_1}} {|\Delta_2|\over{\omega_c(1)}}
 \sinh^{-1}\biggl[{\omega_c(2)\over{|\Delta_2|}}\biggl],
\label{gap_eq}
\end{equation}
or 
$|\Delta_2| \sim (Z_1 / Z_2) |\Delta_1|$, with $|\Delta_1| \cong \omega_c(1)$.
%%${1\over 2}(Z_1/Z_2) [\omega_c(1)/\omega_c(2)] \cong [\Delta_2/\omega_c(2)] \sinh^{-1}[\omega_c(2)/\Delta_2]$.
This solution thereby confirms the instability of the Fermi surfaces to $S^{+-}$ pairing,
where the wavefunction renormalization $Z_1$ on the larger electron-type Fermi surface 
is of moderate size compared to unity,
while the wavefunction renormalization $Z_2$ on the smaller hole-type Fermi surface is large compared to unity.
(See Fig. \ref{FS2}.)

\section {Discussion}
The previous results of electron Fermi surface pockets and faint hole Fermi surface pockets
at the corner of the folded (two-iron) Brillouin zone,
with $S^{+-}$ Cooper pairing that  alternates in sign between them,
is compared below to a local-moment model for electron-doped iron selenide
and to high-temperature iron-selenide superconductors themselves.

\subsection{Comparison with Local-Moment Model}
A local-moment model of the electronic physics in electron-doped iron selenide exists 
that captures many of the
principal features of the above Eliashberg theory\cite{jpr_17}.
It emerges near half filling in the strong correlation limit, $U_0\rightarrow\infty$.
By (\ref{break_iso_symm}),
isospin symmetry is broken strongly along the $I^{(3)}$ axis in such a case.
In particular, doubly occupied orbital states listed in Table \ref{atomic_states} are projected out.
The following Hund-Heisenberg model in terms of the spin operators
${\bm S}_{i,d+}$ and ${\bm S}_{i,d-}$ then accurately describes the spin dynamics\cite{jpr_10}:
\begin{eqnarray}
H_{\rm HH} =
& \sum_i J_0 {\bm S}_{i,d-}\cdot{\bm S}_{i,d+} + \sum_{\langle i,j\rangle} (J_1^{\parallel} {\bm S}_{i,\alpha}\cdot{\bm S}_{j,\alpha} +
J_1^{\perp} {\bm S}_{i,\alpha}\cdot{\bm S}_{j,{\bar\alpha}}) \nonumber \\
& + \sum_{\langle\langle i,j\rangle\rangle} (J_2^{\parallel} {\bm S}_{i,\alpha}\cdot{\bm S}_{j,\alpha} +
J_2^{\perp} {\bm S}_{i,\alpha}\cdot{\bm S}_{j,\bar{\alpha}}).
\label{hund_heisenberg}
\end{eqnarray}
Here the index $\alpha$ is implicitly summed over the iron ${d+}$ and ${d-}$ orbitals.
The  intra-orbital ($\parallel$) and inter-orbital ($\perp$)  Heisenberg exchange coupling constants
are positive, and they satisfy
$J_1^{\parallel} > J_1^{\perp}$ and $J_2^{\parallel} = J_2^{\perp}$.
The above spin Hamiltonian also contains Hund's Rule exchange coupling between the orbitals,
 with a ferromagnetic coupling constant, $J_0 < 0$.
Again, the infinite-$U_0$ limit is taken, which means that the formation of spin singlets per site,
 per $d+$ or $d-$ orbital, is suppressed.
Electron hopping via the Hamiltonian $H_{\rm hop}$ (\ref{hop}) is also added at electron doping,
but in the infinite-$U_0$ limit.
Last, notice  that orbital swap,
$d-\leftrightarrow d+$, is a global symmetry of the Hund-Heisenberg Hamiltonian (\ref{hund_heisenberg}).
It is therefore most natural to consider the case where orbital swap $P_{d,{\bar d}}$
is a global symmetry of the hopping Hamiltonian $H_{\rm hop}$ (\ref{hop})  as well.
This requires the absence of mixing between the $3d_{xz}$ and $3d_{yz}$ orbitals: $t_2^{\perp} = 0$.
The latter restriction for the validity of the two-orbital $t$-$J$ model
emerges from the underlying extended Hubbard model
in the large-$U_0$ limit at half filling\cite{jpr_rm_18}.
  In  such case, for example,
the transverse spin susceptibilities of both models, $\chi_{\perp}$,
coincide only in the limit $t_2^{\perp}/i \rightarrow 0$.
%And in present case of perfect nesting, $t_2^{\parallel} = 0$, the only non-zero hopping matrix
%left is $t_1^{\perp}$.

\begin{figure}
\hspace*{1cm}
\includegraphics[scale=1.00, angle=0]{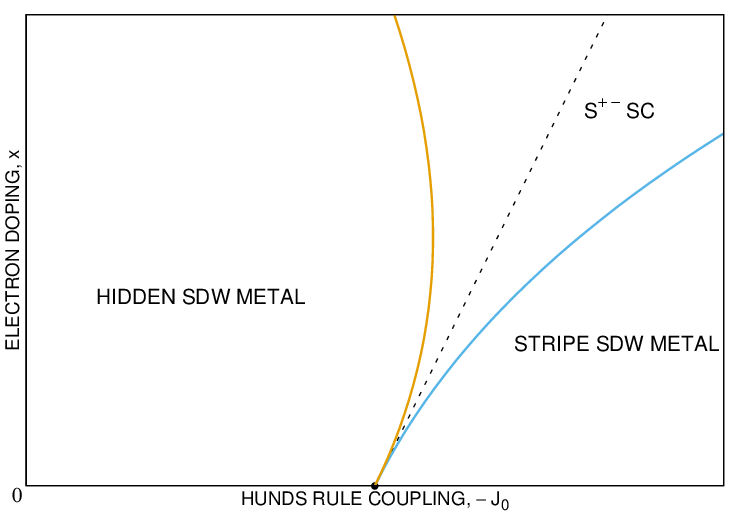}
\caption{Proposed phase diagram for local-moment model of electron-doped iron selenide
 (ref. \cite{jpr_17}).  The latter predicts the phase boundary approaching half filling (dashed line),
which separates the hSDW state from the stripe SDW state.
The intervening $S^{+-}$ superconducting phase is predicted by the Eliashberg Theory
introduced in the main text.}
\label{phase_diagram}
\end{figure}

The author 
%in ref. 
exploited the Schwinger-boson-slave-fermion representation of the correlated electron
to study the above local-moment model\cite{jpr_17}.  
Here, the correlated electron {\it fractionalizes}
into a Schwinger boson that carries spin
and a slave fermion that carries charge.
At half filling, an hSDW of the type depicted by Fig. \ref{sdw_hsdw_states}b
is predicted at $J_1^{\parallel} > J_1^{\perp}$ and at 
%weak enough 
Hund's Rule exchange coupling\cite{jpr_17,jpr_10}, $-J_0$,
below a critical one.
% $-J_{0c}$.
%(See Fig. \ref{phase_diagram}.)
%In particular, the creation operator of the correlated electron is written as
%${\tilde c}_{i,\alpha,s}^{\dagger} = b_{i,\alpha,s} f_{i,\alpha}^{\dagger}$, 
%along with the constraint per site-orbital
%%
%\begin{equation}
%b_{i,\alpha,\uparrow}^{\dagger} b_{i,\alpha,\uparrow} +
%b_{i,\alpha,\downarrow}^{\dagger} b_{i,\alpha,\downarrow} +
%f_{i,\alpha,}^{\dagger} f_{i,\alpha} = 2 s_0.
%\label{constraint}
%\end{equation}
%%
%Here, $b_{i,\alpha,s}^{\dagger}$ and $b_{i,\alpha,s}$ are creation and annihilation operators
%for Schwinger bosons, 
%$f_{i,\alpha}^{\dagger}$ and $f_{i,\alpha}$ are the corresponding operators for the slave fermions,
%and $s_0 = 1/2$ is the electron spin.
%The constraint (\ref{constraint}) 
%is enforced only on average over the bulk within mean field theory for the hSDW state.
In particular,
a quantum-critical point exists at
moderate Hund's Rule coupling $-J_{0c}$,
where the spin-excitation spectrum collapses to zero energy 
at stripe SDW wave numbers $(\pi/a,0)$ and $(0,\pi/a)$.
Specifically,
the QCP occurs at\cite{jpr_17,jpr_10}
\begin{equation}
-J_{0c} = 2 (J_1^{\parallel} - J_1^{\perp}) + 2 t_1^{\perp} x/(1-x)^2 s_0 - 4 J_2^{\parallel}
\label{J_0c}
\end{equation}
in the minimal case where only the $t_1^{\perp}({\hat{\bm x}}) = -t_1^{\perp}({\hat{\bm y}})$
hopping matrix elements are non-zero.
Here, $x$ denotes the concentration of electron doping from half filling,
while $s_0$ denotes the spin of the electron.
The quantum-critical line (\ref{J_0c})
is depicted by the dashed line in Fig. \ref{phase_diagram}.
It is possible to identify the critical normal state
of the previous Eliashberg theory
at half filling 
($\Delta_b$, $\Delta_1$, and $\Delta_2\rightarrow 0$)
with this QCP.

%

%Electron doping of this Mott insulator
% results in electron-type Fermi surface pockets at the corner of the
%%two-iron Brillouin zone\cite{jpr_17}.

Both Schwinger-boson-slave-fermion mean field theory
about the hidden N\'eel state and exact calculations on finite clusters
for the above local-moment model
find evidence for a $d_{xz}$ and a $d_{yz}$ Fermi-surface pocket
at the corner of the two-iron Brillouin zone,
at electron doping\cite{jpr_17}.
This result  agrees with the previous results based on Eliashberg Theory,
which are summarized by Fig. \ref{FS2}.
Yet how does the area of the slave-fermion  Fermi-surface pockets 
compare with that  predicted by the previous Eliashberg theory, Fig. \ref{FS2} ?
Because the slave fermions do not carry spin, we have by charge conservation that
\begin{equation}
[D_+({\rm top})+D_-({\rm bottom})]\delta\mu = 2 [D_+(0) + D_-(0)] \mu_0.
\label{conserve_charge}
\end{equation}
The left-hand and the right-hand sides above correspond, respectively,
to the cases where interactions are turned on (Fig. \ref{FS1}) and turned off (Fig. \ref{FS0}).
In the above Eliashberg theory, (\ref{2_E_eqs_a})-(\ref{2_E_eqs_c}) and (\ref{U2F}),
it has been assumed throughout that $D_+(\varepsilon)\cong D_+({\rm top})$
and that $D_-(\varepsilon)\cong D_-({\rm bottom})$,
however.
Within that approximation,
(\ref{conserve_charge}) thereby yields the susceptibility
$$\chi_{\rm E} = {\delta\mu\over{\mu_0}} \cong 2$$
from the Schwinger-boson-slave-fermion mean field theory.
It agrees with the corresponding result from Eliashberg theory listed in Table \ref{eps_chi},
 at hopping matrix element $t_1^{\parallel} \rightarrow 0$,
in which case the renormalized Fermi-surface pockets 
become perfectly circular as $U_0$ grows large ($x_0\rightarrow 0$).
This coincides with the hopping parameters studied in the local-moment model within the mean-field approximation\cite{jpr_17},
in which case only $t_1^{\perp}$  is non-zero.

And how do the predictions for wavefunction renormalization by the previous Eliashberg theory 
compare with the local-moment model\cite{jpr_17}?
A faint hole band with quasi-particle weight $1/Z_2$ that vanishes at criticality, $\Delta_b\rightarrow 0$,
is predicted by Eliashberg theory. (See Fig. \ref{FS2}.)
It crosses the Fermi level near the corner of the folded (two-iron) Brillouin zone.
At electron doping,
both mean field theory and exact calculations on finite clusters 
find {\it no} evidence for low-energy hole excitations 
%of this type 
in the two-orbital $t$-$J$ model
at momenta $(\pi/a,0)$ and $(0,\pi/a)$. 
% (CHECK!!!)
This is consistent with the previous.
Also, in the limit of large electron spin $s_0$,
 Schwinger-boson-slave-fermion mean field theory yields a
 coherent contribution to the one-particle Greens function equal to
$G_{\rm coh}({\bm k},\omega) = s_0/[\omega+\mu_1  - \varepsilon_-({\bm k})].$
This is also consistent with the appreciable quasi-particle weight $1/Z_1$
predicted by Eliashberg theory at electron doping
for the electron-type Fermi surface pockets at the corner of the two-iron Brillouin zone.

Last, exact calculations of the local-moment model for electron-doped FeSe
 on finite clusters find evidence 
for an $S$-wave Cooper pair at an energy below a continuum of states near the QCP\cite{jpr_17}.
This is consistent with the prediction made above by Eliashberg theory
for an instability of the Fermi surface to $S^{+-}$ superconductivity.
The former exact calculations also find a $D$-wave Cooper pair
at an energy below the continuum of states,
but lying above the $S$-wave Cooper pair.  
The separation in energy between the two pair states collapses to zero at the QCP.

Finally, the Schwinger-boson-slave-fermion mean field theory for the local-moment model
assumes only inter-orbital nearest neighbor hopping between iron atoms\cite{jpr_17},
$t_1^{\perp}({\bm{\hat x}}) = - t_1^{\perp}({\bm{\hat y}})$.
Such hopping of electrons leaves the two sublattices of
 the hidden N\'eel order displayed by Fig. \ref{sdw_hsdw_states}b intact.
Yet intra-orbital hopping of electrons across next-nearest neighbors, $t_2^{\parallel}$,
also leaves the sublattices for hidden N\'eel order intact.
%This observation indicates that 
Switching it on then
% next-nearest neighbor intra-orbital hopping, $t_2^{\parallel} \neq 0$, 
leaves the predictions for the
local-moment model mentioned above\cite{jpr_17} unchanged.
Direct calculation of the Schwinger-boson-slave-fermion mean field theory confirms this claim
for $|t_1^{\perp}| > 2 t_2^{\parallel}$.
Because the results of the local-moment model coincide with those of
the Eliashberg Theory for the extended Hubbard model obtained in the previous section,
we believe that the latter can remain valid
off perfect nesting (\ref{prfct_nstng}), at $t_2^{\parallel}\neq 0$.

\subsection{Comparison with Experiment}
The prediction
displayed by Fig. \ref{FS2}
of  electron-type Fermi surface pockets 
centered at the corner of the two-iron Brillouin zone
agrees with ARPES on electron-doped iron selenide\cite{qian_11,liu_12,niu_15}.
Eliashberg theory also predicts 
the opening of an $S$-wave gap over such Fermi surface pockets,
which also agrees with ARPES on these systems\cite{xu_12,peng_14,lee_14,zhao_16},
as well as with STM\cite{xue_12,fan_15,yan_15}.
Electron-electron interactions are expected to be moderately strong in iron selenide.
This rules out conventional $S$-wave pairing over the electron Fermi surface pockets
in electron-doped iron selenide.
The $S^{+-}$ Cooper pairing that is predicted here
 between the electron Fermi surface pockets and
faint hole Fermi surface pockets at the corner of the folded Brillouin zone
therefore potentially resolves the puzzling observations of isotropic pair gaps
in electron-doped iron selenide.
In particular,
the frequency cutoff for the hidden spin fluctuations 
that appears in the gap equations (\ref{K})-(\ref{gap_eq}) obtained from Eliashberg Theory
is approximately $\omega_c(1) \cong 2 k_{F 1} c_b = 2 (2\pi x_1)^{1/2} c_b/a$
in the case of the anti-bonding band, $n=1$,
where $x_1$ denotes the concentration of each electron pocket.
The Hund-Heisenberg model (\ref{hund_heisenberg}) at half filling
 predicts a hidden-spin-wave velocity
in the hSDW state of\cite{jpr_10}
\begin{equation}
c_b/a = 2 s_1 \Bigl[\Bigl(J_1^{\parallel} - J_1^{\perp}\Bigr)
\Bigl({1\over 2}J_0 + 2 J_1^{\parallel} + 2 J_2^{\perp}\Bigr)\Bigr]^{1/2},
\end{equation}
 at $J_2^{\parallel} = J_2^{\perp}$.
ARPES on electron-doped iron selenide finds a gap in the range $\Delta_1 = 10$-$20$ meV
at zero temperature\cite{xu_12,peng_14,lee_14,zhao_16}.
Let us
set $J_1^{\perp} = 0$ and assume values for the remaining
Heisenberg exchange coupling constants
and the Hund's Rule exchange coupling constant of order\cite{jpr_10}
  $|J_0|, J_1^{\parallel}, J_2^{\parallel}=J_2^{\perp} \sim 100$ meV.
At moderately small electron pockets, $2 (2\pi x_1)^{1/2} \sim 1$,
this then yields a  gap  $\Delta_1 \cong \omega_c(1)$
that is  of order 
the gap  determined by ARPES
for  low ordered moments, $s_1 < 1/2$.

The spectrum of hidden spin fluctuations
centered at the antiferromagnetic wave number ${\bm Q}_{\rm AF} = (\pi/a, \pi/a)$
is what binds together electrons into $S^{+-}$ Cooper pairs
in the present Eliashberg theory.
Recent inelastic neutron scattering studies on intercalated iron selenide\cite{pan_17}
find low-energy magnetic excitations at wave numbers {\it around} ${\bm Q}_{\rm AF}$,
but no low-energy spin excitations {\it at} ${\bm Q}_{\rm AF}$.
Such a ring of low-energy magnetic excitations is in fact consistent with
the low-energy hidden spin fluctuations 
that are exploited by the present Eliashberg theory.
In particular, both the two-orbital local-moment model discussed above
 and the underlying extended Hubbard model
for electron-doped iron selenide described in subsection \ref{xtndd_hbbrd_mdl}
predict that the low-energy hidden spin fluctuations
centered at ${\bm Q}_{\rm AF}$ are not observable
in the true-spin channel of the iron atoms\cite{jpr_rm_18,jpr_20a}.
This leaves a ring of observable spin excitations around the antiferromagnetic
wavevector ${\bm Q}_{\rm AF}$,
in agreement with inelastic neutron scattering\cite{pan_17}.

\subsection{Iron $3d_{xy}$ Orbital, Buried Hole Bands, and Polarization Correction}
Although the present study suggests that the iron $3 d_{xz} / 3 d_{yz}$  orbitals are
the principal ones in electron-doped iron selenide,
ARPES and density-functional theory indicate that the iron $3 d_{xy}$ orbital
also plays an important role\cite{yi_15}.
Indeed, it is quite possible that the two $3d_{xz}/3d_{yz}$ bands and the $3d_{xy}$ band
are approximately half filled  in electron-doped iron selenide\cite{Lee_Wen_08,Yu_Si_13}.
%An iron [Ar]$3d^7 4s^1$ atomic configuration, with 
Doubly occupied $3d_{x^2-y^2}$ and $3d_{x^2+y^2-2z^2}$ orbitals
are consistent with such fillings among the iron $3d$ bands.
%(Cf. ref. \cite{xu_song_wang_17}.)
Electron doping could 
achieve an atomic configuration [Ar]$3d^7 4s^2$ for Fe$^-$,
which in turn is consistent with such occupancies among the $3 d$ orbitals.
%fill the half filled $4s^1$ shell in such an atomic configuration.
A relatively flat and hole-type $3 d_{xy}$ band
can be added to the present Eliashberg Theory
for $3 d_{xz} / 3 d_{yz}$ electrons interacting with hidden spin fluctuations (Fig. \ref{EFD}).
(See ref. \cite{sup_mat}, Fig. S1.)
Because electrons in the $3 d_{xy}$ band
do {\it not} interact with hidden spin fluctuations,
they may be considered to be spectators.
Weak mixing of the two $3 d_{xz} / 3 d_{yz}$ bands with the $3 d_{xy}$ band
results in the expected level repulsion of 
the renormalized  electron/hole Fermi surface pockets shown in Fig. \ref{FS1}.
(See ref. \cite{sup_mat}, Fig. S3.)
This implies that the Lifshitz transition to such a renormalized band structure at half filling
is robust in the presence of weak mixing with the $3 d_{xy}$ band.
Also, within the present Eliashberg Theory,
a direct calculation of the propagator for such  $3 d_{xy}$ spectator electrons
finds that they inherit divergent 
wavefunction renormalization at the Fermi level from the $3 d_{xz} / 3 d_{yz}$ electrons at half filling.  
(See ref. \cite{sup_mat}, Eq. (S20).)
In particular, the vanishing quasi-particle weight
of the $3 d_{xz} / 3 d_{yz}$ electrons at the Fermi level, $Z^{-1}\rightarrow 0$,
implies the vanishing quasi-particle weight of the $3 d_{xy}$ electrons at the Fermi level.
(Cf. ref. \cite{Yu_Si_13}.)
Similar results are obtained when the spin-orbit interaction is included on iron atoms
that are strictly equivalent over the square lattice\cite{sup_mat}.
Last, the tips of the electron Fermi surface pockets ($n = 1$) 
shown in Fig. \ref{FS1} can acquire $3d_{xy}$ orbital character if   the electron/hole
Fermi surface pockets are large enough.
This coincides with predictions made by band-structure calculations
 on alkali-atom intercalated iron selenides\cite{maier_11,mazin_11}.
%It is left to be seen if the rigid shift of the renormalized electronic structure by
%electron doping that is predicted by the two-orbital Eliashberg Theory,
%which is depicted by Fig. \ref{FS2},
% survives the addition of the spectator $3 d_{xy}$ band. 
%That calculation, however, lies outside the scope of the present study.

%\subsection{Buried Hole Band at $\Gamma$-Point}
ARPES on electron-doped iron selenide finds hole-type bands at
 the center of the unfolded Brillouin zone
that lie below the bottom of the electron-type bands at
the corner of the folded Brillouin zone\cite{qian_11,liu_12,lee_14,zhao_16,niu_15}.
Schwinger-boson-slave-fermion mean field theory of
 the local-moment model mentioned previously
 finds evidence for incoherent hole bands
 buried below the Fermi level at the $\Gamma$-point as well\cite{jpr_17}.
The present calculations based on  Eliashberg theory
 do not predict such hole-type bands, however.
% this hole-type band that is buried below the Fermi level, however.
%It is important to mention that
% the anti-bonding band $\varepsilon_-({\bm k})$
%shows hole-type curvature at ${\bm k} = 0$
Figure \ref{FS0} shows bare bonding and anti-bonding bands near the Fermi level
that become degenerate at momenta ${\bm k} = (0,0)$ and $(\pi/a, \pi/a)$
in the unfolded Brillouin zone.
The two bands show opposite curvatures at these points
%This occurs 
when the hybridization between the $3d_{xz}$ and $3 d_{yz}$ orbitals
 lies inside the window
 $t_1^{\parallel} < 2 |t_2^{\perp}| <  |t_1^{\perp}|$.
%(See Fig. \ref{FS0}.)
%The anti-bonding band $\varepsilon_-({\bm k})$
% also lies below the Fermi level there.
%Perfect nesting (\ref{prfct_nstng}) implies
%that an equal and opposite electron-type dispersion exists
%in the bonding band $\varepsilon_{+} ({\bm k})$ at ${\bm k} = (\pi/a, \pi/a)$,
%above the Fermi level.
%The  hole-type dispersion of the anti-bonding ($-$) band
% lies {\it above} the renormalized Fermi level for the hole band, $\mu_1$ , however,
%which lies just above the bottom of that band.
%The momentum dependence of the band shifts  $\mu_n$
%have been ignored in the present Eliashberg theory calculations,
%as well as that of the wavefunction renormalizations $Z_n$,
%and that of the quasi-particle energy gaps $\Delta_n$.
It is possible that including momentum and/or frequency dependence
in the present Eliashberg Theory
opens a gap at the Fermi level at these $\Gamma$-points in momentum space.
%leaves the hole-type dispersion in $\varepsilon_-({\bm k})$
% at ${\bm k} = 0$
% below the renormalized Fermi level there, $\mu_1 (0)$.
Such a calculation lies outside the scope of the present one, however.

Absent from the previous calculations of
 the electron self-energy corrections within Eliashberg Theory, Fig. \ref{EFD},
is the polarization correction\cite{schrieffer_64,scalapino_69}
 to the propagator (\ref{D}) for hidden spinwaves. 
Recall the wavefunction renormalization for electrons near the Fermi surface at half filling:
$Z = \varepsilon_{\rm E} / \Delta_b$.
The  former polarization correction therefore {\it vanishes} at the QCP, where $\Delta_b\rightarrow 0$.
Proximity to the QCP is assumed throughout, which justifies the neglect of the polarization
correction to the propagation of hidden spinwaves.

\section{Conclusions}
We have shown above how low-energy hidden spin fluctuations
 near the wavevector for the checkerboard on the square lattice of iron atoms
 in electron-doped iron selenide
lead to superconductivity, 
with isotropic Cooper pairs that alternate in sign between 
strong electron Fermi surface pockets and faint hole Fermi surface pockets.
(See Fig. \ref{FS2}.)
By contrast with the incipient-band mechanism for $S^{+-}$ superconductivity\cite{linscheid_16},
the latter do  {\it not} coincide with the hole bands buried below the Fermi level
at the center of the Brillouin zone.
%, however. (Cf. ref. \cite{linscheid_16}.)
Both electron and hole Fermi surface pockets lie at the corner of the folded (two-iron) Brillouin zone.
A comparison of the gap that is predicted to open over the electron Fermi surface pockets
with that observed by ARPES\cite{xu_12,peng_14,lee_14,zhao_16}
 is consistent with short-range hidden magnetic order,
with a moderate to weak ordered moment.

Like true spin fluctuations in the case of iron-pnictide materials\cite{mazin_08,kuroki_08,graser_09},
the hidden spin fluctuations studied here are due to nested Fermi surfaces.
%unlike true spin fluctuations,
In the present case,
however, 
 the exchange of hidden spin fluctuations
give rise to significant band shifts.  In particular, Eliashberg theory reveals
that they incite a Lifshitz transition from nested Fermi surfaces at the center and
at the corner of the unfolded (one-iron) Brillouin zone to
nested Fermi surfaces at the corner of the folded Brillouin zone\cite{jpr_rm_18}.
Also, like true spin fluctuations in the case of
iron-pnictide materials\cite{mazin_08,kuroki_08,graser_09,dolgov_09,benfatto_09,ummarino_09},
hidden spin fluctuations give rise to repulsive inter-band interactions between electrons
that favor $S^{+-}$ Cooper pairing between the renormalized Fermi surface pockets.
%Unlike true spin fluctuations,
In the present case, however,
orbital matrix elements result in weak effective inter-band interactions.
%due to the exchange of  hidden spin fluctuations at strong on-site repulsion.
This  justifies the neglect of vertex corrections in Eliashberg theory\cite{schrieffer_64,scalapino_69}.

It has also been recently argued by the author
 that hidden spin fluctuations account for the ring of
low-energy spin excitations at the checkerboard wavevector observed by inelastic
neutron scattering in electron-doped iron selenide\cite{jpr_20a,pan_17}.
This, coupled with the prediction of $S^{+-}$ superconductivity described above,
suggests that hidden spin fluctuations play an important role
in high-temperature iron-selenide superconductors.

\begin{acknowledgments}
The author is indebted to Stefan-Ludwig Drechsler for suggesting
Eliashberg Theory in both the particle-hole and in the particle-particle
channels. (Cf. ref. \cite{ginzburg_kirzhnits_82}, Chapter 5.)
He also thanks Yongtao Cui for useful discussions,
and he acknowledges the hospitality of the 
Kavli Institute for Theoretical Physics.
This work was supported in part by the US Air Force
Office of Scientific Research under grant No. FA9550-17-1-0312
and by the National Science Foundation under Grant No. NSF PHY-1748958.
\end{acknowledgments}

\clearpage

%\appendix
\begin{appendix}

\section{Spin and Isospin Operators}\label{ppndx_isospin_operator}
The spin operator at iron site $i$ and orbital $\alpha$
is the usual contraction of Pauli matrices ${\boldsymbol{\sigma}}$  over spin quantum numbers:
\begin{equation}
{\bm S}_{i,\alpha} = {\hbar\over 2} 
\sum_{s=\uparrow,\downarrow} \sum_{s^{\prime}=\uparrow,\downarrow} c_{i,\alpha,s}^{\dagger} {\boldsymbol{\sigma}}_{s,s^{\prime}} c_{i,\alpha,s^{\prime}}.
\label{S_ia}
\end{equation}
The spin operator at iron site $i$ is then
\begin{equation}
{\bm S}_i = {\bm S}_{i,d+} +  {\bm S}_{i,d-}.
\label{S_i}
\end{equation}
The isospin operator at iron site $i$ and for spin $s$, on the other hand,
 is the contraction of Pauli matrices ${\boldsymbol{\tau}}$ over the two orbital quantum numbers:
\begin{equation}
{\bm I}_{i,s} = {1\over 2}
\sum_{\alpha={d+},{d-}}\sum_{\alpha^{\prime}={d+},{d-}} c_{i,\alpha,s}^{\dagger} {\boldsymbol{\tau}}_{\alpha,\alpha^{\prime}} c_{i,\alpha^{\prime},s}.
\label{I_is}
\end{equation}
The isospin operator at iron site $i$ is then
\begin{equation}
{\bm I}_i = {\bm I}_{i,\uparrow} + {\bm I}_{i,\downarrow}.
\label{I_i}
\end{equation}
Last, the operator for the tensor product of the spin with isospin is
the contraction of
 ${\boldsymbol{\sigma}} {\boldsymbol{\tau}}$
over both the spin and isospin quantum numbers:
\begin{equation}
({\bm S}\otimes{\bm I})_{i} = {\hbar\over 4}
\sum_{s=\uparrow,\downarrow} \sum_{\alpha={d+},{d-}}
\sum_{\alpha^{\prime}={d+},{d-}}\sum_{s^{\prime}=\uparrow,\downarrow}
 c_{i,\alpha,s}^{\dagger} {\boldsymbol\sigma}_{s,s^{\prime}} {\boldsymbol{\tau}}_{\alpha,\alpha^{\prime}} c_{i,\alpha^{\prime},s^{\prime}}.
\label{SI}
\end{equation}

\section{Orbital Matrix Element}\label{ppndx_m_e}
The operators that create
the eigenstates (\ref{plane_waves}) of the
electron hopping Hamiltonian, $H_{\rm hop}$, are
\begin{equation}
c_s^{\dagger}(n,{\bm k}) = {\cal N}^{-1/2} \sum_i \sum_{\alpha=0,1}
(-1)^{\alpha n} e^{i(2\alpha-1)\delta(\bm k)} e^{i{\bm k}\cdot{\bm r}_i} c_{i,\alpha,s}^{\dagger},
\label{ck}
\end{equation}
where $\alpha = 0$ and $1$ index the $d-$ and $d+$ orbitals,
and where $n=1$ and $2$ index the anti-bonding and bonding orbitals
$(-i) d_{y(\delta)z}$ and $d_{x(\delta)z}$.
The inverse of the above is then
\begin{equation}
c_{i,\alpha,s}^{\dagger} = {\cal N}^{-1/2} \sum_{\bm k} \sum_{n=1,2}
(-1)^{\alpha n} e^{-i(2\alpha-1)\delta({\bm k})} e^{-i{\bm k}\cdot{\bm r}_i} c_s^{\dagger}(n,{\bm k}).
\label{ci}
\end{equation}
Plugging (\ref{ci}) and its hermitian conjugate into the expression
for the hidden electron spin operator,
\begin{eqnarray}
{\bm S} (\pi,{\bm q}) = {1\over 2}\sum_s\sum_{s^{\prime}}
\sum_i\sum_{\alpha} (-1)^{\alpha} e^{i{\bm q}\cdot{\bm r}_i}
c_{i,\alpha,s}^{\dagger} {\boldsymbol{\sigma}}_{s,s^{\prime}} c_{i,\alpha,s^{\prime}},
\label{S}
\end{eqnarray}
yields the form
\begin{eqnarray}
{\bm S} (\pi,{\bm q}) = {1\over 2}\sum_{s}\sum_{s^{\prime}}\sum_{\bm k}\sum_{n,n^{\prime}}
{\cal M}_{n,{\bm k};n^{\prime},{\bm k}^{\prime}} \, c_s^{\dagger}(n^{\prime},{\bm k}^{\prime})
{\boldsymbol{\sigma}}_{s,s^{\prime}} c_{s^{\prime}} (n,{\bm k}),\nonumber\\
\label{s}
\end{eqnarray}
with the matrix element\cite{jpr_rm_18}
\begin{equation}
{\cal M}_{n,{\bm k};n^{\prime},{\bm k}^{\prime}} =
\begin{cases}
-i\, \sin[\delta({\bm k})-\delta({\bm k}^{\prime})] & {\rm for} \quad n^{\prime} = n ,\\
\cos[\delta({\bm k})-\delta({\bm k}^{\prime})] & {\rm for} \quad n^{\prime} \neq n .
\end{cases}
\label{A_MM}
\end{equation}
Here, ${\bm k}^{\prime} = {\bm k} - {\bm q}$.
Now replace ${\bm k}^{\prime}$ above with
${\bar{\bm k}}^{\prime} = {\bm k}^{\prime} + {\bm Q}_{\rm AF}$.
Using the identity
\begin{equation}
\delta({\bm k}^{\prime} + {\bm Q}_{\rm AF}) = \pm {\pi \over 2} - \delta({\bm k}^{\prime})
\end{equation}
yields the equivalent expression\cite{jpr_rm_18}
\begin{equation}
{\cal M}_{n,{\bm k};n^{\prime},{\bar{\bm k}}^{\prime}} =
\begin{cases}
\pm i\, \cos[\delta({\bm k})+\delta({\bm k}^{\prime})] & {\rm for} \quad n^{\prime} = n ,\\
\pm \sin[\delta({\bm k})+\delta({\bm k}^{\prime})] & {\rm for} \quad n^{\prime} \neq n .
\end{cases}
\label{m_e}
\end{equation}
Here, ${\bm k}^{\prime} = {\bm k} - {\bm q} - {\bm Q}_{\rm AF}$.

\section{Definite Integrals Approaching Criticality}\label{ppndx_def_int}
The following definite integrals appear in the solution of the Eliashberg equations (\ref{e_eqs})
at half filling, at criticality:
\begin{equation}
I(y) = \int_0^{\infty} dx {1\over{y  + \cosh\, x}},
\label{i}
\end{equation}
and
\begin{equation}
J(y) =
\int_0^{\infty} dx\, {\rm ln} \Biggl(1 + {y\over{\cosh\, x}}\Biggr).
\label{j}
\end{equation}
The first one (\ref{i}) can be evaluated directly by using the definition
$\cosh x = {1\over 2}z + {1\over 2}z^{-1}$, with $z = e^x$. 
Changing variables leads to the expression
\begin{equation}
I(y) = \int_1^{\infty} dz {2\over{z^2+2yz+1}}.
\end{equation}
Factorizing the denominator into $(z-z_+)(z-z_-)$, with 
$z_{\pm} = -y\pm\sqrt{y^2-1}$, and resolving the integrand into partial fractions
yields
\begin{equation}
I(y) = \int_1^{\infty} dz {1\over{z_+ - z_-}}\Biggl({1\over{z-z_+}}-{1\over{z-z_-}}\Biggl).
\end{equation}
Hence, we arrive at the closed-form expression
\begin{equation}
I(y) = {1\over{\sqrt{y^2-1}}} {\rm ln}\Biggl({1+y+\sqrt{y^2-1}\over{1+y-\sqrt{y^2-1}}}\Biggr).
\label{I_closed}
\end{equation}
Simplifying the argument of the logarithm above yields the equivalent expression
\begin{equation}
I(y) = {1\over{\sqrt{y^2-1}}} {\rm ln}\Biggl({1\over{y-\sqrt{y^2-1}}}\Biggr).
\label{I_closed_prime}
\end{equation}
And concerning the second definite integral (\ref{j}),
notice that ({\it i}) $d J /d y = I$ and ({\it ii}) $J(0) = 0$.  The expression
\begin{equation}
J(y) = {\pi^2\over 8} + {1\over 2}[{\rm ln}(y-\sqrt{y^2-1})]^2
\end{equation}
satisfies both conditions.  It therefore coincides with the definite integral (\ref{j}).

Further,
the constant $F$ that appears in the linear response at half filling to electron doping
 can also be evaluated in closed form.
Expression (\ref{f}) for it can be re-expressed as
\begin{equation}
F =  {\rm lim}_{\Delta_b\rightarrow 0}{1\over 2} {\epsilon_{\rm E}\over{\Delta_b}}
\Biggl[\int_0^{x_2} dx\,{\rm sech}\, x
+
{\partial\over{\partial y}}\int_0^{x_2} dx (1 + y\, {\rm sech}\, x)^{-1}\Biggr],
\end{equation}
where $x_2 = \cosh^{-1}(\omega_{\rm uv}/\Delta_b)$.
But $(1 + y\, {\rm sech}\, x)^{-1} = 1 - y (y + \cosh x)^{-1}$, which yields the identity
$$\int_0^{x_2} dx (1 + y\, {\rm sech}\, x)^{-1} 
= x_2 - y \int_0^{x_2} dx (y + \cosh x)^{-1}.$$
Substituting it above then yields the closed-form expression
\begin{equation}
F = {1\over 2} {\epsilon_{\rm E}\over{\Delta_b}}
\Bigl[{\pi\over 2} - I(y) - y I^{\prime}(y)\Bigl]
\end{equation}
as $\Delta_b \rightarrow 0$,
 where $I^{\prime}(y)$ is the derivative of (\ref{I_closed_prime}).

\end{appendix}

% Merge with supplemental material
%
% Supplemental Material for paper: August 14, 2020.

\pagebreak
\widetext
\begin{center}
\textbf{\large Supplemental Material: Superconductivity by Hidden Spin Fluctuations
 in Electron-Doped Iron Selenide}
\end{center}

\bigskip
\begin{center}
\text{Jose P. Rodriguez}
\end{center}

\medskip
\begin{center}
\it{Department of Physics and Astronomy,}
\end{center}
\begin{center}
\it{California State University at Los Angeles, Los Angeles, CA 90032}
\end{center}

%%% Prefix an "S" to all equations, figures, tables and reset the counter %%%
\setcounter{equation}{0}
\setcounter{figure}{0}
\setcounter{table}{0}
\setcounter{page}{1}
\setcounter{section}{0}
\makeatletter
\renewcommand{\theequation}{S\arabic{equation}}
\renewcommand{\thefigure}{S\arabic{figure}}
\renewcommand{\bibnumfmt}[1]{[S#1]}
\renewcommand{\citenumfont}[1]{S#1}
%%% Prefix an "S" to all equations, figures, tables and reset the counter %%%

\section{Add $3d_{xy}$ Orbital to Eliashberg Theory with $3d_{xz}/3d_{yz}$ Orbitals}
A Lifshitz transition from the unrenormalized Fermi surfaces displayed by Fig. 1 in the paper
to the renormalized Fermi surfaces displayed by Fig. 5 in the paper
is predicted by an
Eliashberg Theory for electrons in the principal $3d_{xz}/3d_{yz}$ orbitals
interacting with hidden spin fluctuations at half filling.
That analysis is found in  section IV.A of the paper.
Let us add a third $3d_{xy}$ orbital.
It then  becomes important to recall that the heights of the selenium atoms
above and below the square lattice of iron atoms make a checkerboard pattern.
Lee and Wen pointed out\cite{S_Lee_Wen_08}, however,
 that an isolated layer of iron selenide
is invariant under the glide-reflection symmetries
$T(a{\bm{\hat x}}) P_z$ and $T(a{\bm{\hat y}}) P_z$,
where $T(a{\bm{\hat x}})$ and $T(a{\bm{\hat y}})$ are unit translations along the principal axes
of the square lattice of iron atoms,
and where $P_z$ is a reflection about that square lattice.
% the $x$-$y$ plane.
This symmetry permits the introduction of  {\it pseudo momentum} quantum numbers, ${\bm{\tilde k}}$.
Specifically, plane waves within the tight-binding approximation are given by
\begin{equation}
|{\bm{\tilde k}},\alpha\rangle\rangle =
N_{\rm Fe}^{-1/2} \sum_{m,n} e^{i {\bm{\tilde k}}\cdot{\bm R}(m,n)}
[T(a{\bm{\hat x}}) P_z]^m [T(a{\bm{\hat y}}) P_z]^n |\alpha\rangle ,
\label{pseudo}
\end{equation}
where $R(m,n) = m a {\bm{\hat{x}}} + n a {\bm{\hat{y}}}$ is an iron site,
and where $|\alpha\rangle$ is the boundstate for orbital $\alpha$.
In the paper, and henceforth in this section and in the next one, 
all momentum quantum numbers $\bm k$
coincide with pseudo momentum ${\bm{\tilde k}}$.
Following Lee and Wen\cite{S_Lee_Wen_08},
assume an  energy spectrum for the $3d_{xy}$ electrons that disperses as
\begin{equation}
\varepsilon_{xy}({\bm k}) = -2t_1^{xy} ( \cos \, k_x a + \cos \, k_y a )
-2t_2^{xy} ( \cos \, k_+ a + \cos \, k_- a ) ,
\label{e_xy}
\end{equation}
where $t_1^{xy}$ and $t_2^{xy}$ are real nearest neighbor and next-nearest neighbor
hopping matrix elements, and where $k_{\pm} = k_x \pm k_y$.
Also assume 
that the $3d_{xz}/3d_{yz}$ orbitals mix with the $3 d_{xy}$ orbital,
with hopping matrix elements in momentum space of the form
\begin{equation}
\varepsilon_{xz,xy} ({\bm k}) = - 2 i t_{xz,xy} \sin k_x a 
\quad {\rm and} \quad
\varepsilon_{yz,xy} ({\bm k}) = - 2 i t_{yz,xy} \sin k_y a .
\label{mix_xyz}
\end{equation}
Here, $t_{xz,xy}$ and $t_{yz,xy}$ are real nearest neighbor hopping matrix elements that
satisfy $t_{xz,xy} = t_{yz,xy}$ by reflection symmetry.  Last, 
assume  a difference in energy of $\Delta E$ 
between the $3d_{xy}$ orbital and the degenerate $3d_{xz}/3d_{yz}$ orbitals.

\begin{figure}
\includegraphics[scale=0.65, angle=0]{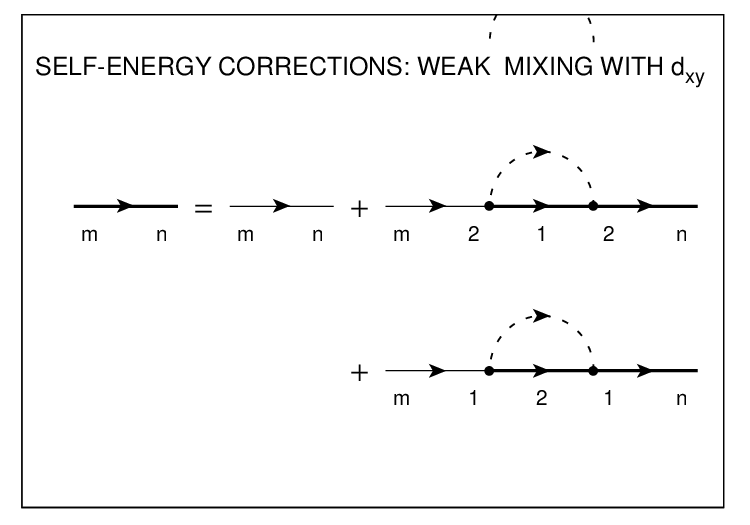}
\caption{Feynman diagrams for the original Eliashberg theory introduced in section III of the paper,
but weakly coupled to the $3d_{xy}$ band, $n = 3$. The band indices $m$ and $n$ run through
$1$, $2$, and $3$. See Eq. (\ref{S_1/G}) for the matrix inverse of the
electron propagator.}
\label{SFDI}
\end{figure}

We shall now reconsider the analysis found in section IV.A of the paper
 of the Eliashberg Theory for 
electrons in $3d_{xz}/3d_{yz}$ orbitals at half filling
that interact with hidden spin fluctuations, but with the $3d_{xy}$ orbital
described above added to it.  As in section IV.A of the paper, again assume that 
the superconducting gap is null at half filling.  The electron propagator
is then a $3\times 3$ matrix Greens function, with indices
$1$ for the anti-bonding ($-$) band, $d_{y(\delta)z}$,
$2$ for the bonding ($+$) band, $d_{x(\delta)z}$, and
$3$ for the $d_{xy}$ band.
In the absence of interactions,
the matrix inverse of the bare electron propagator
is then given by
\begin{equation}
G_0^{-1} =
\begin{bmatrix}
\omega-\varepsilon_{-} & 0 & -\varepsilon_{y(\delta)z,xy} \\
0 & \omega-\varepsilon_{+} & -\varepsilon_{x(\delta)z,xy} \\
-\varepsilon_{xy,y(\delta)z} & -\varepsilon_{xy,x(\delta)z} & \omega-\varepsilon_{xy}-\Delta E
\end{bmatrix} .
\label{S_1/G0}
\end{equation}
Above, the off-diagonal Hamiltonian matrix elements are given by
\begin{subequations}
\begin{align}
\label{mix_orbs_a}
\varepsilon_{x(\delta)z,xy} = (\cos \delta)\varepsilon_{xz,xy}-(\sin \delta)\varepsilon_{yz,xy} = \varepsilon_{xy,x(\delta)z}^* ,\\
\label{mix_orbs_b}
\varepsilon_{y(\delta)z,xy} = (\sin \delta)\varepsilon_{xz,xy}+(\cos \delta)\varepsilon_{yz,xy} = \varepsilon_{xy,y(\delta)z}^*  .
\end{align}
\end{subequations}
Because hidden spin fluctuations interact exclusively with the degenerate $3d_{xz}/3d_{yz}$ orbitals,
self-energy corrections connected to the $3d_{xy}$ orbital are null:
$\Sigma_{3,n} = 0 = \Sigma_{m,3}$.
Next,
we shall henceforth confine ourselves to the regime of weak mixing between
the $3d_{xz}/3d_{yz}$ and $3d_{xy}$ orbitals: 
$\varepsilon_{y(\delta)z,xy}, \varepsilon_{x(\delta)z,xy} \rightarrow 0$.
We shall thereby neglect the contribution of such mixing to
self-energy corrections among the $n=1$ and $n=2$ bands:
e.g.,  $\Sigma_{1,2} \cong 0$.  Figure \ref{SFDI} displays
the Feynman diagrams for the new Eliashberg equations within this approximation.
The matrix inverse of the electron propagator
within such an Eliashberg Theory therefore has the form
\begin{equation}
G^{-1} \cong
\begin{bmatrix}
Z\omega-(\varepsilon_{-}+\nu) & 0 & -\varepsilon_{y(\delta)z,xy} \\
0 & Z\omega-(\varepsilon_{+}-\nu) & -\varepsilon_{x(\delta)z,xy} \\
-\varepsilon_{xy,y(\delta)z} & -\varepsilon_{xy,x(\delta)z} & \omega-\varepsilon_{xy}-\Delta E
\end{bmatrix} ,
\label{S_1/G}
\end{equation}
where $Z$ and $\nu$ are the wavefunction renormalization and the band shift
computed in section IV.A of the paper.
Approaching quantum criticality, $\Delta_b\rightarrow 0$,
and as the interaction with hidden spin fluctuations grows strong,
$U(\pi)\rightarrow\infty$,
recall that $Z$ diverges at the Fermi level,
while $\nu$ approaches the upper band edge of $\varepsilon_+({\bm k})$.
Thus, a Lifshitz transition to the renormalized
Fermi surfaces displayed by Fig. 5 in the paper is predicted.

\begin{figure}
\includegraphics[scale=0.65, angle=0]{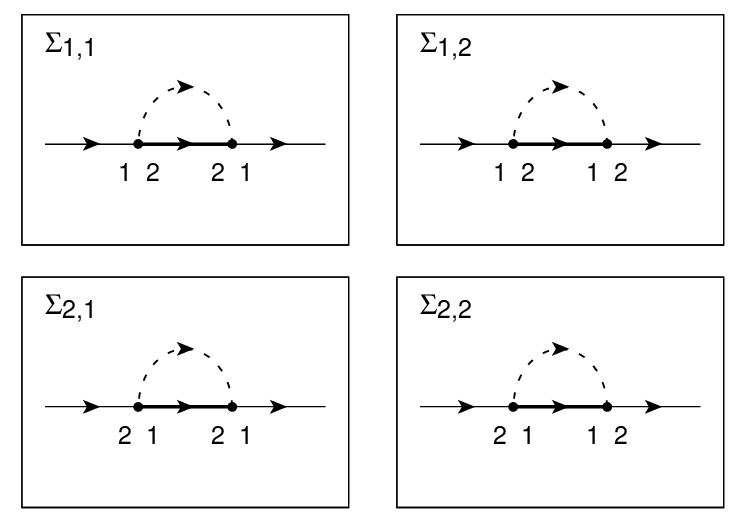}
\caption{All possible self-energy corrections due to 
interactions with hidden spin fluctuations
in the presence of mixing between the $3d_{xz}/3d_{yz}$ bands,
 $1$ and $2$, and the $3d_{xy}$ band, $3$.}
\label{SFDII}
\end{figure}

The approximate $3$-orbital Eliashberg theory encoded by the
right-hand side of (\ref{S_1/G}) neglects  contributions
to the self-energy corrections of the $3d_{xz}/3d_{yz}$ orbitals
due to mixing with the $3d_{xy}$ orbital.
We can check the validity of this appoximation by computing the Greens function
via Cramer's Rule for the matrix inverse:
\begin{equation}
G_{i,j} = (-1)^{i-j} |g^{-1} (j,i)| / |G^{-1}| .
\label{Cramer_Rule}
\end{equation}
Above, $g^{-1} (j,i)$ is the {\it minor} $2\times 2$ matrix at row $j$ and column $i$ of the 
matrix $G^{-1}$,
while $|G^{-1}|$ denotes the determinant of $G^{-1}$.
In particular, for the diagonal component $G_{1,1}$, the determinant of the minor matrix is
\begin{equation}
|g^{-1}(1,1)| = [Z\omega-(\varepsilon_{+}-\nu)](\omega-\varepsilon_{xy}-\Delta E)-|\varepsilon_{x(\delta)z,xy}|^2.
\label{minor_11}
\end{equation}
The determinant of $G^{-1}$, on the other hand, is
\begin{eqnarray}
|G^{-1}| &=& [Z\omega-(\varepsilon_{-}+\nu)][Z\omega-(\varepsilon_{+}-\nu)](\omega-\varepsilon_{xy}-\Delta E)
\nonumber \\
&& -|\varepsilon_{y(\delta)z,xy}|^2 [Z\omega-(\varepsilon_{+}-\nu)]
-|\varepsilon_{x(\delta)z,xy}|^2 [Z\omega-(\varepsilon_{-}+\nu)] .
\label{det_inv_g}
\end{eqnarray}
Cramer's Rule (\ref{Cramer_Rule}) then yields the result
\begin{equation}
G_{1,1} \cong  {1\over{Z\omega - (\varepsilon_{-}+\nu)}} + 
{|\varepsilon_{y(\delta)z,xy}|^2\over{[Z\omega-(\varepsilon_{-}+\nu)]^2 (\omega-\varepsilon_{xy}-\Delta E)}}
\label{G_11}
\end{equation}
to lowest non-trivial order in the mixing with the $3d_{xy}$ orbital.  Similar calculations yield the result
\begin{equation}
G_{2,2} \cong  {1\over{Z\omega - (\varepsilon_{+}-\nu)}} + 
{|\varepsilon_{x(\delta)z,xy}|^2\over{[Z\omega-(\varepsilon_{+}-\nu)]^2 (\omega-\varepsilon_{xy}-\Delta E)}}.
\label{G_22}
\end{equation}
Finally, the determinant of the minor matrix $g^{-1}(2,1)$ is 
$|g^{-1}(2,1)| = - \varepsilon_{y(\delta)z,xy} \varepsilon_{xy,x(\delta)z}$.  
Cramer's Rule (\ref{Cramer_Rule}) then yields
\begin{equation}
G_{1,2} \cong {\varepsilon_{y(\delta)z,xy} \varepsilon_{xy,x(\delta)z}\over{
[Z\omega - (\varepsilon_{-}+\nu)] [Z\omega - (\varepsilon_{+}-\nu)] (\omega-\varepsilon_{xy}-\Delta E)}}
\label{G_12}
\end{equation}
to lowest order in the mixing with the $3d_{xy}$ orbital.
Last, $G_{2,1} = G_{1,2}^*$.
By Fig. \ref{SFDII},
we  conclude that the contributions to the self-energy corrections among the $3d_{xz}/3d_{yz}$
orbitals due to the $3d_{xy}$ orbital are second order in the mixing. 
By comparison with (\ref{S_1/G}),
they can therefore be neglected in the regime of weak mixing with the $3d_{xy}$ orbital.
This confirms the previous assumption that $\Sigma_{1,2} \cong 0$.

\section{Effects and Properties of $3 d_{xy}$ Orbital}
\begin{figure}
\includegraphics[scale=0.65, angle=0]{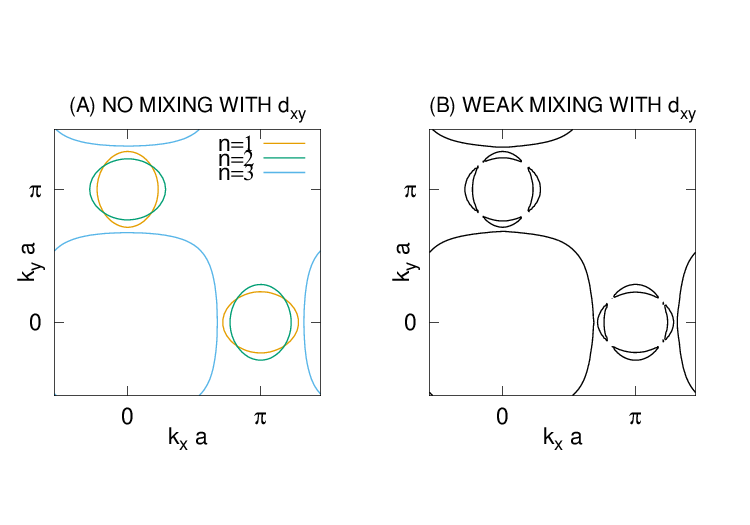}
\caption{Fermi surfaces of the $3d_{xz}/3d_{yz}$ bands ($n=1, 2$) and
of the $3 d_{xy}$ band ($n=3$), with and without weak mixing.
The $3d_{xz}/3d_{yz}$ bands are identical to Fig. 5 in the paper,
while the $3 d_{xy}$ band is hole-type and relatively flat,
with nearest neighbor and next-nearest neighbor hopping parameters
$t_1^{xy} = -40$ meV and $t_2^{xy} = -28$ meV. The splitting between the
$3 d_{xy}$ and the $3d_{xz}/3d_{yz}$ 
orbitals is set to $\Delta E = 20$ meV.
As in the paper, the momentum $(k_x,k_y)$ coincides with {\it pseudo momentum}.
(See ref. \cite{S_Lee_Wen_08} and the text.)}
\label{SFSII}
\end{figure}

The form of the inverse Greens function (\ref{S_1/G}) implies level repulsion of the degenerate
$3d_{xz}/3d_{yz}$ bands because of mixing with the $3 d_{xy}$ band.  In particular,
the Fermi surfaces are determined by the characteristic equation
$|G^{-1}(\omega)| = 0$.  
Directly exanding the latter (\ref{det_inv_g}) yields the third-order polynomial in frequency
\begin{equation}
|G^{-1}| = a_3 \omega^3 + a_2 \omega^2 + a_1 \omega + a_0 ,
\label{polynomial}
\end{equation}
with coefficients
\begin{subequations}
\begin{align}
\label{plnml_3}
a_3 &= Z^2 ,\\
\label{plnml_2}
a_2 &= - (\varepsilon_{xy}+\Delta E) Z^2 - (\varepsilon_{+} + \varepsilon_{-}) Z ,\\
\label{plnml_1}
a_1 &= (\varepsilon_{xy}+\Delta E)(\varepsilon_{+}+\varepsilon_{-})Z
-(|\varepsilon_{y(\delta)z,xy}|^2+|\varepsilon_{x(\delta)z,xy}|^2) Z + (\varepsilon_{-}+\nu)(\varepsilon_{+}-\nu),\\
\label{plnml_0}
a_0 &= -(\varepsilon_{-}+\nu)(\varepsilon_{+}-\nu)(\varepsilon_{xy}+\Delta E)
+(\varepsilon_{+}-\nu)|\varepsilon_{y(\delta)z,xy}|^2 + (\varepsilon_{-}+\nu)|\varepsilon_{x(\delta)z,xy}|^2 .
\end{align}
\end{subequations}
Above, the magnitude squared of the mixing matrix elements are given explicitly by
\begin{subequations}
\begin{align}
|\varepsilon_{x(\delta)z,xy}|^2 = {1\over 2}(1 + \cos\,2\delta)|\varepsilon_{xz,xy}|^2 -
(\sin\, 2\delta) |\varepsilon_{xz,xy} \varepsilon_{yz,xy}| +
{1\over 2}(1 - \cos\,2\delta)|\varepsilon_{yz,xy}|^2 ,\\
|\varepsilon_{y(\delta)z,xy}|^2 = {1\over 2}(1 - \cos\,2\delta)|\varepsilon_{xz,xy}|^2 +
(\sin\, 2\delta) |\varepsilon_{xz,xy} \varepsilon_{yz,xy}| +
{1\over 2}(1 + \cos\,2\delta)|\varepsilon_{yz,xy}|^2 ,
\end{align}
\label{mixing}
\end{subequations}
with $\cos\,2\delta$ and $\sin\, 2\delta$ given by expressions (6a) and (6b) in the paper.
Notice, as expected, that
$$|\varepsilon_{x(\delta)z,xy}|^2 + |\varepsilon_{y(\delta)z,xy}|^2 =
|\varepsilon_{xz,xy}|^2 + |\varepsilon_{yz,xy}|^2 .$$
Figure \ref{SFSII}b displays the resulting Fermi surfaces, with hopping matrix elements
among the $3 d_{xz} / 3 d_{yz}$ orbitals that are identical to those in
Fig. 5 of the paper for the renormalized band structure:
$t_1^{\parallel} = 100$ meV, $t_1^{\perp} = 500$ meV, $t_2^{\parallel} = 0$, and $t_2^{\perp}/i = 100$ meV.
The energy shift  between the two $3 d_{xz} / 3 d_{yz}$ bands
 is set by the staggered chemical potential $\nu = 1.7$ eV.
A relatively flat hole-type dispersion for the $3 d_{xy}$ band is taken\cite{S_Lee_Wen_08},
 with hopping matrix elements
$t_1^{xy} = -40$ meV and $t_2^{xy} = -28$ meV, along with $3d_{xz}/3d_{yz}$-$3d_{xy}$ mixing
$t_1^{xz,xy} = 20\, {\rm meV}\, = t_1^{yz,xy}$.  Last, the energy splitting between the 
$3d_{xy}$ and $3d_{xz}/3d_{yz}$ orbitals is tuned to $\Delta E = 20$ meV,
at which point the system of three bands is half filled.
Figure \ref{SFSII}a shows the corresponding Fermi surfaces in the absence of mixing with the $3d_{xy}$ band:
$t_1^{xz,xy} = 0 = t_1^{yz,xy}$.
The system of three bands is slightly electron doped in such case.
Comparison of Figs. \ref{SFSII}a and \ref{SFSII}b reveals the expected level repulsion of
electron/hole Fermi surface pockets because of mixing with the $3d_{xy}$ band.
At half filling,
the renormalized bands
 $\varepsilon_{-} ({\bm k}) + \nu$ and $\varepsilon_{+} ({\bm k}) - \nu$
inter-penetrate at energies near the Fermi level,
 forming hybridized energy dispersions similar in shape to
the {\it universal joint} in a drive shaft.

And what is the effect of the wavefunction renormalization $Z$ of the $3d_{xz}/3d_{yz}$ bands
on the $3d_{xy}$ band?  To answer this question, let us first write the determinant of the
matrix inverse (\ref{S_1/G}) as
$|G^{-1}(\omega)| = Z^2 (\omega - \omega_1)(\omega - \omega_2)(\omega - \omega_3)$,
where $\omega_1$, $\omega_2$, and $\omega_3$ are the three roots of the charactersistic equation
$|G^{-1}(\omega)| = 0$.  Near the Fermi surface of the $3d_{xy}$ band shown by Fig. \ref{SFSII}b,
$\omega_3({\bm k}) = 0$,
the former determinant can therefore be approximated by
$|G^{-1}(\omega)| = A_3 (\omega-\omega_3)$,
with a constant prefactor
\begin{equation}
A_3 =  {\partial\over{\partial\omega}}|G^{-1}|_3
    =  3 a_3 \omega_3^2 + 2 a_2 \omega_3 + a_1 = a_1.
\label{a_3}
\end{equation}
Yet the determinant of the $2\times 2$ minor matrix $g^{-1}(3,3)$ of (\ref{S_1/G}) is
\begin{equation}
|g^{-1}(3,3)| = [Z\omega - (\varepsilon_{-}+\nu)] [Z\omega - (\varepsilon_{+}-\nu)].
\label{det_mnr_33}
\end{equation}
Cramer's Rule (\ref{Cramer_Rule}) thereby yields the following diagonal component for the electron propagator in the
$3d_{xy}$ band near the Fermi surface, $\omega_3({\bm k}) = 0$:
\begin{equation}
G_{3,3}(\omega) = {Z_3^{-1}\over{\omega-\omega_3}}
\quad{\rm with}\quad
Z_3 = {a_1\over{(\varepsilon_{-}+\nu)(\varepsilon_{+}-\nu)}}.
\label{Z_3}
\end{equation}
Last, the characteristic equation at the Fermi surface is equivalent to $a_0 = 0$.
Study of the expression (\ref{plnml_0}) for the coefficient $a_0$ yields the identity
\begin{equation}
\varepsilon_{xy}+\Delta E = {|\varepsilon_{y(\delta)z,xy}|^2\over{\varepsilon_{-}+\nu}}+
{|\varepsilon_{x(\delta)z,xy}|^2\over{\varepsilon_{+}-\nu}} .
\label{e_xy_fs}
\end{equation}
%
% to lowest non-trivial order in the mixing with the $3 d_{xy}$ orbital.
After substituting it into the expression (\ref{plnml_1}) for the coefficient $a_1$,
(\ref{Z_3}) then ultimately yields
the following result for the inherited wavefunction renormalization:
\begin{equation}
Z_3 = 1 + \biggl[{|\varepsilon_{y(\delta)z,xy}|^2\over{(\varepsilon_{-}+\nu)^2}}+
{|\varepsilon_{x(\delta)z,xy}|^2\over{(\varepsilon_{+}-\nu)^2}}\biggr] Z .
\label{Z_3_fnl}
\end{equation}
The $3 d_{xy}$ band therefore inherits the divergent wavefunction renormalization of the $3d_{xz}/3d_{yz}$ bands
at the Fermi level,
$\omega\rightarrow 0$:
 $Z\rightarrow\infty$ implies $Z_3\rightarrow\infty$.

And how do the above results depend on changes in the nature of the three Fermi surfaces 
in the absence of mixing with the $3d_{xy}$ orbital
that are shown in Fig. \ref{SFSII}a ?
Increasing the size of the hole-type Fermi surface ($n=3$) for the $3d_{xy}$ electrons
is achieved by increasing the energy splitting $\Delta E$  between the 
$3d_{xy}$ and $3d_{xz}/3d_{yz}$ orbitals, while
increasing the size of the electron/hole Fermi surface pockets ($n=1, 2$)
for the $3d_{xz}/3d_{yz}$ electrons
is achieved by lowering the on-site repulsive energy $U(\pi)$. [See Eq. (49) in the paper.]
Either of these changes in parameters can result in $3d_{xy}$ orbital character at the tips of the
electron Fermi surface pockets, $n = 1$.
(Cf. Fig. \ref{SFSII}a.)
It    coincides with predictions made by band-structure calculations in the case of
alkali-metal intercalated iron selenides\cite{S_maier_11,S_mazin_11}.

\section{Spin-Orbit Coupling}
Consider next adding spin-orbit coupling at each iron atom:
\begin{equation}
H_{\rm SO} = \lambda_{\rm SO} {\bm L}\cdot{\bm S} 
       = \lambda_{\rm SO}\Bigl( {1\over 2}L_{+} S_{-} + {1\over 2}L_{-} S_{+} + L_z S_z \Bigr).
\label{H_SO}
\end{equation}
%For the sake of simplicity,
%take the limit of {\it equivalent} iron atoms,
%in which case
% the electrostatic background potential
%is perfectly periodic over the square lattice of iron atoms in a single layer of FeSe.
%In particular,
%turn off the nearest-neighbor hopping between the 
%$3d_{xz}/3d_{yz}$ orbitals and the $3d_{xy}$ orbital\cite{S_Lee_Wen_08}: $t_{xz,xy} = 0 = t_{yz,xy}$.
The principal iron orbitals at the electron Fermi surface pockets 
in FeSe are the degenerate  $3d_{xz}/3d_{yz}$ orbitals and the $3d_{xy}$ orbital.
Henceforth, we therefore shall project out the 
$3d_{x^2-y^2}$ and the $3d_{2z^2-x^2-y^2}$ orbitals\cite{S_agterberg_17,S_eugenio_vafek_18}.
Two irreducible Hilbert spaces thereby emerge under the spin-orbit Hamiltonian (\ref{H_SO}):
\begin{equation}
\{|d_{y(\delta)z},\uparrow\rangle , |d_{x(\delta)z},\uparrow\rangle , |d_{xy},\downarrow\rangle\}
\quad {\rm and} \quad
\{|d_{y(\delta)z},\downarrow\rangle , |d_{x(\delta)z},\downarrow\rangle , |d_{xy},\uparrow\rangle\}.
\label{irreducibles}
\end{equation}
Above, $x(\delta)$ and $y(\delta)$ are the orbital coordinates measured with respect to new axes
rotated by an angle $-\delta$ about the $z$ axis.
Matrix elements of $H_{\rm SO}$ between the two Hilbert spaces above are null.
On the other hand, matrix elements of $H_{\rm SO}$ within each one of the two Hilbert
spaces above are easily computed, and these are listed in Tables \ref{so_me_up} and \ref{so_me_dn}.

\begin{table}\hspace*{3cm}
\begin{tabular}{|c|c|c|c|}
\hline
$\langle m|H_{\rm SO}| n\rangle$ &\ $|d_{y(\delta)z},\uparrow\rangle$\ &\ $|d_{x(\delta)z},\uparrow\rangle$\ &\ $|d_{xy},\downarrow\rangle$\ \\
\hline
$\langle d_{y(\delta)z},\uparrow|$ & $0$ & $+i\lambda_{\rm SO}/2$ & $-e^{-i\delta}\lambda_{\rm SO}/2$ \\
$\langle d_{x(\delta)z},\uparrow|$ & $-i\lambda_{\rm SO}/2$ & $0$ & $+i e^{-i\delta}\lambda_{\rm SO}/2$  \\
$\langle d_{xy},\downarrow|$ & $-e^{+i\delta}\lambda_{\rm SO}/2$ & $-i e^{+i\delta}\lambda_{\rm SO}/2$ & $0$  \\
\hline
\end{tabular}
\caption{Matrix elements for spin-orbit coupling, Eq. (\ref{H_SO}), in the ``up'' irreducible Hilbert space.
See Eq. (\ref{irreducibles}).}
\label{so_me_up}
\end{table}
\begin{table}\hspace*{3cm}
\begin{tabular}{|c|c|c|c|}
\hline
$\langle m|H_{\rm SO}| n\rangle$ &\ $|d_{y(\delta)z},\downarrow\rangle$\ &\ $|d_{x(\delta)z},\downarrow\rangle$\ &\ $|d_{xy},\uparrow\rangle$\ \\
\hline
$\langle d_{y(\delta)z},\downarrow|$ & $0$ & $-i\lambda_{\rm SO}/2$ & $+e^{+i\delta}\lambda_{\rm SO}/2$ \\
$\langle d_{x(\delta)z},\downarrow|$ & $+i\lambda_{\rm SO}/2$ & $0$ & $+i e^{+i\delta}\lambda_{\rm SO}/2$  \\
$\langle d_{xy},\uparrow|$ & $+e^{-i\delta}\lambda_{\rm SO}/2$ & $-i e^{-i\delta}\lambda_{\rm SO}/2$ & $0$  \\
\hline
\end{tabular}
\caption{Matrix elements for spin-orbit coupling, Eq. (\ref{H_SO}), in the ``down'' irreducible Hilbert space.
See Eq. (\ref{irreducibles}).}
\label{so_me_dn}
\end{table}

In the previous sections, all momentum quantum numbers $\bm k$
coincide with pseudo momentum\cite{S_Lee_Wen_08} ${\bm{\tilde k}}$. 
Specifically, plane waves within the tight-binding approximation 
have the form (\ref{pseudo}).
Yet does pseudo momentum remain a good quantum number when
the spin-orbit interaction (\ref{H_SO})
at each iron atom is included?
Observe that $P_z^{-1} L_x P_z = -L_x$, that $P_z^{-1} L_y P_z = -L_y$,
and that $P_z^{-1} L_z P_z = +L_z$.  
Angular momentum then does not commute with a reflection about the $x$-$y$ plane,
hence pseudo momentum is no longer a good quantum number when 
the spin-orbit interaction (\ref{H_SO}) is included.

The square lattice of spin-orbit interactions (\ref{H_SO}) does, however, commute with $T(a{\bm{\hat x}})$
and with $T(a{\bm{\hat y}})$.
This   means that conventional crystal momentum ${\bm k}$ remains a good quantum number,
with conventional tight-binding plane waves
\begin{equation}
|{\bm k},\alpha\rangle\rangle =
N_{\rm Fe}^{-1/2} \sum_{m,n} e^{i {\bm k}\cdot{\bm R}(m,n)}
[T(a{\bm{\hat x}})]^m [T(a{\bm{\hat y}})]^n |\alpha\rangle .
\label{real}
\end{equation}
If we turn off the contribution due to electrostatic interactions
to hopping  between the $3d_{xz}/3d_{yz}$ orbitals and the $3d_{xy}$ orbital,
$t_{xz,xy} = 0 = t_{yz,xy}$,
then the on-site irreducible Hilbert spaces (\ref{irreducibles}) under the spin-orbit interaction
can be extended to the following sets of plane-wave states:
\begin{equation}
\{|{\bm k},d_{y(\delta)z},\uparrow\rangle\rangle , |{\bm k},d_{x(\delta)z},\uparrow\rangle\rangle , |{\bm k},d_{xy},\downarrow\rangle\rangle\}
\quad {\rm and} \quad
\{|{\bm k},d_{y(\delta)z},\downarrow\rangle\rangle , |{\bm k},d_{x(\delta)z},\downarrow\rangle\rangle , |{\bm k},d_{xy},\uparrow\rangle\rangle\}.
\label{irreducible_plane_waves}
\end{equation}
Observe, now, that $P_z|d_{x(\delta)z}\rangle = - |d_{x(\delta)z}\rangle$ 
and  $P_z|d_{y(\delta)z}\rangle = - |d_{y(\delta)z}\rangle$ ,
while  $P_z|d_{xy}\rangle = + |d_{xy}\rangle$.
By (\ref{pseudo}),
this means that the conventional crystal momentum for electrons
 in the $3d_{xz}/3d_{yz}$ orbitals is shifted
with respect to the pseudo momentum by the checkerboard wavenumber ${\bm Q}_{\rm AF}$,
while that the conventional crystal momentum for electrons  in the $3d_{xy}$ orbital coincides
with the pseudo momentum.
In conclusion, the inverse Greens function for the Eliashberg Theory in the paper,
at null superconducting gaps,
has the form
\begin{equation}
G^{-1} \cong
\begin{bmatrix}
Z\omega-(\varepsilon_{-}+\nu) & -\varepsilon_{y(\delta)z,x(\delta)z} & -\varepsilon_{y(\delta)z,xy} \\
-\varepsilon_{x(\delta)z,y(\delta)z} & Z\omega-(\varepsilon_{+}-\nu) & -\varepsilon_{x(\delta)z,xy} \\
-\varepsilon_{xy,y(\delta)z} & -\varepsilon_{xy,x(\delta)z} & \omega-{\bar\varepsilon}_{xy}-\Delta E
\end{bmatrix} ,
\label{inverse_g}
\end{equation}
where ${\bar\varepsilon}_{xy}({\bm k}) = \varepsilon_{xy}({\bm k}+{\bm Q}_{\rm AF})$,
and where the off-diagonal matrix elements above, $\varepsilon_{\alpha,\beta}$, coincide
with those for the spin-orbit interaction that are listed in Tables \ref{so_me_up} and \ref{so_me_dn}.

The determinant of the inverse Greens function (\ref{inverse_g}) is given by
\begin{eqnarray}
|G^{-1}| &=& [Z\omega-(\varepsilon_{-}+\nu)][Z\omega-(\varepsilon_{+}-\nu)](\omega-{\bar\varepsilon}_{xy}-\Delta E)
\nonumber \\
&& -|\varepsilon_{y(\delta)z,xy}|^2 [Z\omega-(\varepsilon_{+}-\nu)]
-|\varepsilon_{x(\delta)z,xy}|^2 [Z\omega-(\varepsilon_{-}+\nu)] \nonumber \\
&&-|\varepsilon_{y(\delta)z,x(\delta)z}|^2 (\omega - {\bar\varepsilon}_{xy} - \Delta E)  \nonumber \\
&& -\varepsilon_{y(\delta)z,x(\delta)z} \varepsilon_{x(\delta)z,xy} \varepsilon_{xy,y(\delta)z}
-\varepsilon_{x(\delta)z,y(\delta)z} \varepsilon_{y(\delta)z,xy} \varepsilon_{xy,x(\delta)z} .
\label{det_1/g}
\end{eqnarray}
Then by Cramer's Rule (\ref{Cramer_Rule}), the Greens functions among the $3d_{xz}/3d_{yz}$
orbitals are given by diagonal components
$G_{1,1} \cong [Z\omega -(\varepsilon_{-}+\nu)]^{-1}$ and
$G_{2,2} \cong [Z\omega -(\varepsilon_{+}-\nu)]^{-1}$,
and by off-diagonal components
$G_{1,2} \cong \varepsilon_{y(\delta)z,x(\delta)z} /
([Z\omega -(\varepsilon_{-}+\nu)] [Z\omega -(\varepsilon_{+}-\nu)])$ and $G_{2,1} = G_{1,2}^*$,
up to first order in the spin-orbit coupling, $\lambda_{\rm SO}$.
Substituting the latter
into the off-diagonal self-energy corrections depicted by Fig. \ref{SFDII} yields that
the  portion of the spin-orbit interaction, $\lambda_{\rm SO} L_z S_z$,
that is represented by the off-diagonal matrix elements 
$\varepsilon_{x(\delta)z,y(\delta)z}$ and $\varepsilon_{y(\delta)z,x(\delta)z}$
receives substantial renormalization within Eliashberg Theory.
On the other hand,
substitution of the former into Fig. \ref{SFDII}
 suggests that the divergent wave function renormalization
and the Lifshitz transition that are predicted in the absence of the $3d_{xy}$ orbital by Eliashberg Theory
survive the addition of the $3d_{xy}$ orbital and of the spin-orbit interaction.

\begin{figure}
\includegraphics[scale=0.65, angle=0]{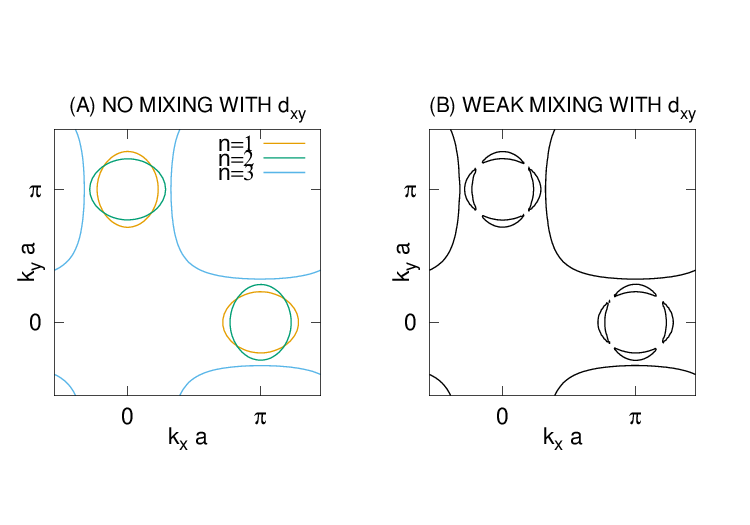}
\caption{Fermi surfaces of the $3d_{xz}/3d_{yz}$ bands ($n=1, 2$) and
of  the $3 d_{xy}$ band ($n=3$),
weakly mixing 
via the spin-orbit interactions at each  iron atom.
The $3d_{xz}/3d_{yz}$ bands are identical to those in Fig. 5 of the paper.
Also, the $3 d_{xy}$ band is identical to that in Fig. \ref{SFSII},
but the momentum $(k_x, k_y)$ coincides with the crystal momentum
shifted by the checkerboard wavevector $(\pi/a,\pi/a)$.}
%The spin-orbit coupling energy is set to
%$\lambda_{\rm SO} = 14$ meV.
\label{SFSIII}
\end{figure}

And as in the previous case of the background electrostatic potential,
do the electrons at the Fermi surface of the $3d_{xy}$ band, $\omega_3({\bm k})$,
inherit the infinite wavefunction renormalization
from the $3d_{xz}/3d_{yz}$ bands at the Fermi surface  at  half filling?
To answer this question, we first
expand the expression above for the determinant (\ref{det_1/g}).
This yields the third-order polynomial (\ref{polynomial}),
with coefficients that are given by
\begin{subequations}
\begin{align}
\label{polynomial_3}
a_3 =& Z^2 ,\\
\label{polynomial_2}
a_2 =& - ({\bar\varepsilon}_{xy}+\Delta E) Z^2 - (\varepsilon_{+} + \varepsilon_{-}) Z ,\\
\label{polynomial_1}
a_1 =& ({\bar\varepsilon}_{xy}+\Delta E)(\varepsilon_{+}+\varepsilon_{-})Z
-(|\varepsilon_{y(\delta)z,xy}|^2+|\varepsilon_{x(\delta)z,xy}|^2) Z \nonumber \\
& -|\varepsilon_{y(\delta)z,x(\delta)z}|^2 + (\varepsilon_{-}+\nu)(\varepsilon_{+}-\nu),\\
\label{polynomial_0}
a_0 =& -(\varepsilon_{-}+\nu)(\varepsilon_{+}-\nu)({\bar\varepsilon}_{xy}+\Delta E)
+(\varepsilon_{+}-\nu)|\varepsilon_{y(\delta)z,xy}|^2 + (\varepsilon_{-}+\nu)|\varepsilon_{x(\delta)z,xy}|^2 \nonumber \\
& + ({\bar\varepsilon}_{xy}+\Delta E) |\varepsilon_{y(\delta)z,x(\delta)z}|^2
-\varepsilon_{y(\delta)z,x(\delta)z} \varepsilon_{x(\delta)z,xy} \varepsilon_{xy,y(\delta)z}
-\varepsilon_{x(\delta)z,y(\delta)z} \varepsilon_{y(\delta)z,xy} \varepsilon_{xy,x(\delta)z} .
\end{align}
\end{subequations}
The off-diagonal matrix elements above, $\varepsilon_{\alpha,\beta}$,
are listed in Tables \ref{so_me_up} and \ref{so_me_dn}.
Following the previous analysis (\ref{a_3}),
the determinant at frequencies $\omega$ approaching the Fermi surface, $\omega_3({\bm k}) = 0$,
is given by $|G^{-1}(\omega)| = a_1 (\omega - \omega_3)$.
And in the present case, 
the $2\times 2$ minor matrix $g^{-1}(3,3)$ has determinant
\begin{equation}
|g^{-1}(3,3)| = [Z\omega - (\varepsilon_{-}+\nu)] [Z\omega - (\varepsilon_{+}-\nu)] 
- |\varepsilon_{y(\delta)z,x(\delta)z}|^2.
\label{dtrmnnt_mnr_33}
\end{equation}
By Cramer's Rule (\ref{Cramer_Rule}),
the wavefunction renormalization at the Fermi surface for $3d_{xy}$ electrons is then
\begin{equation}
Z_3 = {a_1\over{(\varepsilon_{-}+\nu)(\varepsilon_{+}-\nu) - |\varepsilon_{y(\delta)z,x(\delta)z}|^2}} .
\label{Z_3_SO}
\end{equation}
Now recall that the Fermi surface is determined by the characteristic equation $a_0 = 0$.
Using expression (\ref{polynomial_0}) for the coefficient $a_0$
thereby yields the identity (\ref{e_xy_fs}) to lowest non-trivial order in the mixing with the $3d_{xy}$ orbital,
with the exception that $\varepsilon_{xy}$ is replaced by ${\bar\varepsilon}_{xy}$.
After substituting it into expression (\ref{polynomial_1}) for the coefficient $a_1$,
(\ref{Z_3_SO}) in turn yields expression (\ref{Z_3_fnl}) for the wavefunction renormalization at weak mixing with
the $3 d_{xy}$ orbital, in the regime of large wavefunction renormalization, $Z \gg 1$.
The wavefunction renormalization $Z_3^{-1}$
is therefore positive, and it vanishes as $Z\rightarrow\infty$.

Figure \ref{SFSIII}b shows the Fermi surfaces set by the characteristic equation
$|G^{-1}| = 0$ at $\omega = 0$; i.e., $a_0({\bm k}) = 0$.  The $3d_{xz}/3d_{yz}$ and $3d_{xy}$ bands in isolation
are identical to those shown by Fig. \ref{SFSII}a,
but the momentum $(k_x,k_y)$ instead coincides with the crystal momentum
shifted by the checkerboard wavevector $(\pi/a,\pi/a)$.
Here, the spin-orbit coupling energy is set to $\lambda_{\rm SO} = 14$ meV.
Notice the expected level repulsion shown by the $3d_{xz}/3d_{yz}$ Fermi surface pockets
at $(\pi/a,0)$ and $(0,\pi/a)$  in Fig. \ref{SFSIII}b.
The electronic filling fraction is close to half filling.


\clearpage


\begin{thebibliography}{}

%%\bibitem{new_sc} Y. Kamihara, T. Watanabe, M. Hirano, and H. Hosono,
%%J. Am. Chem. Soc. {\bf 130}, 3296 (2008).

\bibitem{qian_11}
T. Qian, X.-P. Wang, W.-C. Jin, P. Zhang, P. Richard, G. Xu, X. Dai, Z. Fang, J.-G. Guo, X.-L. Chen, H. Ding,
``Absence of a Holelike Fermi Surface for the Iron-Based K$_{0.8}$Fe$_{1.7}$Se$_2$
 Superconductor Revealed by Angle-Resolved Photoemission Spectroscopy'',
 Phys. Rev. Lett. {\bf 106}, 187001 (2011).

\bibitem{xu_12}
M. Xu, Q. Q. Ge, R. Peng, Z. R. Ye, Juan Jiang, F. Chen,
 X. P. Shen, B. P. Xie, Y. Zhang, A. F. Wang, X. F. Wang, X. H. Chen, and D. L. Feng,
``Evidence for an S-Wave Superconducting Gap in K$_x$Fe$_{2-y}$Se$_2$ from Angle-Resolved Photoemission'',
 Phys. Rev. B {\bf 85}, 220504(R) (2012).

\bibitem{zheng_11}
B. Zeng, B. Shen, G. F. Chen, J. B. He, D. M. Wang, C. H. Li, and H. H. Wen,
``Nodeless Superconductivity of Single-Crystalline K$_x$Fe$_{2-y}$Se$_2$
Revealed by the Low-Temperature Specific Heat'',
Phys. Rev. B {\bf 83}, 144511 (2011).

\bibitem{yu_11}
Weiqiang Yu, L. Ma, J. B. He, D. M. Wang, T.-L. Xia, G. F. Chen, and Wei Bao,
``$^{77}$Se NMR Study of the Pairing Symmetry and the Spin Dynamics in K$_y$Fe$_{2-x}$Se$_2$'',
Phys. Rev. Lett. {\bf 106}, 197001 (2011).

\bibitem{xue_12} Q.-Y. Wang, Z. Li, W.-H. Zhang, Z.-C. Zhang, J.-S. Zhang, W. Li,
H. Ding, Y.-B. Ou, P. Deng, K. Chang, J. Wen, C.-L. Song, K. He, J.-F. Jia, S.-H. Ji,
Y. Wang, L. Wang, X. Chen, X. Ma, Q.-K. Xue,
``Interface-Induced High-Temperature Superconductivity in Single Unit-Cell FeSe Films on SrTiO$_3$'',
 Chin. Phys. Lett. {\bf 29}, 037402 (2012).

\bibitem{zhang_14} W.-H. Zhang, Y. Sun, J.-S. Zhang, F.-S. Li, M.-H. Guo, Y.-F. Zhao, H.-M. Zhang,
J.-P. Peng, Y. Xing, H.-C. Wang, T. Fujita, A. Hirata, Z. Li, H. Ding, C.-J. Tang, M. Wang,
Q.-Y. Wang, K. He, S.-H. Ji, X. Chen, J.-F. Wang, Z.-C. Xia, L. Li, Y.-Y. Wang, J. Wang,
L.-L. Wang, M.-W. Chen, Q.-K. Xue, and X.-C. Ma,
``Direct Observation of High-Temperature Superconductivity in One-Unit-Cell FeSe Films'',
 Chin. Phys. Lett. {\bf 31}, 017401 (2014).

\bibitem{deng_14} L.Z. Deng, B. Lv, Z. Wu, Y.Y. Xue, W.H. Zhang, F.S. Li, L.L. Wang, X.C. Ma,
 Q.K. Xue, and C.W. Chu,
``Meissner and Mesoscopic Superconducting States in 1–4 Unit-Cell FeSe Films'',
 Phys. Rev. B {\bf 90}, 214513 (2014).

\bibitem{ge_15} J.-F. Ge, Z.-L. Liu, C. Liu, C.-L. Gao, D. Qian,
Q.-K. Xue, Y. Liu, J.-F. Jia,
``Superconductivity Above 100 K in Single-Layer FeSe Films on Doped SrTiO$_3$'',
 Nat. Mater. {\bf 14}, 285 (2015).

\bibitem{liu_12} D. Liu, W. Zhang, D. Mou, J. He, Y.-B. Ou,Q.-Y. Wang, Z. Li, L. Wang,
L. Zhao, S. He, Y. Peng, X. Liu, C. Chaoyu, L. Yu, G. Liu, X. Dong, J. Zhang, C. Chen,
Z. Xu, J. Hu, X. Chen, Z. Ma, Q. Xue and X.J. Zhou,
``Electronic Origin of High-Temperature Superconductivity in Single-Layer FeSe Superconductor'',
 Nat. Comm. {\bf 3}, 931 (2012).

\bibitem{peng_14} R. Peng, X.P. Shen, X. Xie, H.C. Xu, S.Y. Tan, M. Xia,
T. Zhang, H.Y. Cao, X.G. Gong, J.P. Hu, B.P. Xie, D. L. Feng,
``Measurement of an Enhanced Superconducting Phase and a Pronounced Anisotropy of
 the Energy Gap of a Strained FeSe Single Layer in FeSe/Nb: SrTiO$_3$/KTaO$_3$
 Heterostructures Using Photoemission Spectroscopy'',
Phys. Rev. Lett. {\bf 112}, 107001 (2014).
% arXiv:1310.3060 .

\bibitem{lee_14} J.J. Lee, F.T. Schmitt, R.G. Moore, S. Johnston, Y.-T. Cui, W. Li,
M. Yi, Z.K. Liu, M. Hashimoto, Y. Zhang, D.H. Lu,
T.P. Devereaux, D.-H. Lee and Z.-X. Shen,
``Interfacial Mode Coupling as the Origin of the Enhancement of $T_c$ in FeSe Films on SrTiO$_3$'',
 Nature {\bf 515}, 245 (2014).

\bibitem{fan_15} Q. Fan, W.H. Zhang, X. Liu, Y.J. Yan, M.Q. Ren, R. Peng,
 H.C. Xu, B.P. Xie, J.P. Hu, T. Zhang, and  D.L. Feng,
``Plain S-Wave Superconductivity in Single-Layer FeSe on SrTiO$_3$ Probed by Scanning Tunneling Microscopy'',
 Nat. Phys. {\bf 11}, 946 (2015).

\bibitem{zhao_16} L. Zhao, A. Liang, D. Yuan, Y. Hu, D. Liu, J. Huang,
S. He, B. Shen, Y. Xu, X. Liu, L. Yu, G. Liu, H. Zhou, Y. Huang, X. Dong, 
F. Zhou, Z. Zhao, C. Chen, Z. Xu, X.J. Zhou, 
``Common Electronic Origin of Superconductivity in (Li,Fe)OHFeSe 
Bulk Superconductor and Single-Layer FeSe/SrTiO$_3$ Films'',
Nat. Comm. {\bf 7}, 10608 (2016).
% arXiv:1505.06361 .

\bibitem{niu_15} X.H. Niu, R. Peng, H.C. Xu, Y.J. Yan, J. Jiang, D.F. Xu, T.L. Yu,
Q. Song, Z.C. Huang, Y.X. Wang, B.P. Xie, X.F. Lu, N.Z. Wang, X.H. Chen, Z. Sun,
and D.L. Feng,
``Surface Electronic Structure and Isotropic Superconducting Gap in 
(Li$_{0.8}$Fe$_{0.2}$)OHFeSe'',
 Phys. Rev. B {\bf 92}, 060504(R) (2015).

\bibitem{yan_15} Y.J. Yan, W.H. Zhang, M.Q. Ren, X. Liu, X.F. Lu, N. Z. Wang,
X.H. Niu, Q. Fan, J. Miao, R. Tao, B.P. Xie, X.H. Chen, T. Zhang, D.L. Feng, 
``Surface Electronic Structure and Evidence of Plain S-Wave Superconductivity in (Li$_{0.8}$Fe$_{0.2}$)OHFeSe'',
Phys. Rev. B {\bf 94}, 134502 (2016).
%arXiv:1507.02577 .

%\bibitem{miyata_15} Y. Miyata, K. Nakayama, K. Suawara, T. Sato, and T. Takahashi,
%``High-Temperature Superconductivity in Potassium-Coated Multilayer FeSe Thin Films'',
%Nat. Mater. {\bf 14}, 775 (2015).

%\bibitem{wen_16} C.H.P. Wen, H.C. Xu, C. Chen, Z.C. Huang, X. Lou, Y.J. Pu, Q. Song,
%B.P. Xie, M. Abdel-Hafiez, D.A. Chareev, A.N. Vasiliev, R. Peng, and D.L. Feng,
%``Anomalous Correlation Effects and Unique Phase Diagram of 
%Electron-Doped FeSe Revealed by Photoemission Spectroscopy'',`
%Nat. Comm. {\bf 7}, 10840, (2016).

%\bibitem{lei_16} B. Lei, J.H. Cui, Z.J. Xiang, C. Shang, N.Z. Wang, G.J. Ye, X.G. Luo,
%T. Wu, Z. Sun, and X.H. Chen,
%``Evolution of High-Temperature Superconductivity from a Low-$T_c$
%Phase Tuned by Carrier Concentration in FeSe Thin Flakes'',
% Phys. Rev. Lett. {\bf 116}, 077002 (2016).

%\bibitem{hosono_16} K. Hanzawa, H. Sato, H. Hiramatsu, T. Kamiya, and H. Hosono,
%``Electric Field-Induced Superconducting Transition of Insulating FeSe Thin Film at 35 K'',
%Proc. Nat. Acad. Sci. {\bf 113}, 3986 (2016).

\bibitem{budko_prb_13} S.L. Bud'ko, M. Sturza, D.Y. Chung, M.G. Kanatzidis and P.C. Canfield,
``Heat Capacity Jump at $T_c$ and Pressure Derivatives of Superconducting Transition Temperature in 
the Ba$_{1-x}$K$_x$Fe$_2$As$_2$ ($0.2 \leq x \leq 1.0$) Series'',
 Phys. Rev. B {\bf 87}, 100509(R) (2013).

\bibitem{sato_prl_09} T. Sato, K. Nakayama, Y. Sekiba, P. Richard, Y.-M. Xu, S. Souma,
T. Takahashi, G.F. Chen, J.L. Luo, N.L. Wang and H. Ding,
``Band Structure and Fermi Surface of an Extremely Overdoped Iron-Based Superconductor KFe$_2$As$_2$'',
Phys. Rev. Lett. {\bf 103}, 047002 (2009).

\bibitem{maier_11} T.A. Maier, S. Graser, P.J. Hirschfeld, D.J. Scalapino,
``D-Wave Pairing from Spin Fluctuations in the K$_x$Fe$_{2-y}$Se$_2$ Superconductors'',
 Phys. Rev. B {\bf 83}, 100515(R) (2011).

\bibitem{wang_11} F. Wang, F. Yang, M. Gao, Z.-Y. Lu, T. Xiang, D.-H. Lee,
``The Electron Pairing of K$_x$Fe$_{2-y}$Se$_2$'',
 Europhys. Lett. {\bf 93}, 57003 (2011).

\bibitem{mazin_11} I.I. Mazin,
``Symmetry Analysis of Possible Superconducting States in K$_x$Fe$_y$Se$_2$ Superconductors'',
 Phys. Rev. B {\bf 84}, 024529 (2011).

\bibitem{Lee_Wen_08} P.A. Lee and X.-G. Wen,
``Spin-Triplet P-Wave Pairing in a Three-Orbital Model for Iron Pnictide Superconductors'',
 Phys. Rev. B {\bf 78}, 144517 (2008).

%\bibitem{cvetkovic_vafek_13} V. Cvetkovic and O. Vafek,
%``Space Group Symmetry, Spin-Orbit Coupling, and the Low-Energy Effective Hamiltonian
% for Iron-Based Superconductors'', Phys. Rev. B {\bf 88}, 134510 (2013).

\bibitem{agterberg_17} D.F. Agterberg, T. Shishidou, J. O'Halloran, P.M.R. Brydon, and M. Weinert,
``Resilient Nodeless D-Wave Superconductivity in Monolayer FeSe'',
Phys. Rev. Lett. {\bf 119}, 267001 (2017).

\bibitem{eugenio_vafek_18} P.M. Eugenio and O. Vafek,
``Classification of Symmetry Derived Pairing at the M Point in FeSe'',
Phys. Rev. B {\bf 98}, 014503 (2018).

\bibitem{khodas_chubukov_12} M. Khodas and A.V. Chubukov,
``Interpocket Pairing and Gap Symmetry in Fe-Based Superconductors with Only Electron Pockets'',
 Phys. Rev. Lett. {\bf 108}, 247003 (2012).

\bibitem{BMS_12} E. Berg, M.A. Metlitski and S. Sachdev,
``Sign-Problem-Free Quantum Monte Carlo of the Onset of Antiferromagnetism in Metals'',
 Science {\bf 338}, 1606 (2012).

\bibitem{jpr_rm_18} J.P. Rodriguez and R. Melendrez,
``Fermi Surface Pockets in Electron-Doped Iron Superconductor by Lifshitz Transition'',
 J. Phys. Commun. {\bf 2}, 105011 (2018);
``Corrigendum: Fermi Surface Pockets in Electron-Doped Iron Superconductor by Lifshitz Transition'',
J. Phys. Commun. {\bf 3}, 019501 (2019).

\bibitem{xu_muller_sachdev_08} C. Xu, M. M\"uller, and S. Sachdev,
``Ising and Spin Orders in the Iron-Based Superconductors'',
Phys. Rev. B {\bf 78}, 020501(R) (2008).

\bibitem{kang_fernandez_16} J. Kang and R.M. Fernandes,
``Superconductivity in FeSe Thin Films Driven by the Interplay between Nematic
 Fluctuations and Spin-Orbit Coupling'',
Phys. Rev. Lett. {\bf 117}, 217003 (2016).

\bibitem{jpr_17} J.P. Rodriguez,
``Isotropic Cooper Pairs with Emergent Sign Changes in a Single-Layer Iron Superconductor'',
 Phys. Rev. B {\bf 95}, 134511 (2017).

\bibitem{raghu_08} S. Raghu, Xiao-Liang Qi, Chao-Xing Liu, D.J. Scalapino, Shou-Cheng Zhang,
``Minimal Two-Band Model of the Superconducting Iron Oxypnictides'',
Phys. Rev. B {\bf 77}, 220503(R) (2008).

\bibitem{jpr_mana_pds_14} J.P. Rodriguez, M.A.N. Araujo, P.D. Sacramento,
``Emergent Nesting of the Fermi Surface from
 Local-Moment Description of Iron-Pnictide High-$T_c$ Superconductors'',
Eur. Phys. J. B {\bf 87}, 163 (2014).

\bibitem{2orb_Hbbrd} M. Daghofer, A. Moreo, J.A. Riera, E. Arrigoni, D.J. Scalapino, and E. Dagotto,
``Model for the Magnetic Order and Pairing Channels in Fe Pnictide Superconductors'',
Phys. Rev. Lett. {\bf 101}, 237004 (2008);
A. Moreo, M. Daghofer, J.A. Riera, and E. Dagotto,
``Properties of a Two-Orbital Model for Oxypnictide Superconductors: Magnetic Order,
B$_{2 g}$ Spin-Singlet Pairing Channel, and its Nodal Structure'',
Phys. Rev. B {\bf 79}, 134502 (2009).

\bibitem{riseborough_12} P.S. Riseborough, B. Coqblin, S.G. Magalh\~aes,
``Phase Transition Arising from the Underscreened Anderson Lattice Model: 
A Candidate Concept for Explaining Hidden Order in URu$_2$Si$_2$'',
Phys. Rev. B {\bf 85}, 165116 (2012).

\bibitem{jpr_10} J.P. Rodriguez,
``Magnetic Excitations in Ferropnictide Materials Controlled by a Quantum Critical Point into Hidden Order'',
 Phys. Rev. B {\bf 82}, 014505 (2010).

\bibitem{anderson_52} P.W. Anderson,
``An Approximate Quantum Theory of the Antiferromagnetic Ground State'',
Phys. Rev. {\bf 86}, 694 (1952).

\bibitem{halperin_hohenberg_69} B.I. Halperin and P.C. Hohenberg,
``Hydrodynamic Theory of Spin Waves'',
 Phys. Rev. {\bf 188}, 898 (1969).

\bibitem{forster_75} D. Forster, {\it Hydrodynamic Fluctuations, Broken Symmetry, and Correlation Functions}
 (Benjamin/Cummings), Reading, MA, 1975).

\bibitem{jpr_20a} J.P. Rodriguez,
``Spin Resonances in Iron-Selenide High-$T_c$ Superconductors by Proximity to Hidden Spin Density Wave'',
Phys. Rev. B {\bf 102}, 024521 (2020).
% arXiv:2002.01732 .

\bibitem{eliashberg_60} G.M. Eliashberg,
``Interactions between Electrons and Lattice Vibrations in a Superconductor'',
 Sov. Phys. JETP {\bf 11}, 696 (1960).

\bibitem{eliashberg_61} G.M. Eliashberg,
``Temperature Green's Function for Electrons in a Superconductor'',
 Sov. Phys. JETP {\bf 12}, 1000 (1961).

\bibitem{schrieffer_64} J.R. Schrieffer, {\it Theory of Superconductivity}
(Benjamin, New York, 1964).

\bibitem{scalapino_69} D.J. Scalapino, in {\it Superconductivity}, v. 1, ed. R.D. Parks
(Dekker, New York, 1969).

\bibitem{nambu_60} Y. Nambu,
``Quasi-Particles and Gauge Invariance in the Theory of Superconductivity'',
 Phys. Rev. {\bf 117}, 648 (1960).

\bibitem{gorkov_58} L.P. Gorkov, Zh. Eksperim. i Teor. Fiz. {\bf 34}, 735 1958;
``About the Energy Spectrum of Superconductors'',
Sov. Phys. JETP {\bf 7}, 505 (1958).

\bibitem{mazin_08} I.I. Mazin, D.J. Singh, M.D. Johannes, and M.H. Du,
``Unconventional Superconductivity with a Sign Reversal in the Order Parameter of LaFeAsO$_{1-x}$F$_x$'',
Phys. Rev. Lett. {\bf 101}, 057003 (2008).

\bibitem{kuroki_08} K. Kuroki, S. Onari, R. Arita, H. Usui, Y. Tanaka, H. Kontani, and H. Aoki,
``Unconventional Pairing Originating from the Disconnected Fermi Surfaces of Superconducting
LaFeAsO$_{1-x}$F$_x$'',
Phys. Rev. Lett. {\bf 101}, 087004 (2008).

\bibitem{graser_09} S. Graser, T.A. Maier, P.J. Hirschfeld, and D.J. Scalapino,
``Near-Degeneracy of Several Pairing Channels in Multiorbital Models for the Fe Pnictides'',
New J. Phys. {\bf 11}, 025016 (2009).

\bibitem{linscheid_16} A. Linscheid, S. Maiti, Y. Wang, S. Johnston, and P.J. Hirschfeld,
``High-$T_c$ via Spin Fluctuations from Incipient Bands: 
Application to Monolayers and Intercalates of FeSe'',
Phys. Rev. Lett. {\bf 117}, 077003 (2016).

\bibitem{morel_anderson_62} P. Morel and P.W. Anderson,
``Calculation of the Superconducting State Parameters with Retarded Electron-Phonon Interaction'',
 Phys. Rev. {\bf 125}, 1263 (1962).

\bibitem{mcmillan_68} W.L. McMillan,
``Transition Temperature of Strong-Coupled Superconductors'',
 Phys. Rev. {\bf 167}, 331 (1968).

\bibitem{ginzburg_kirzhnits_82} {\it High-Temperature Superconductivity}, 
edited by V.L. Ginzburg and D.A. Kirzhnits (Consultants Bureau, New York, 1982).

\bibitem{carbotte_90} J.P. Carbotte,
``Properties of Boson-Exchange Superconductors'',
 Rev. Mod. Phys. {\bf 62}, 1027 (1990).

\bibitem{pan_17}
 B. Pan, Y. Shen, D. Hu, Y. Feng, J.T. Park, A.D. Christianson, Q. Wang, Y. Hao, H. Wo, Z. Yin,
 T.A. Maier and J. Zhao,
``Structure of Spin Excitations in Heavily Electron-Doped Li$_{0.8}$Fe$_{0.2}$ODFeSe Superconductors'',
 Nat. Commun. {\bf 8}, 123 (2017).

\bibitem{yi_15} M. Yi, Z-K Liu, Y. Zhang, R. Yu, J.-X. Zhu, J.J. Lee, R.G. Moore,
F.T. Schmitt, W. Li, S.C. Riggs, J.-H. Chu, B. Lv, J. Hu, M. Hashimoto, S.-K. Mo,
Z. Hussain, Z.Q. Mao, C.W. Chu, I.R. Fisher, Q. Si, Z.-X. Shen, and D.H. Lu,
``Observation of Universal Strong Orbital-Dependent Correlation Effects in Iron Chalcogenides'',
Nat. Comm. {\bf 6}, 7777 (2015).

\bibitem{Yu_Si_13}
R. Yu and Q. Si,
``Orbital-Selective Mott Phase in Multiorbital Models for Alkaline Iron Selenides K$_{1-x}$Fe$_{2-y}$Se$_2$'',
Phys. Rev. Lett. {\bf 110}, 146402 (2013).

%\bibitem{xu_song_wang_17} M. Xu, X. Song and H. Wang,
%``Substrate and Band Bending Effects on Monolayer FeSe on SrTiO$_3$(001)'',
%Phys. Chem. Chem. Phys. {\bf 19}, 7964 (2017).

\bibitem{sup_mat} See Supplemental Material:
 I. Add $3d_{xy}$ Orbital to Eliashberg Theory with $3d_{xz}/3d_{yz}$ Orbitals;
II. Effects and Properties of $3 d_{xy}$ Orbital;
III. Spin-Orbit Coupling.

\bibitem{dolgov_09} O.V. Dolgov, I.I. Mazin, D. Parker, A.A. Golubov,
``Interband Superconductivity: Contrasts between BCS and Eliashberg Theory'',
Phys. Rev. B{\bf 79}, 060502(R) (2009);
O.V. Dolgov, I.I. Mazin, D. Parker, A.A. Golubov,
``Erratum: Interband Superconductivity: Contrasts between Bardeen-Cooper-Schrieffer and
 Eliashberg Theories [Phys. Rev. B 79, 060502(R) (2009)]'',
Phys. Rev. B {\bf 80}, 219901(E) (2009).

\bibitem{benfatto_09} L. Benfatto, E. Cappelluti, and C. Castellani,
``Spectroscopic and Thermodynamic Properties in a Four-Band Model for Pnictides'',
Phys. Rev. B {\bf 80}, 214522 (2009).

\bibitem{ummarino_09} G. A. Ummarino, M. Tortello, D. Daghero, and R. S. Gonnelli
``Three-Band s$\pm$ Eliashberg Theory and the Superconducting Gaps of Iron Pnictides''
Phys. Rev. B {\bf 80}, 172503 (2009).


\end{thebibliography}

\begin{thebibliography}{}

\bibitem{S_Lee_Wen_08} P.A. Lee and X.-G. Wen,
``Spin-Triplet P-Wave Pairing in a Three-Orbital Model for Iron Pnictide Superconductors'',
 Phys. Rev. B {\bf 78}, 144517 (2008).

\bibitem{S_maier_11} T.A. Maier, S. Graser, P.J. Hirschfeld, D.J. Scalapino,
``D-Wave Pairing from Spin Fluctuations in the K$_x$Fe$_{2-y}$Se$_2$ Superconductors'',
 Phys. Rev. B {\bf 83}, 100515(R) (2011).

\bibitem{S_mazin_11} I.I. Mazin,
``Symmetry Analysis of Possible Superconducting States in K$_x$Fe$_y$Se$_2$ Superconductors'',
 Phys. Rev. B {\bf 84}, 024529 (2011).

%\bibitem{S_cvetkovic_vafek_13} V. Cvetkovic and O. Vafek,
%``Space Group Symmetry, Spin-Orbit Coupling, and the Low-Energy Effective Hamiltonian
% for Iron-Based Superconductors'', Phys. Rev. B {\bf 88}, 134510 (2013).

\bibitem{S_agterberg_17} D.F. Agterberg, T. Shishidou, J. O'Halloran, P.M.R. Brydon, and M. Weinert,
``Resilient Nodeless D-Wave Superconductivity in Monolayer FeSe'',
Phys. Rev. Lett. {\bf 119}, 267001 (2017).

\bibitem{S_eugenio_vafek_18} P.M. Eugenio and O. Vafek,
``Classification of Symmetry Derived Pairing at the M Point in FeSe'',
Phys. Rev. B {\bf 98}, 014503 (2018).

\end{thebibliography}
\end{document}